\documentclass[
aps,%
11pt,%
final,%
notitlepage,%
oneside,%
twocolumn,%
nobibnotes,%
nofootinbib,%
superscriptaddress,%
noshowpacs,%
centertags]%
{revtex4}

\begin{document}

\title{On Possibility of Detection of Variable Sources Using
the Data \\ of  ``Cold'' Surveys Carried Out on RATAN-600}

\author{\firstname{E.~K.}~\surname{Majorova}}
\affiliation{\saoname}

\author{\firstname{O.~P.}~\surname{Zhelenkova}}
\affiliation{\saoname}


\begin{abstract}
In this study we attempt to assess the possibility of detection of
variable sources using the data of the 7.6-cm wavelength surveys
carried out on the RATAN-600 radio telescope in the period from
1980 through 1994. Objects selected according to certain criteria
from the RCR catalog are used to construct the calibration curves
and to estimate the accuracy of the resulting calibration curves
and determine the r.m.s. errors for the measured source flux
densities. To check the calibration sources for the presence of
variable objects, quantitative estimates are performed for a
number of parameters that characterize variability, in particular,
for the long-term variability index $V$ and the $\chi^2$
(chi-square) probability $p$. The long-term variability index was
found to be positive for 14 out of approximately 80 calibration
sources, possibly indicating that these sources are variable. The
most likely candidate variables are the three sources with the
\mbox{$\chi^2$ probability $ p> 0.95 $}. Five sources have
$\chi^2$ probabilities in the \mbox {$ 0.85 <p <0.95 $} interval,
and the remaining six in the \mbox {$ 0.6 <p <0.8 $} interval.
Nine out of 14 objects are possibly variable in the optical range.
The light curves and spectra are determined for possible variable
sources and a number of ``non-variable'' objects. We plan to use
the results of this study in our future searches for variable
radio sources using the data of the ``Cold'' surveys.
\end{abstract}

\maketitle

\section{INTRODUCTION}

The problem of searching for variability of cosmic objects was
already formulated during the preparation phase of the first deep
search surveys on the \mbox{RATAN-600} radio telescope, namely the
``Cold''~\cite{h:Majorova_n} and Zelenchuk
surveys~\cite{a0:Majorova_n}. The samples of radio sources
obtained as a result of the Zelenchuk survey at  3.9 and
7.5~GHz~\mbox{\cite{a1:Majorova_n,a2:Majorova_n,a3:Majorova_n}}
formed the basis for the first studies of variable sources on
\mbox{RATAN-600}. The results of the analysis of their statistical
properties can be found in~\cite{g1:Majorova_n}.

Starting from 1998, long-term sets of multifrequency observations
have been carried out on the Northern sector of the radio
telescope to study variable objects. The duration of continuous
daily observations of the same sources ranged from one to three
months. These studies targeted mostly discrete bright radio
sources with flat spectra. Such sources exhibit variations on time
scales ranging from tens of minutes to several decades. The
results of these long-term studies were reported in many
publications by the researchers from the Sternberg Astronomical
Institute, the Special Astrophysical Observatory of the Russian
Academy of Sciences, and the Astro Space Center of the Lebedev
Physical Institute of the Russian Academy of Sciences,
e.g.~\cite{k1:Majorova_n,k2:Majorova_n,g2:Majorova_n,g3:Majorova_n,g4:Majorova_n,s1:Majorova_n,V:Majorova_n}.

In this paper we analyze the possibility of discovery of variable
radio sources based on the data of the deep surveys carried out on
the Northern sector of the RATAN-600 radio telescope from 1980
through 1999.

To this end, we use a sample of calibration sources selected by
certain criteria to construct the calibration curves and perform
detailed estimates of the flux density measurement errors.

We use several criteria, including statistical ones, for
quantitative estimates of the possible variability of the objects
studied, and construct light curves for suspected variable sources
and a number of  ``non-variable'' objects.

\section{DEEP SURVEYS ON RATAN-600}

In 1980 the first 3.94~GHz deep blind survey was performed on the
Northern sector of \mbox{RATAN-600} within the framework of the
``Cold'' experi-\linebreak
ment~\mbox{\cite{h:Majorova_n,pa2:Majorova_n}} at the declination
of the SS\,433 source. Practically at the same time the
multifrequency  Zelenchuk
survey~\cite{a1:Majorova_n,a2:Majorova_n} was carried out with a
flat reflector on the Southern sector.

Starting from 1998 the multiwavelength \linebreak \mbox{($\lambda
= 1$--$55$~cm)} RZF zenith
survey~\cite{p1:Majorova_n,kbu:Majorova_n} was carried out on the
Northern sector. This survey was carried out in 9 and 17 sections
since 2001 and 2006, respectively.

A radio-source catalog (the RC catalog) with a detection threshold
of $10$~mJy~\cite{p3:Majorova_n,p4:Majorova_n} was produced based
on the data of the ``Cold'' survey. To refine the flux densities
and coordinates of the RC catalog sources, several more observing
runs were carried out on the Northern sector of the radio
telescope  at the same frequency and at the same declination
\mbox{($Dec_{1980}=4^{\circ}57'$)}.

The results of the reduction of these observations were reported
in~\cite{kbu:Majorova_n,bu:Majorova_n,so1:Majorova_n}\footnote{Bursov~\cite{kbu:Majorova_n}
gives a complete bibliography of papers published on the
subject.}. Soboleva et al.~\cite{so2:Majorova_n} reported the
results obtained using newly reduced records of the ``Cold-80''
experiment in the interval of right ascensions \mbox{$7^{\rm h}
\le  RA  < 17^{\rm h}$}. The list of objects found in this strip
and identified with the objects of the NVSS
catalog~\cite{co1:Majorova_n} can be found in the RCR (RATAN Cold
Refined) catalog.\footnote{The spectra of the  RCR catalog sources
are available at \url{http://www.sao.ru/hq/len/RCR/}.}

The reduction of the data of these surveys revealed that the flux
densities of a number of objects vary from one observing run to
another. The authors of the above studies averaged the flux
densities over all the observing runs, since identifying variable
radio sources was not among their tasks. These averaged flux
densities and their errors are reported in the \mbox{RCR
catalog}~\cite{so2:Majorova_n}.

In this paper we try to analyze whether it is possible to discover
variable radio sources in the search surveys.

To solve this problem, we use the data of the \mbox{7.6-cm}
surveys carried out in 1980, 1988, 1993, and 1994 at the
declination of $Dec_{1980}=4^{\circ}57'$  in the \mbox{$7^{\rm h}
\le  RA  < 17^{\rm h}$} strip. The detection thresholds (or the
average $\overline{3\sigma}$ values) in these surveys were equal
to $ 8.0 \pm0.5 $~mJy for the  1980 survey; $ 10.6\pm1.3 $~mJy for
the 1988 survey; $ 10.4\pm3.7$~mJy for the 1993 survey; $
9.6\pm1.2$~mJy for the 1994 survey; $ 13.5\pm5.5$~mJy for the
1994 ($H=51^{\circ}09'$) survey, $11.1\pm2.0$~mJy for the
($H=51^{\circ}22'$) survey\footnote{In 1994 the antenna was set
not only to the declination of  the SS\,433 source, but also to
$4^{'}$ above ($H=51^{\circ}22'$) or below ($H=51^{\circ}09'$)
this declination.}~\cite{so2:Majorova_n} ($H$ is the elevation to
which the antenna was set during the survey).

Here we do not analyze the data of the 1990, 1991, and 1999
surveys carried out at the same wavelength and declination because
of their lower sensitivity. We will return to these surveys later.

The use of surveys to study the variability of radio sources has a
certain advantage due to the fact that in the process of the
survey  the antenna is focused onto a certain elevation $H$
(declination  $Dec_{0}$ of the central survey section) and its
configuration remains practically unchanged during the
observations.

This reduces the errors due to the repositioning of the antenna,
which is especially important for the determination of flux
densities of faint sources. Studies of variable sources carried
out in the mode described by Gorshkov et al.~\cite{V:Majorova_n}
involve repeated repositioning of the antenna to different areas
of the sky.

Another advantage of search surveys is that due to the specificity
of the power beam pattern (PBP) of RATAN-600 its field
simultaneously covers many sources in a single run of the sky
strip.\footnote{More than 30\,000 radio sources cross the PBP of
\mbox{RATAN-600} through the area within the sheet envelope in a
single crossing of the sky at
$\lambda\,7.6$~cm~\cite{co1:Majorova_n}.}

The number of sources crossing the PBP that can be identified in records increases
with the sensitivity of the telescope and integration time. Integration time is determined
by the number of repeated transits of the given sky strip (i.e., the number of scans).

The number of transits of the observed sky strip in the surveys
considered varied from 20 to 35 depending on the survey and hour
of observation. In the programs described by Gorshkov et
al.~\cite{V:Majorova_n} each source was observed three to six
times.

Thus repeated scanning of the same sky strip in the surveys not
only increases the number of objects, but also makes it possible
to study fainter sources compared to the mode described by
Gorshkov et al.~\cite{V:Majorova_n}.

Note that the data of the considered surveys can be used to study
the long-term variability of radio sources on time scales of
several years, which is known to be due to the nonstationary
processes in active galactic nuclei.

\section{SELECTION OF CALIBRATION SOURCES}

\begin{figure}[tb]
\onelinecaptionsfalse
\centerline{
\vbox{
\hbox{
\includegraphics[angle=0,width=0.35\textwidth,clip]{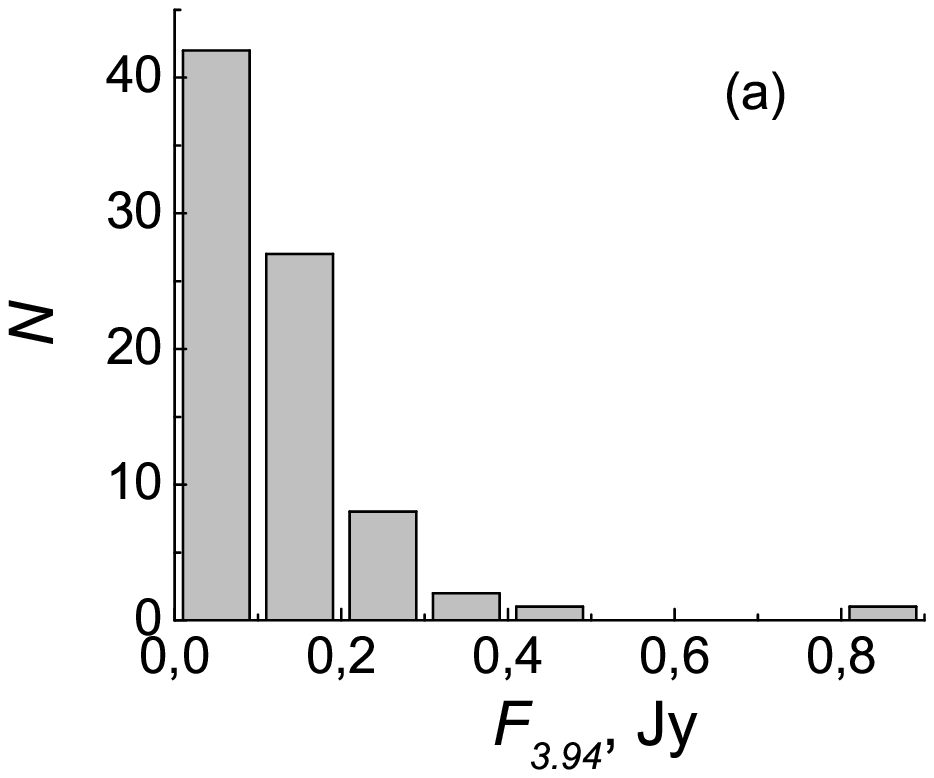}
}
\hbox{
\includegraphics[angle=0,width=0.35\textwidth,clip]{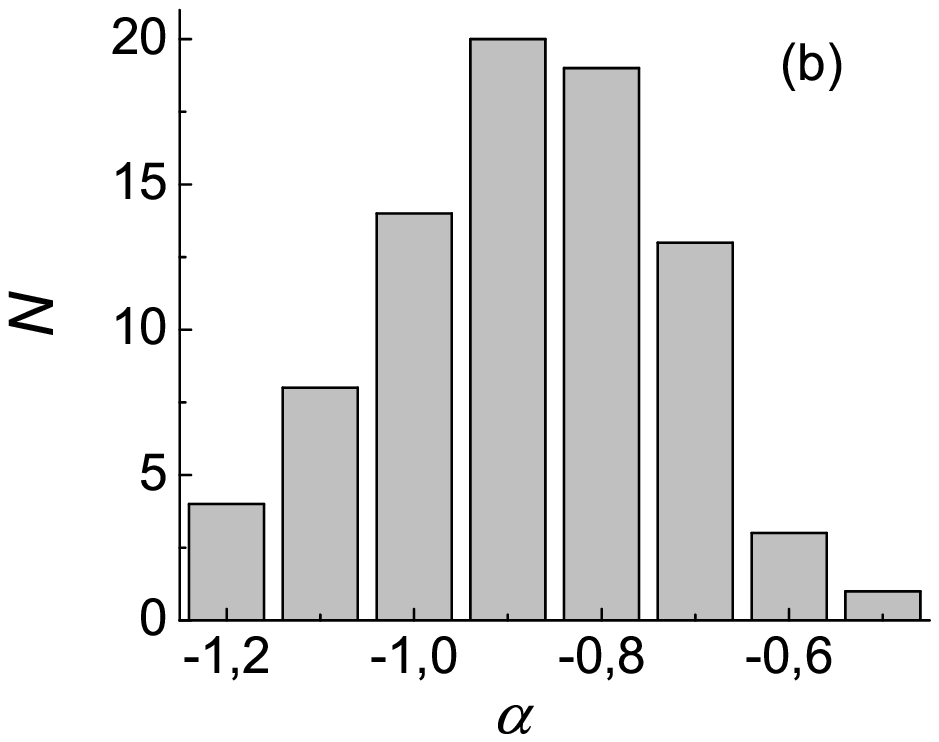}
}
\hbox{
\includegraphics[angle=0,width=0.37\textwidth,clip]{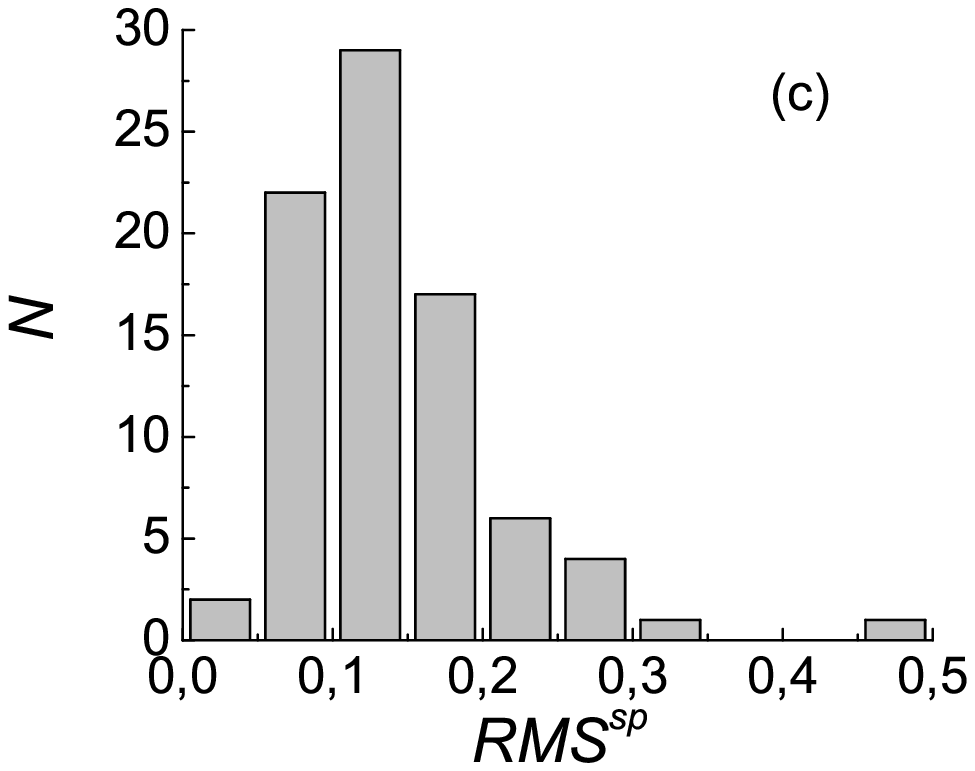}
} } } \setcaptionmargin{0mm} \captionstyle{normal}
\caption{Histograms of the flux densities (a), spectral indices
(b), and the r.m.s. error of the scatter of data points in the
spectra (c) for calibration sources.} \label{fig1:Majorova_n}
\end{figure}

\begin{figure}[bt]
\onelinecaptionsfalse \centerline{ \vbox{
 \hbox{
\includegraphics[angle=0,width=0.37\textwidth,clip]{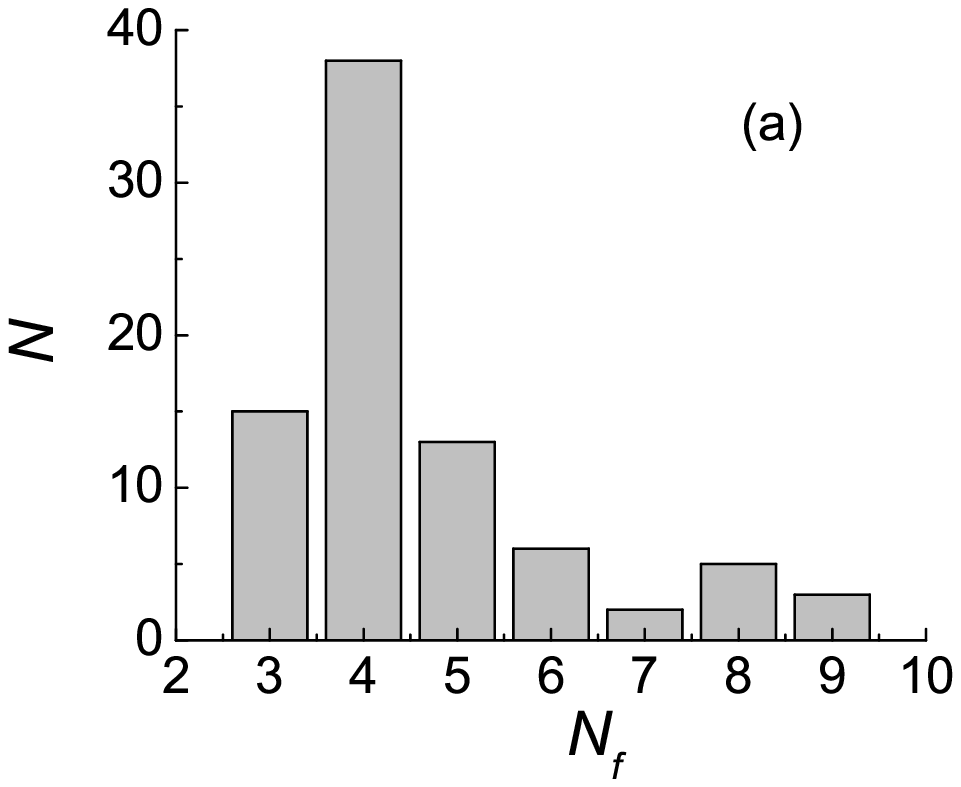}
}
\hbox{
\includegraphics[angle=0,width=0.37\textwidth,clip]{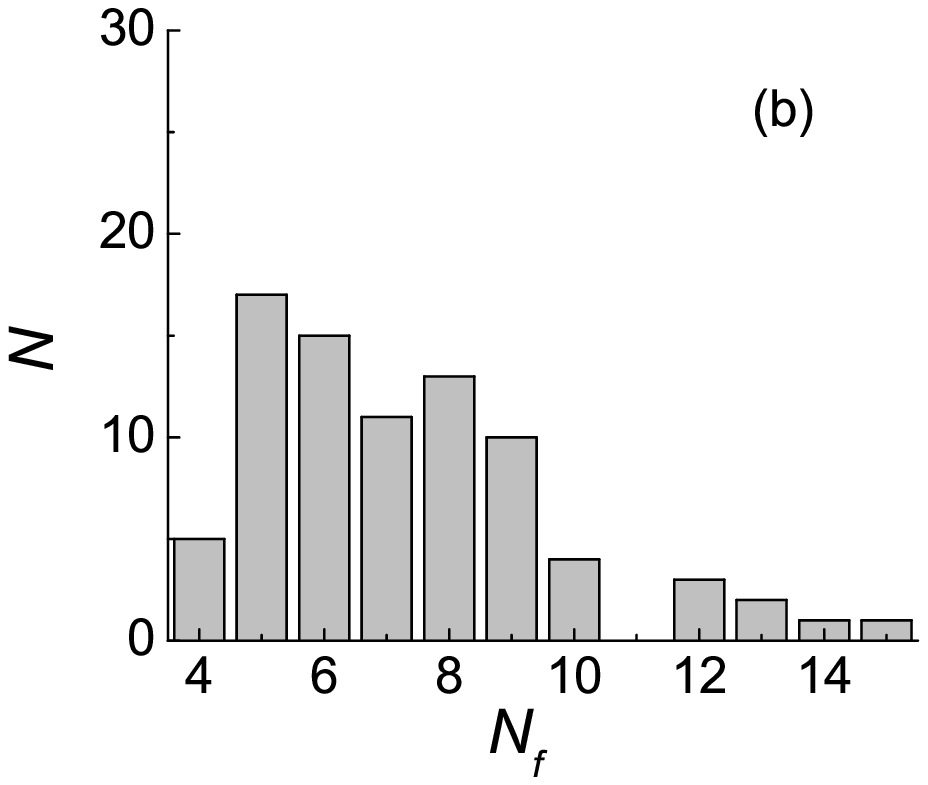}
} } } \setcaptionmargin{0mm} \captionstyle{normal}
\caption{Histograms of the number of frequencies for which the
flux density data are available for the spectra of radio sources.
The left-hand panel shows the distribution based on the NED data
exclusively, and the right-hand panel shows the distribution based
on all the available data collected from different catalogs
including the data of RATAN-600 surveys, and on the estimates
based on the maps of the VLSS  and GB6 surveys. }
\label{fig2:Majorova_n}
\end{figure}

The principal aim of this work is to derive the calibration curves
that can be used to compute the source flux densities and to
estimate the flux density errors.

To derive these calibration curves, we selected RCR radio sources
with steep and well-studied spectra with available flux density
data at several frequencies. We selected sufficiently bright
objects with minimal scatter  of data points in their spectra.

Radio sources with steep spectra seldom exhibit variations at
frequencies greater than 1~GHz. However, such variability is
observed in objects where a compact component is found, which is
responsible for flux density
variations~\cite{Sp:Majorova_n,Al:Majorova_n}. Our sample does not
include known variable sources, which have mostly flat spectra.

We selected a total of 75 sources with flux densities $F_{3.94}>
40$~mJy and six more sources with \linebreak \mbox{$F_{3.94}
\sim30$}~mJy. ($F_{3.94}$ is the flux density at  3.94~GHz.) Note
that the number of calibration sources somewhat changed from one
survey to another.

Figure~\ref{fig1:Majorova_n} shows the histograms of the following
properties of calibration sources: flux densities $F_{3.94}$~(a),
spectral indices $\alpha$ (b), and the relative r.m.s. scatter of
data points, $RMS^{\,sp}$, on their spectra (c).

The r.m.s. (root mean square) error $RMS^{\,sp}$ of the scatter of
data points on the spectrum relative to the approximating curve is
normalized to the 3.94~GHz flux density of the source. We fitted
the approximating curve (or parabola) using the least squares
method.

Most of the selected sources have spectral
indices\footnote{$\alpha_{3.94}$ is the spectral index at
$f=3.94$~GHz.} $\alpha_{3.94} < -0.75$ ($F_{f} \sim f^{\alpha}$)
and r.m.s. errors of the scatter of data points on the spectrum
\mbox{$RMS^{\,sp} < 20\%$}.

The average  $RMS^{\,sp}$ value for the entire sample of
calibration sources was
\mbox{$\overline{RMS^{\,sp}}=0.12\pm0.06$}. According to the data
of the used catalogs, the source flux density errors at different
frequencies lie in the interval from 6\% to 28\%. The average flux
density error for the entire sample of calibration sources is
$15\% \pm0.03\%$.

Most of the calibration sources appear double on the FIRST radio
maps, and a minor fraction of them are point sources, identified
both with galaxies and quasars.

Figure~\ref{fig2:Majorova_n} shows the histograms of the number of
frequencies for which the data on the flux densities in the
spectra of radio sources are available.

The histogram in the left panel takes into account only the data
available in the NED database~\cite{ned:Majorova_n}, and that in
the right panel, all the available data collected from various
catalogs including the RATAN-600 surveys and our
estimates~\cite{so2:Majorova_n} based on VLSS
maps~\cite{cohen:Majorova_n} and GB6
surveys~\cite{gregory:Majorova_n}. It is evident from the
histograms that the NED data for the selected calibration sources
are available at four or more frequencies, and with other catalogs
taken into account, the data coverage increases to include  five
to nine frequencies for the overwhelming majority of the sources.

\section{CONSTRUCTION OF THE CALIBRATING CURVES AND ESTIMATION OF THE SOURCE FLUX DENSITY ERRORS}

\begin{figure*}
\onelinecaptionsfalse \centerline{ \vbox{ \hbox{
\includegraphics[angle=0,width=0.4\textwidth,clip]{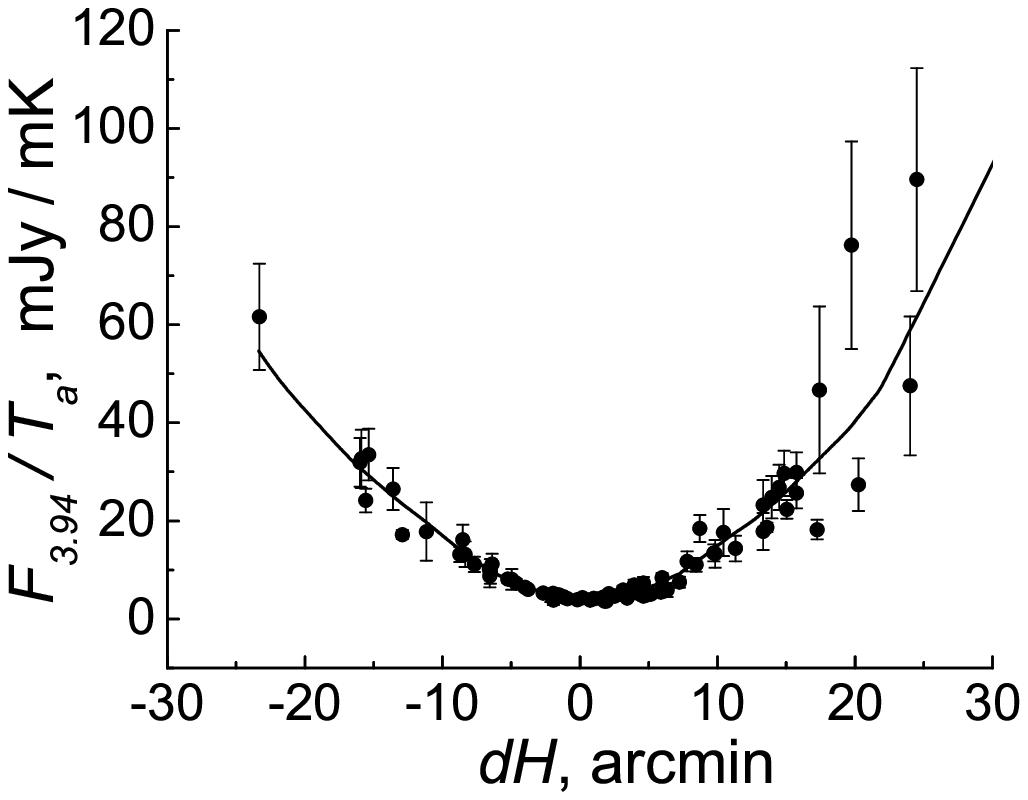}
\includegraphics[angle=0,width=0.4\textwidth,clip]{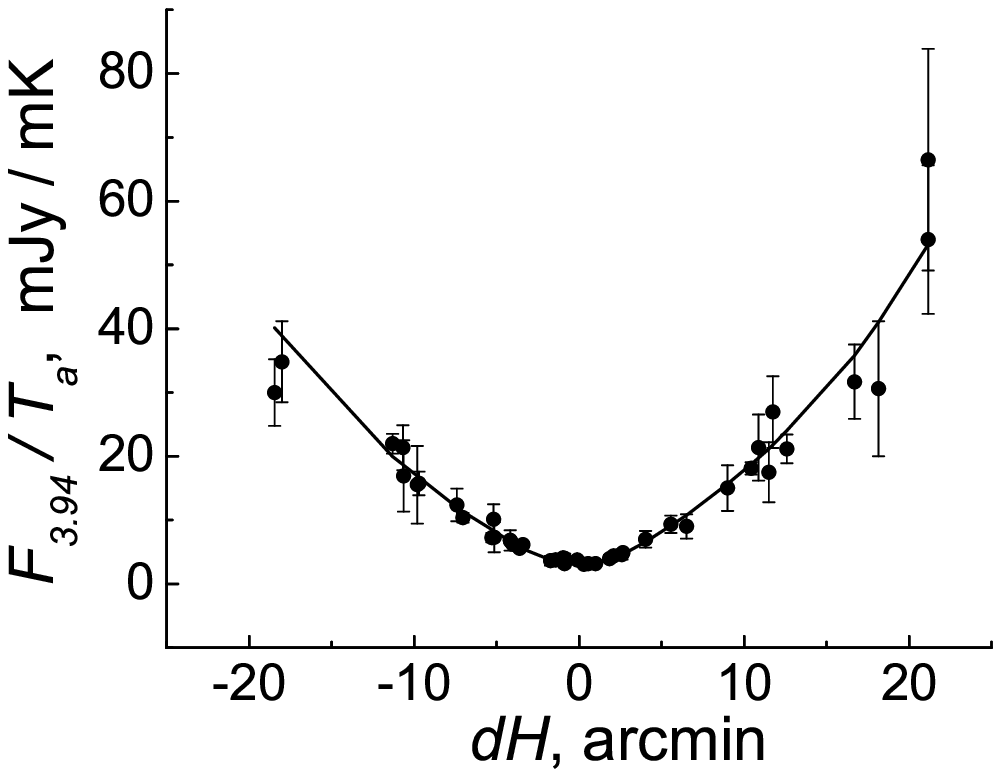}
} \hbox{
\includegraphics[angle=0,width=0.4\textwidth,clip]{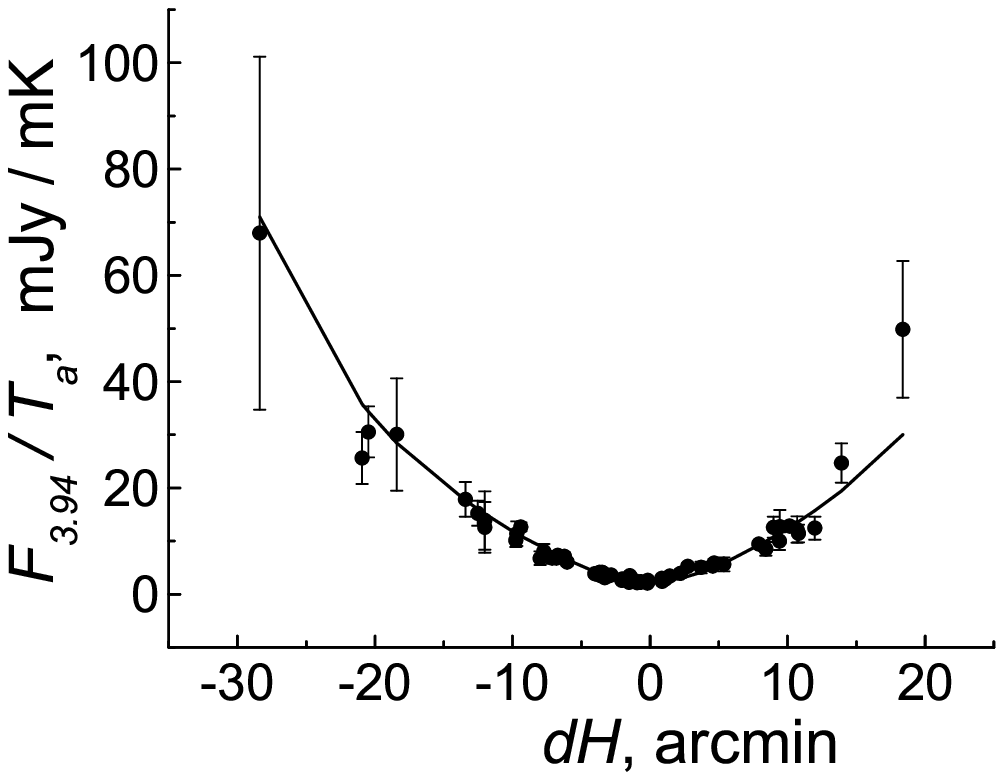}
\includegraphics[angle=0,width=0.4\textwidth,clip]{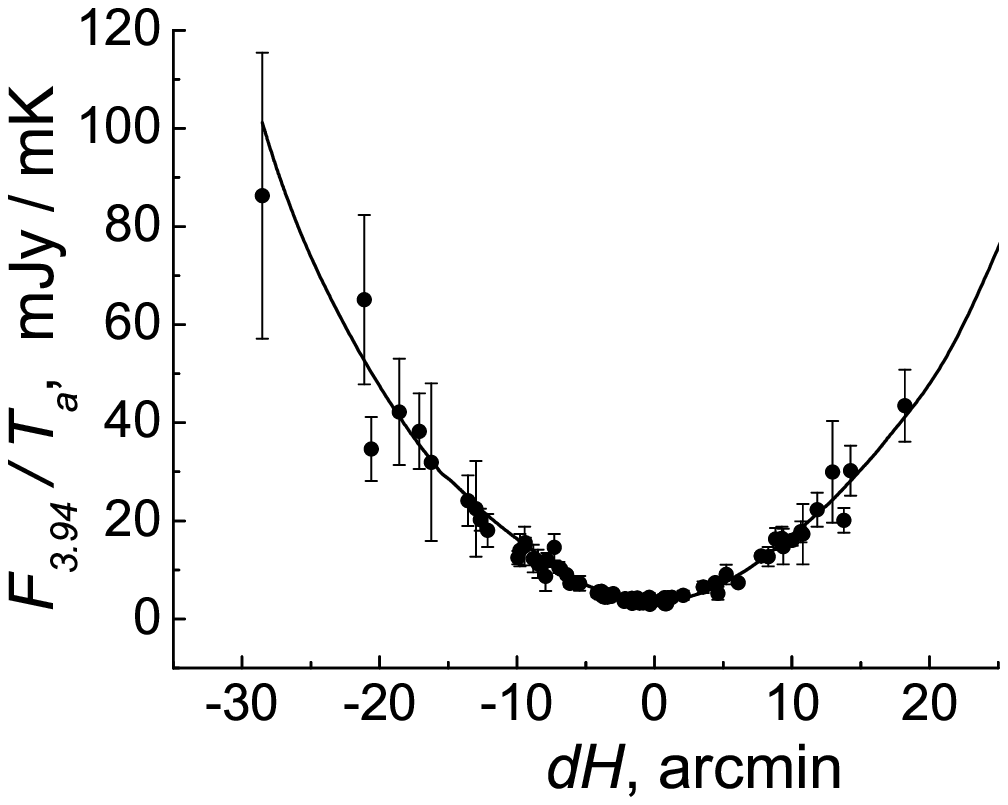}
} } } \setcaptionmargin{0mm} \captionstyle{normal} \caption{
Dependence of the $F_{3.94}/T_{a}$ ratio on $dH$ (circles) based
on the data for the calibration sources from the 1980, 1988, 1993,
and 1994 surveys (from left to right and from top to bottom) and
the computed  $A/k_{P\!B\!P}(dH)$ curves (the solid lines). }
\label{fig3:Majorova_n}
\end{figure*}

Let us recall some of the features of the observations on the
RATAN-600 radio telescope whose PBP differs significantly from
that of a parabolic
dish~\mbox{\cite{e1:Majorova_n,e2:Majorova_n,e3:Majorova_n,m1:Majorova_n,m2:Majorova_n}}.
In the mode of single-sector observations the PBP broadens  with
increasing angular distance from its central section.
Correspondingly, the farther the source is from the central
section, the broader is the response width and the weaker the
signal.

One-dimensional scans are superpositions of the sources that have
crossed different horizontal sections of the power beam pattern.

We repeated the reduction of the selected sources. Our initial
data consisted of the averaged records of several-day long
observations that have already been subjected to primary
reduction~\cite{kbu:Majorova_n}. After background
subtraction\footnote{The background was computed with an 80-s
``smoothing window'' to prevent suppression of the signal from the
sources located far from the central
section~\cite{so2:Majorova_n}.} the sources were identified on the
averaged scans using the Gaussian analysis. We performed the
entire procedure using the standard software for the reduction of
radio astronomical observations~\cite{vo:Majorova_n}.

The temporal calibration was based on the strong sources, with the
use of the data from the \linebreak NVSS catalog. For each source
identified in the record we determined its antenna temperature
${T_{a}}^{i}$, halfwidth $HPBW^{i}$ of the Gaussian fit, and the
right ascension $RA^{i}$.

In our analysis of the data we used the information on the
declination offset $dH$ of the source relative to the central
section of the survey and the computed~\cite{m1:Majorova_n}
$HPBW(dH)$ dependences, where $HPBW$ is the halfwidth of the
vertical PBP, \linebreak \mbox{$dH=\Delta Dec = Dec^{i} -
Dec^{0}$}, $Dec^{i}$ is the declination of the $i$-th source, and
$Dec^{0}$ is the declination of the central section of the survey.
A comparison of the source halfwidths $HPBW^{i}$, determined from
the Gaussian analysis, and the $HPBW(dH)$ dependences tested
experimentally by Majorova and Trushkin~\cite{m2:Majorova_n} and
Majorova and Bursov~\cite{m3:Majorova_n}, allowed us to control
the reliability of the extraction of these objects.

We then constructed for each survey the dependences of
${F_{3.94}}^{i}/{T_{a}}^{i}$ on $dH$. Here ${F_{3.94}}^{i}$ is the
\mbox{3.94-GHz} flux density of the calibration source and
${T_{a}}^{i}$ is its antenna temperature. We determined
${F_{3.94}}^{i}$ from the approximating curve of the spectrum of
the corresponding source, and ${T_{a}}^{i}$ from the Gaussian
analysis of the averaged survey record. The circles in
Fig.~\ref{fig3:Majorova_n} show the ${F_{3.94}}^{i}/{T_{a}}^{i}$
ratios based on the data of the 1980, 1988, 1993, and 1994 surveys
(from left to right and from top to bottom). The solid lines show
the computed calibrating curves $A/k_{P\!B\!P}(dH)$, where
$k_{P\!B\!P}(dH)$ is the pattern factor. It is equal to the
vertical PBP  $F_{v}$ of the telescope if the primary feed is
located at the focus of the antenna, or to the dependence of the
maximum value of the PBP at different horizontal sections on the
offset of this section relative to the central section in the case
of nonzero transversal off-focus offset of the feed.

We computed the pattern factor  $k_{P\!B\!P}(dH)$ for each survey
using the algorithms described by Majorova~\cite{m1:Majorova_n}.
Its value indicates to what extent the response to the source
weakens with increasing distance from the central section of the
survey (or the central section of the PBP).

We computed the $k_{P\!B\!P}(dH)$ taking into account the
transversal offset of the primary feed (horn). The greatest
off-focus offset of the horn was used during the  ``Cold'' survey
in 1980, and in 1988 the horn was located at the focus of the
antenna.

Unlike  Soboleva et al.~\cite{so2:Majorova_n} and
Majorova~\cite{m4:Majorova_n}, in this paper we compute the
pattern factor  $k_{P\!B\!P}(dH)$ for the  1980 survey taking into
account both the transversal offset of the horn and the horn
offset along the direction making an angle of  $50\degr$ to the
horizon and also a small longitudinal offset. Additional off-focus
offsets were applied in the process of the  ``Cold`` experiment in
order to reduce the noise temperature of the antenna. Taking these
offsets into account in our computations of the $k_{P\!B\!P}(dH)$
factor allowed us to match the computed and experimental data and,
in particular, reveal the roughly $1^{'}$ offset of the
experimental vertical PBP from the corresponding computed
one~\cite{m4:Majorova_n}.

\pagebreak

For each survey we chose the $A$ factor\footnote{The $A$ factor is
equal to the \mbox{$A=2k/S_{e\!f\!f}$} ratio, where $k$ is the
Boltzmann constant and $S_{e\!f\!f}$ is the effective area of the
radio telescope.} that minimized the standard error $RMS^{k}$ of
the scatter of experimental data points
${F_{3.94}}^{i}/{T_{a}}^{i}$ relative to the computed calibrating
curve $A/k_{P\!B\!P}(dH)$.
\begin{eqnarray}
RMS^{k} =
\sqrt{\frac{1}{N}\sum_{i}^N\left.\left(\frac{{F_{3.94}}^{i}/{T_{a}}^{i}-A/k_{P\!B\!P}}{A/k_{P\!B\!P}}\right.\right)^2},
\notag
\end{eqnarray}
where $N$ is the number of sources used to construct the calibrating curve for the survey considered.

In this study we somewhat deviated from the technique used by
Bursov~\cite{bu:Majorova_n}, Soboleva et
al.~\cite{so2:Majorova_n}, and Majorova and
Bursov~\cite{m3:Majorova_n}. In those papers the calibrating
curves are the curves fitted to the experimental
${F_{3.94}}^{i}/{T_{a}}^{i}$ data points using the least squares
method.

Majorova~\cite{m4:Majorova_n} showed that the experimental
vertical PBPs  $F_{v}(dH)= k_{P\!B\!P}(dH) =
A/{F_{3.94}}^{i}/{T_{a}}^{i}$ of the radio telescope based on the
data of the 1980--1999 surveys agree well with the computed PBPs.
We therefore used the $A/k_{P\!B\!P}$ ratio as the calibration
curve. We computed the $k_{P\!B\!P}$ factor taking into account
the observing conditions and chose the $A$ factor that would
minimize the $RMS^{k}$. We found that in this case the $RMS^{k}$
error averaged over the entire range of $dH$ is smaller than the
error of the scatter of experimental data points relative to the
least-squares fitted curve  (a second- or fourth-order
polynomial).

A comparison of the calibrating curves derived in this study for
the 1988 survey and those by Bursov~\cite{bu:Majorova_n} shows
that they practically coincide in the $-10^{'} < dH < 10^{'}$
interval. The curves diverge at greater absolute values of $dH$
and at $dH\sim20^{'}$ the ${F_{3.94}}^{i}/{T_{a}}^{i}$ ratios of
Bursov~\cite{bu:Majorova_n} exceed our estimates by a factor of
1.4.

This may be due both to the set of calibration sources and to the
adopted reduction technique, in particular, the background
computation and its subtraction. The latter factor is especially
critical for estimates of the parameters of the sources located
far from the central section. Subtraction of the background
computed with a  ``smoothing window'' of about 20~s results in
underestimated antenna temperatures for distant sources and,
consequently, in the increase of the ${F_{3.94}}^{i}/{T_{a}}^{i}$
ratio with increasing of $dH$.

There was yet another reason why we used the computed
$A/k_{P\!B\!P}(dH)$ dependences instead of the approximating
curves: it was done to avoid the  influence of variable sources,
which may happen to be among the calibration sources.

Table~1 lists the average relative standard errors $RMS^{k}$
($\overline{RMS^{k}}$) computed for the 1980, 1988, 1993, and 1994
surveys. We performed averaging over the entire range of the $dH$
angles considered and the intervals \mbox{$ -15^{'} < dH <
15^{'}$}, \mbox{$-10^{'} < dH < 10^{'}$}, and  \mbox{$-5^{'} < dH
< 5^{'}$}. The $\overline{RMS^{k}}$ values averaged over the
sample of sources whose recorded antenna temperatures exceed
$10\sigma_{s}$ are also listed in the table.

Note that the $\overline{RMS^{k}}$ values for the entire range of
the $dH$ angles are close to the standard error of the scatter of
data points of the experimental PBP relative to the computed PBP
obtained by Majorova~\cite{m4:Majorova_n} for a sample of sources
with flux densities $F_{3.94} > 50$~mJy. The $\overline{RMS^{k}}$
value for the  1980 survey is smaller than the estimate reported
by Majorova~\cite{m4:Majorova_n}, which can be explained by the
allowance for additional off-focus offsets of the horn in the
computation of the pattern factor $k_{P\!B\!P}(dH)$.

We estimated the relative standard errors of the
${F_{3.94}}^{i}/{T_{a}}^{i}$ ratio and its confidence intervals
(Fig.~\ref{fig3:Majorova_n}) using the relative standard errors of
the scatter of data points, $RMS^{\,sp}$, in the spectra of
sources and the relative standard errors
$RMS^{Ta}=\sigma_{s}/{T_{a}}^{i}$ of the inferred antenna
temperatures, where $\sigma_{s}$ is the dispersion of noise in the
sky strip transit records in the considered survey.

\begin{figure}[tb]
\onelinecaptionsfalse
\centerline{
\vbox{
\hbox{
\includegraphics[angle=0,width=0.37\textwidth,clip]{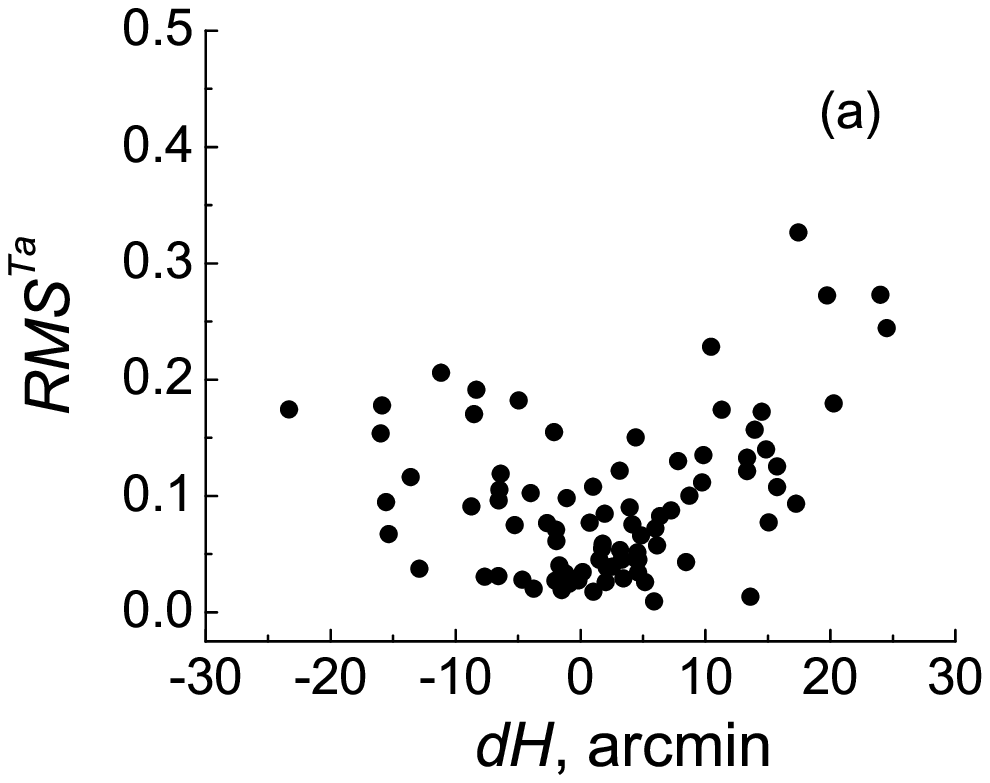}
}
\hbox{
\includegraphics[angle=0,width=0.37\textwidth,clip]{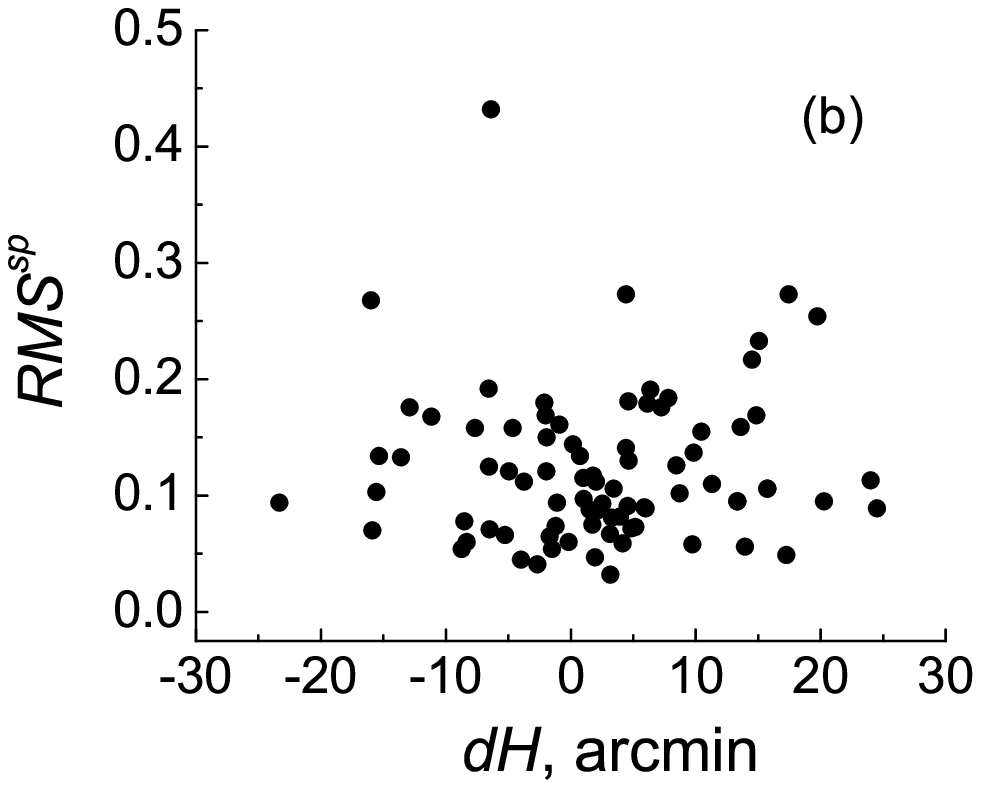}
}
\hbox{
\includegraphics[angle=0,width=0.37\textwidth,clip]{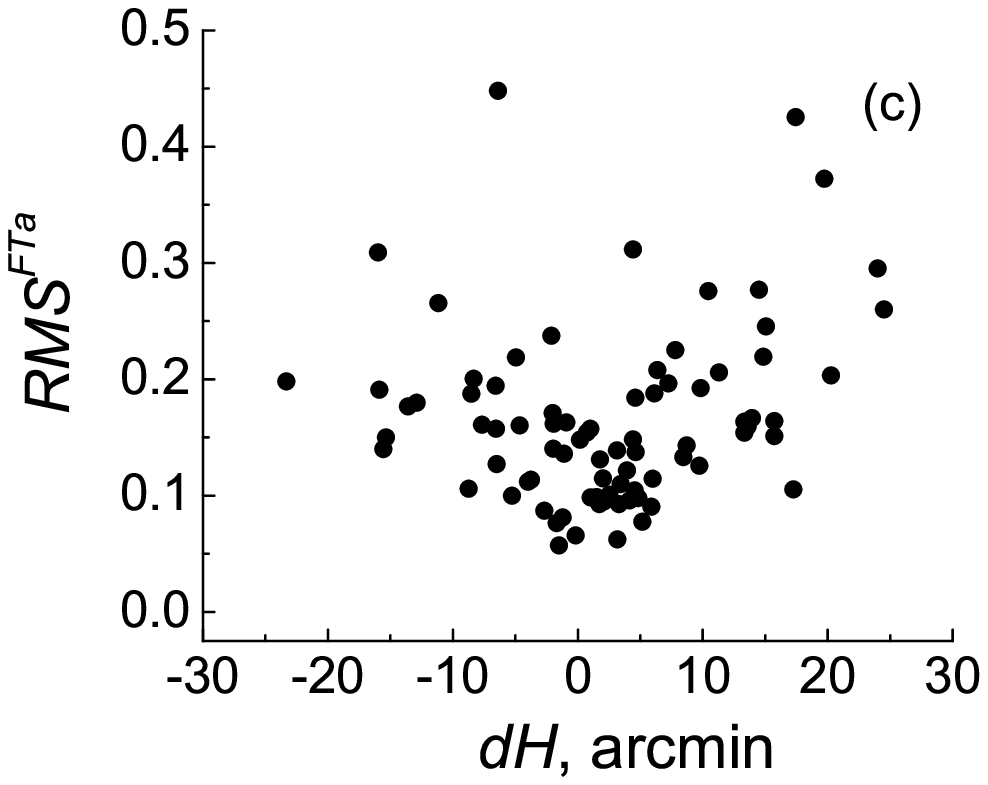}
} } } \setcaptionmargin{0mm} \captionstyle{normal} \caption{ The
$RMS^{Ta}(dH)$ (a), $RMS^{\rm \,sp}(dH)$ (b), and $RMS^{FTa}(dH)$
(c) dependences for the calibration sources based on the data of
the 1980 survey. } \label{fig4:Majorova_n}
\end{figure}

Note that if the standard errors $RMS^{\,sp}$  of the scatter of
data points are sufficiently uniformly distributed with respect to
angle $dH$ then the errors $RMS^{Ta}$ of the antenna temperatures
depend significantly on the distance of the source from the
central section of the survey. We illustrate this point in
Fig.~\ref{fig4:Majorova_n}, where we show the dependences of
$RMS^{Ta}$ on $dH$ (a) and $RMS^{\,sp}$ on $dH$ (b) based on the
sample of calibration sources observed in the 1980 survey. The
relative standard errors of the ${F_{3.94}}^{i}/{T_{a}}^{i}$ ratio
($RMS^{FTa}$) also increase with increasing of the angle $dH$
absolute value (Fig.~\ref{fig4:Majorova_n}c). Table~2 lists the
averaged relative standard errors of the
${F_{3.94}}^{i}/{T_{a}}^{i}$ ratio. We averaged these standard
errors over the  $dH$ intervals indicated in the first column of
the table.

\begin{table}
\setcaptionmargin{0mm} \onelinecaptionsfalse \captionstyle{normal}
\caption{Averaged relative standard errors $RMS^{k}$
($\overline{RMS^{k}}$)}
\medskip
\begin{tabular}{c|c|c|c|c}
\hline
                  & 1980               & 1988            & 1993          & 1994    \\
\hline
 $-30^{'}<dH<30^{'}$ & 0.197           & 0.123           & 0.175         & 0.139              \\
 $-15^{'}<dH<15^{'}$ & 0.139           & 0.107           & 0.154         & 0.132              \\
 $-10^{'}<dH<10^{'}$ & 0.141           & 0.100           & 0.153         & 0.132              \\
 $-~5^{'}<dH<~5^{'}$ & 0.109           & 0.078           & 0.164         & 0.135              \\
 $T_{a} > 10\sigma$  & 0.127           & 0.096           & 0.165         & 0.127              \\
\hline
\end{tabular}
\end{table}

\begin{table}
\setcaptionmargin{0mm} \onelinecaptionsfalse \captionstyle{normal}
\caption{Averaged relative standard errors of the
$F_{i}/{T_{a}}_{i}$ ($\overline{RMS^{FTa}}$) ratio}
\medskip
\begin{tabular}{c|c|c|c|c}
\hline
                  & 1980               & 1988           & 1993         & 1994    \\
\hline

 $-30^{'}<dH<30^{'}$ & 0.165           & 0.186          & 0.181        & 0.187              \\
 $-15^{'}<dH<15^{'}$ & 0.156           & 0.178          & 0.171        & 0.181              \\
 $-10^{'}<dH<10^{'}$ & 0.142           & 0.170          & 0.156        & 0.167              \\
 $-~5^{'}<dH<~5^{'}$ & 0.128           & 0.132          & 0.144        & 0.145              \\
 $T_{a} > 10\sigma$  & 0.128           & 0.131          & 0.139        & 0.131              \\
\hline
\end{tabular}
\end{table}

A comparison of the data listed in Tables~1 and 2 shows that the
errors of the scatter of experimental ${F_{3.94}}^{i}/{T_{a}}^{i}$
data points relative to the $A/k_{P\!B\!P}(dH)$
($\overline{RMS^{k}}$) calibrating curve are smaller than or
comparable to the averaged relative standard errors of the
${F_{3.94}}^{i}/{T_{a}}^{i}$ ratio. The only exceptions were the
$\overline{RMS^{k}}$ errors for the 1980 survey averaged over the
entire $dH$ interval and the $\overline{RMS^{k}}$ errors for the
1993 survey averaged over the $-5^{'} < dH < 5^{'}$ interval. In
the case of the 1980 survey the exclusion of the sole source with
the largest deviation from the computed curve, J\,103938+051031,
reduces the $\overline{RMS^{k}}$ to 0.168, which is comparable to
the $\overline{RMS^{FTa}}$ value in the $-30^{'} < dH < 30^{'}$
interval.

Our analysis of the dependences shown in
Fig.~\ref{fig3:Majorova_n} and the data listed in Table~1 lead us
to conclude that  the ${F_{3.94}}^{i}/{T_{a}}^{i}$ ratios of most
of the considered calibration sources are close to the
$A/k_{P\!B\!P}$ values for the corresponding $dH$ angles and the
difference between the two quantities is within  the confidence
interval of the  ${F_{3.94}}^{i}/{T_{a}}^{i}$ ratio. These sources
mostly lie in the \mbox{$dH = \pm15^{'}$--$\pm17^{'}$} interval.

At greater distances of the sources from the central section of
the survey ($|dH|
> 17^{'}$) deviations of the experimental data points from the
$A/k_{P\!B\!P}$ curve increase and so do the errors of the
${F_{3.94}}^{i}/{T_{a}}^{i}$ ratio. This effect is most
conspicuous in the 1980 survey. These deviations may be due to
both the accuracy of identification in records of the sources
located far from the central section, and to the pattern effects.
Neither can we rule out the possibility that our sample may
contain variable objects.

The averaged standard errors of the scatter of experimental data
points relative to the computed  $\overline{RMS^{k}}$ curve are
minimal in the  1988 (for \linebreak \mbox{$-15^{'} < dH <
15^{'}$}) and 1980 (for \mbox{$-5^{'} < dH < 5^{'}$}) surveys:
they are equal to 8\% and 11\%, respectively. In the
\mbox{$-15^{'} < dH < 15^{'}$} interval the $\overline{RMS^{k}}$
errors for the 1980, 1993, and 1994 surveys are equal
approximately to 14\%, 15\%, and 13\%.

\begin{figure*}[tb]
\onelinecaptionsfalse \centerline{ \vbox{ \hbox{
\includegraphics[angle=0,width=0.40\textwidth,clip]{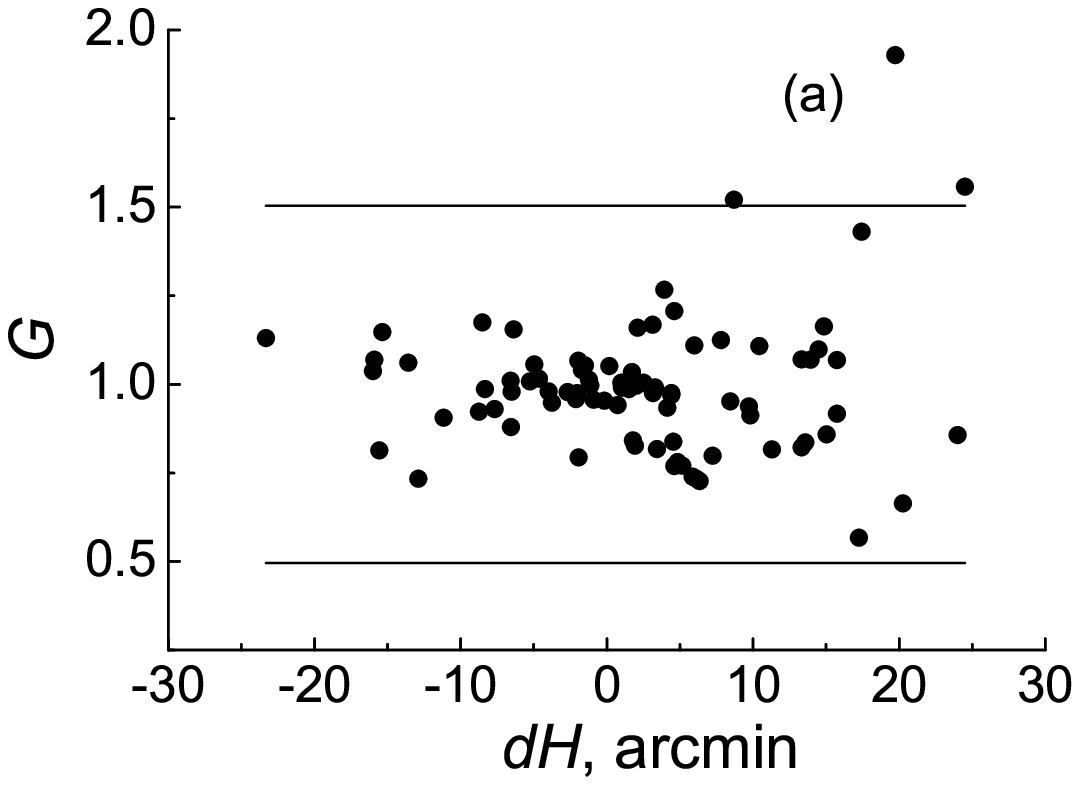}
\includegraphics[angle=0,width=0.40\textwidth,clip]{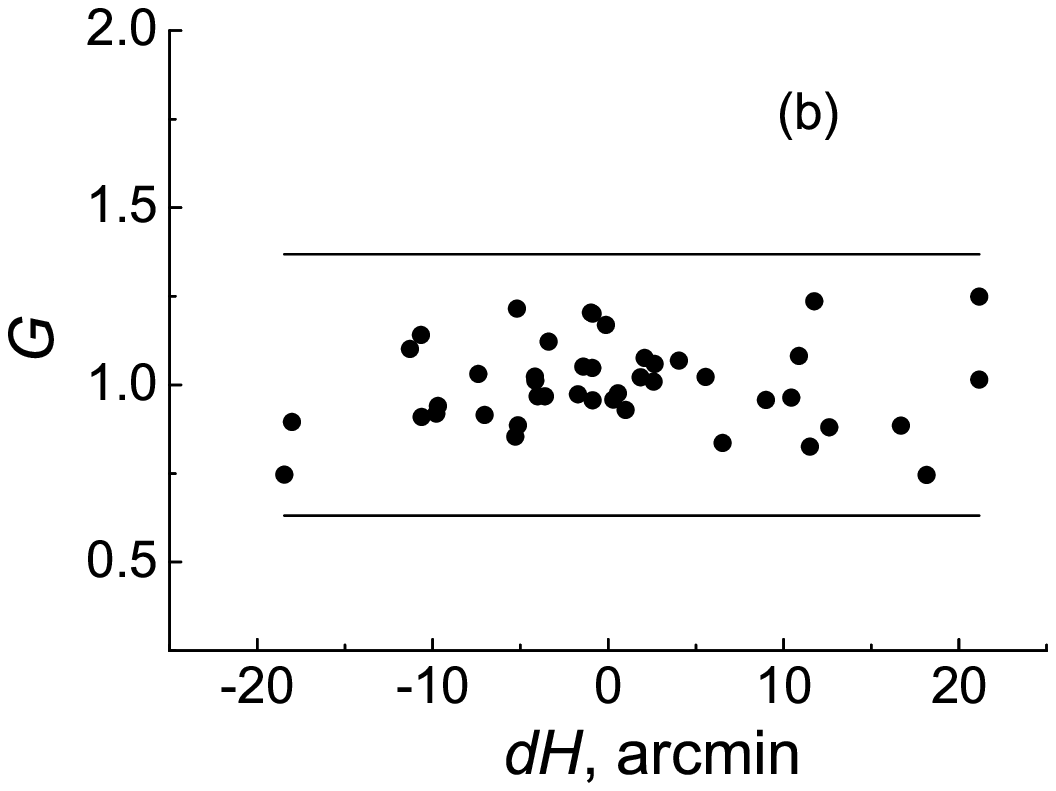}
}
\hbox{
\includegraphics[angle=0,width=0.40\textwidth,clip]{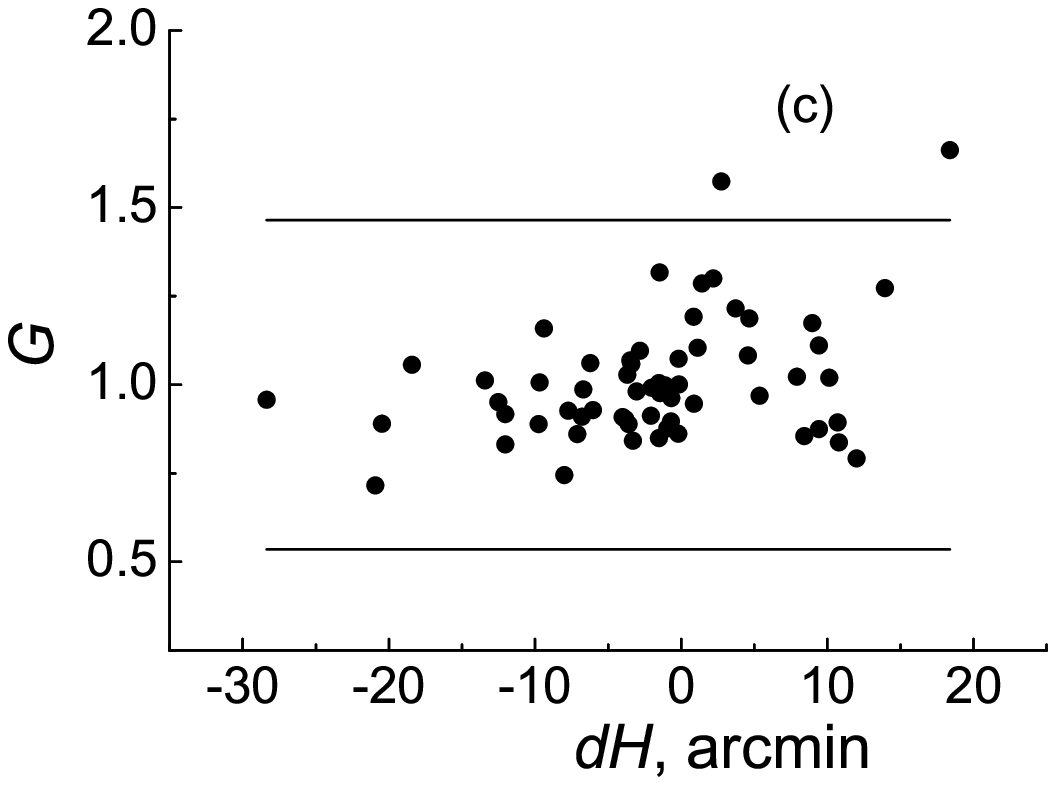}
\includegraphics[angle=0,width=0.40\textwidth,clip]{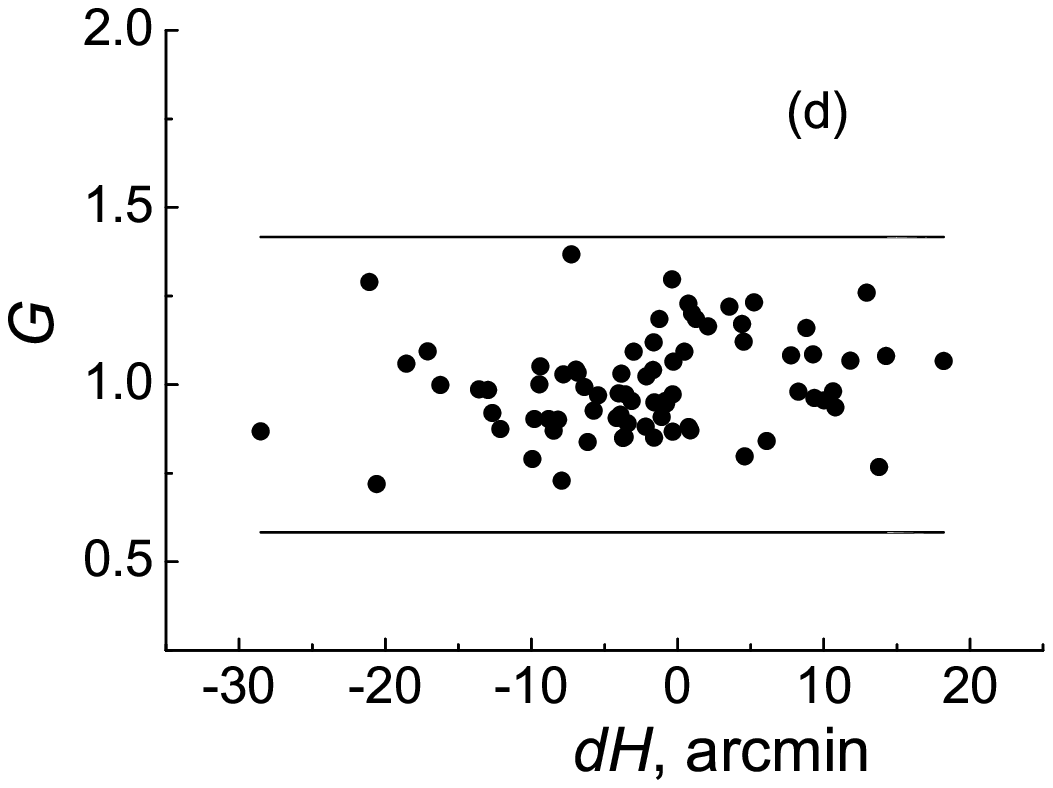}
} } } \setcaptionmargin{0mm} \captionstyle{normal} \caption{
Dependences of the $G = (F_{i}/{T_{a}}_{i})/(A/k_{P\!B\!P})$ ratio
on $dH$ according to the data of the  1980, 1988, 1993, and 1994
surveys (from left to right and from top to bottom). The
horizontal lines correspond to $G = 1.\pm3\overline{RMS^{k}}$. }
\label{fig5:Majorova_n}
\end{figure*}

In conclusion, we show in Fig.~\ref{fig5:Majorova_n} the
dependences of the $G = ({F_{3.94}}^{i}/{T_{a}}^{i})/
(A/k_{P\!B\!P})$ ratio on $dH$ (from left to right and from top to
bottom for the 1980, 1988, 1993, and 1994 surveys, respectively).
The horizontal lines correspond to $G = 1\pm3\overline{RMS^{k}}$.

It is evident from these plots that the deviations of the
experimental data points from the computed curves do not exceed
$3\overline{RMS^{k}}$ except for three points in the 1980 survey
and two points in the 1993 survey. They are sufficiently uniformly
distributed over the entire range of variation of the $dH$ angles.
As we pointed out above, the smallest deviations of the
experimental data points from the computed curves are found in the
1988 survey.

The sources whose ${F_{3.94}}^{i}/{T_{a}}^{i}$ ratios deviate by
more than $\pm3\overline{RMS^{k}}$ from the computed curve may be
variable. These sources are J\,103938+051031, J\,110246+045916,
J\,114220+045459, J\,121852+\linebreak051447, and
J\,142104+050843.

\section{ANALYSIS OF THE SAMPLE OF CALIBRATION SOURCES FOR THE PRESENCE OF VARIABLE SOURCES}

While selecting sources with steep spectra for the construction of
the calibrating curves we tried to reduce the likelihood of
contamination of our sample by variable sources. However, we
cannot completely rule out the presence of such objects in our
list.

\begin{table*}
\setcaptionmargin{0mm} \onelinecaptionstrue \captionstyle{normal}
\caption{The $V, V_{F}$, and $V_{R}$ coefficients}
\medskip
\begin{tabular}{c|@{~}c@{~}|@{~}c@{~}|@{~}c@{~}|r|r|c|r|r|c}
\hline
~~~~~~~$RA_{2000}$~~~~~$DEC_{2000}$ & $V$ & $V_{F}$ & $V_{R}$ & \multicolumn{1}{c|}{$\overline{F}$,} & \multicolumn{1}{c|}{$\sigma^{set}$,} & $RMS^{\,set}$ & \multicolumn{1}{c|}{$dH_{1}$,} & \multicolumn{1}{c|}{$dH_{2}$,} & $\alpha$ \\
RCR & & & & \multicolumn{1}{c|}{mJy} & \multicolumn{1}{c|}{mJy} & & \multicolumn{1}{c|}{arcmin} & \multicolumn{1}{c|}{arcmin} & \\
 (1) & (2) & (3) & (4) & \multicolumn{1}{c|}{(5)} & \multicolumn{1}{c|}{(6)} & (7) & \multicolumn{1}{c|}{(8)} & \multicolumn{1}{c|}{(9)} & (10) \\
\hline
J\,103938.62+051031.3  & 0.264  &3.13&2.53&184 &70  &0.381   &$19.74 $ &$13.78 $ &$-0.68$ \\
J\,155148.09+045930.5  & 0.125  &2.53&1.75&79  &25  &0.312   &$6.13  $ &$2.20  $ &$-1.17$ \\
J\,142104.21+050845.0  & 0.092  &2.33&1.79&183 &53  &0.293   &$17.25 $ &$11.83 $ &$-0.79$ \\
J\,132448.14+045758.8  & 0.088  &1.85&1.50&65  &13  &0.233   &$7.23  $ &$1.26  $ &$-1.03$ \\
J\,135137.56+043542.0  & 0.084  &2.13&1.59&392 &99  &0.262   &$-15.36$ &$-18.46$ &$-0.89$ \\
J\,110246.51+045916.7  & 0.082  &2.18&1.57&102 &26  &0.260   &$5.56  $ &$2.75  $ &$-0.81$ \\
J\,074239.34+050704.3  & 0.077  &2.17&1.56&350 &74  &0.211   &$-9.40 $ &$-12.90$ &$-0.85$ \\
J\,121328.89+050009.9  & 0.076  &1.76&1.51&72  &15  &0.213   &$6.52  $ &$-0.39 $ &$-1.07$ \\
J\,101515.53+045305.6  & 0.061  &2.16&1.41&124 &18  &0.147   &$-0.88 $ &$-3.73 $ &$-1.04$ \\
J\,112437.45+045618.8  & 0.057  &1.95&1.45&466 &84  &0.180   &$5.88  $ &$-0.17 $ &$-0.87$ \\
J\,104551.72+045552.9  & 0.035  &1.51&1.31&157 &17  &0.111   &$5.17  $ &$2.09  $ &$-0.99$ \\
J\,134243.57+050431.5  & 0.008  &1.70&1.35&973 &134 &0.138   &$13.58 $ &$7.76  $ &$-0.72$ \\
J\,140730.77+044934.9  & 0.007  &1.53&1.55&82  &18  &0.216   &$-7.12 $ &$-7.28 $ &$-0.75$ \\
J\,121852.16+051449.4  & 0.007  &1.48&1.72&237 &60  &0.267   &$21.17 $ &$18.38 $ &$-0.67$ \\
\hline
\end{tabular}
\end{table*}

To test our calibration sources for variability, we performed a
number of quantitative estimates and, in particular, estimated the
coefficients $V_{R}$~\cite{VR:Majorova_n},
$V_{F}$~\cite{VF:Majorova_n}, and the long-term variability index
$V$~\cite{V:Majorova_n}.

We computed the coefficients using the following formulas:
\begin{eqnarray}
        V_{R} = F_{i}/F_{j},
\label{2:Majorova_n}
\end{eqnarray}

\begin{eqnarray}
    V_{F} = \frac{F_{i}-F{j}}{\sqrt{(\sigma_{i}^2+\sigma_{j}^2)}},
\label{3:Majorova_n}
\end{eqnarray}

\begin{eqnarray}
    V = \frac{(F_{i}-\sigma_{i})-(F_{j}+\sigma_{j})}{(F_{i}-\sigma_{i})+(F_{j}+\sigma_{j})},
\label{4:Majorova_n}
\end{eqnarray}

\noindent where $F_{i}$ and $F_{j}$ are the flux densities of a
given source measured in cycle $i$ and $j$ surveys, respectively,
and $\sigma_{i}$ and $\sigma_{j}$ are the absolute standard errors
of the inferred flux densities \mbox{($i, j = 80, 88, 93, 94$)}.

The latter two criteria take into account the flux density errors,
and they can therefore be considered to be more reliable for
testing sources for variability.

We computed the flux densities using the following formula:
\begin{eqnarray}
    F = \frac{A}{k_{P\!B\!P}}\,T_{a}.
\label{1:Majorova_n}
\end{eqnarray}

\noindent Here we used the antenna temperatures of the sources
$T_{a}$ determined from the averaged records of the $i$-th year
survey and the corresponding computed $A/k_{P\!B\!P}(dH)$ curves.

We computed the absolute ($\sigma_{i}$) and relative ($RMS_{i}$)
standard errors of the determination of the source flux density in
the $i$-th survey using the following formulas:
\begin{equation}
   RMS_{i}=\sqrt{(RMS^{k})^2+(RMS^{Ta})^2}, \\
\label{5:Majorova_n}
\end{equation}

\begin{equation}
  \sigma_{i} =F_{i}\,RMS_{i}.
\label{6:Majorova_n}
\end{equation}

We computed the $V_{R}$, $V_{F}$, and $V$ coefficients for all the
calibration sources whose flux densities are determined in at
least three surveys. In our computations we used the standard
errors $RMS^{k}$ listed in Table~1. We suspected a source to be
variable if it had a positive long-term variability index ($V>0$).
For such sources the flux density difference determined in
different surveys differed by more than the sum of standard errors
in these surveys.

In the entire sample of calibration sources, 14~objects had a
positive  $V$ index for at least one pair of surveys. Table~3
lists the coefficients $V$, $V_{R}$, and $V_{F}$ for these objects
(columns 2, 3, and 4), their average flux densities $\overline{F}$
(column 5), and the standard deviations $\sigma^{set}$ from the
mean value (column~6).
\begin{equation}
\overline{F}= \frac{1}{n}\sum_{i}^n F_{i}, \\
\label{7:Majorova_n}
\end{equation}
\begin{equation}
\sigma^{set}=\sqrt{\frac{1}{n}\sum_{i}^n(F_{i}-\overline{F})^2},
\label{8:Majorova_n}
\end{equation}
where $n$ is the number of surveys in which the source flux
densities have been determined.

\begin{figure*}[tb]
\onelinecaptionsfalse
\centerline{
\vbox{
\hbox{
\includegraphics[angle=0,width=0.37\textwidth,clip]{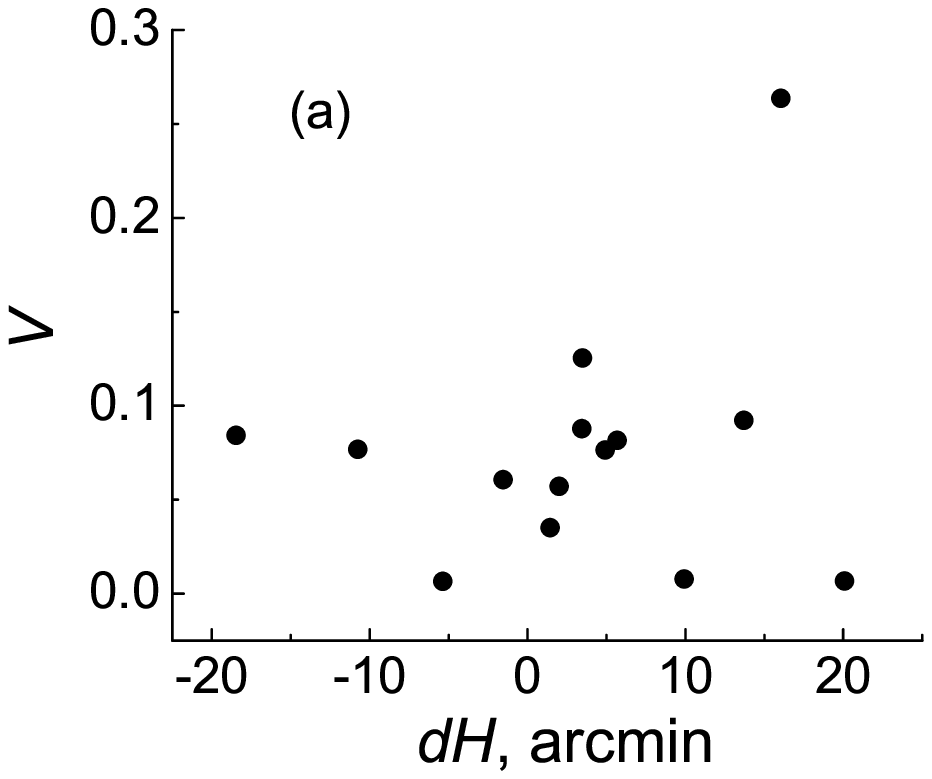}
\includegraphics[angle=0,width=0.37\textwidth,clip]{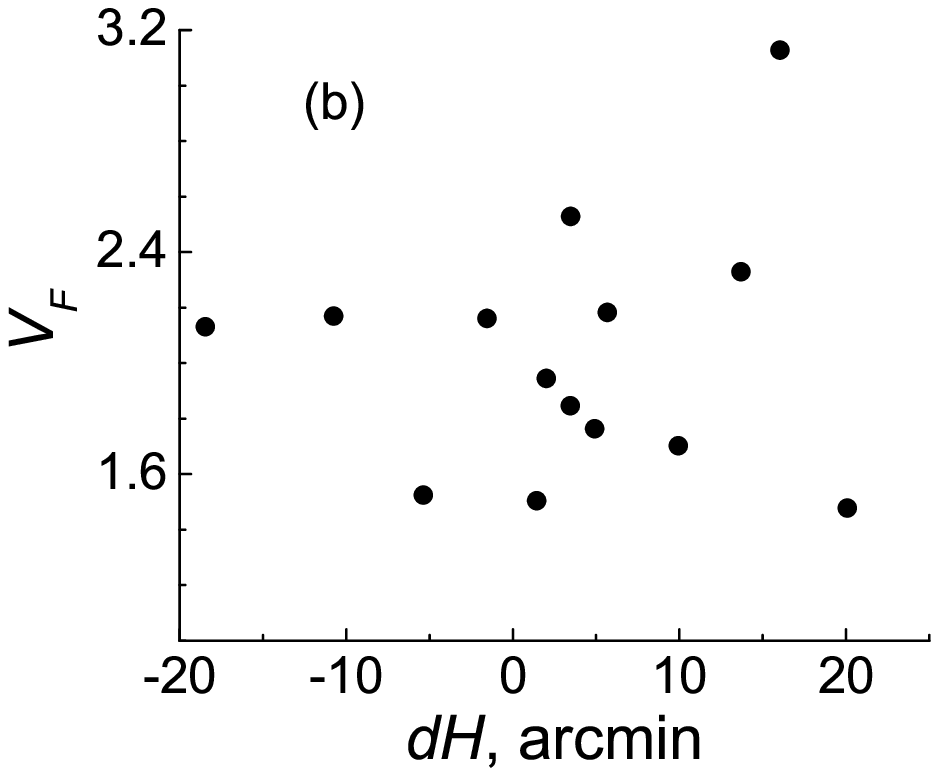}
}
\hbox{
\includegraphics[angle=0,width=0.37\textwidth,clip]{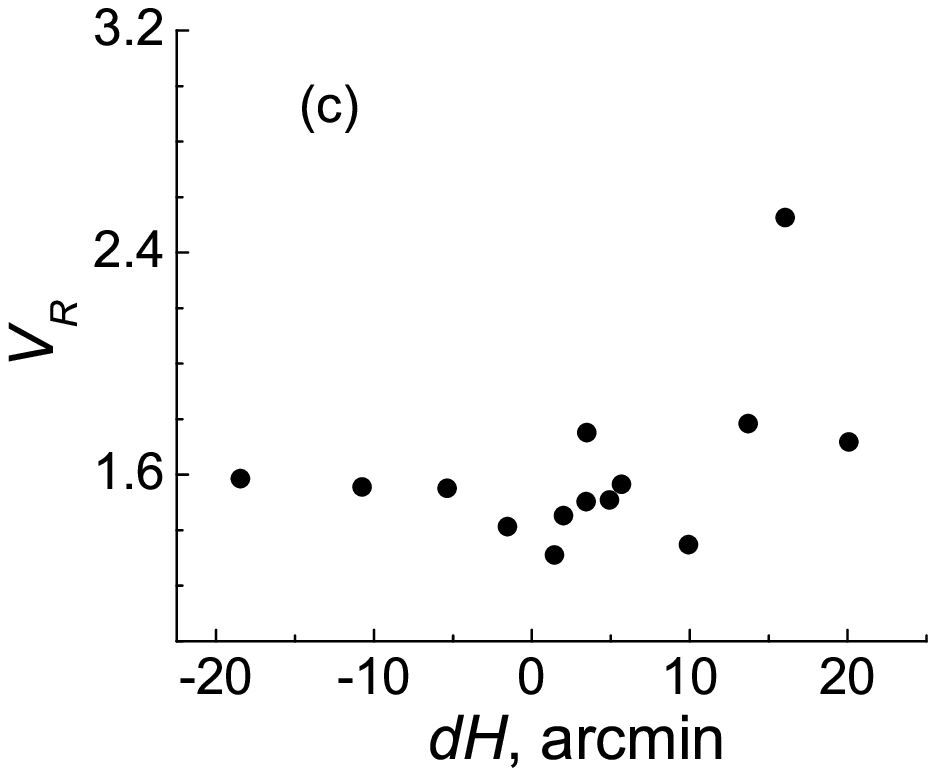}
\includegraphics[angle=0,width=0.37\textwidth,clip]{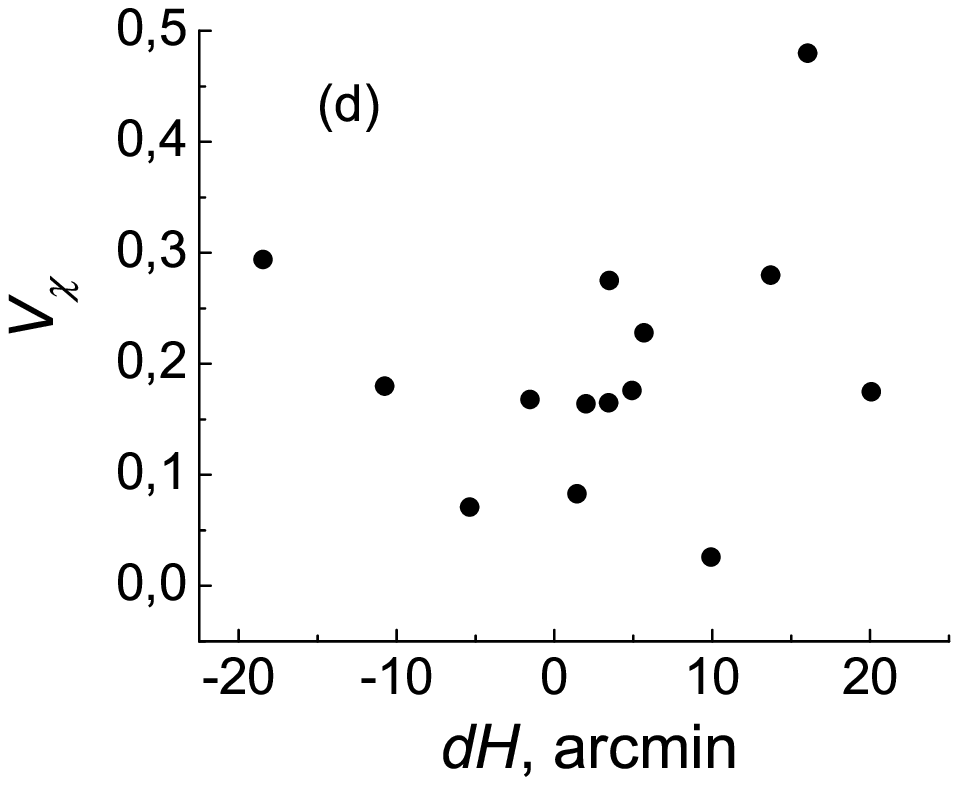}
} } } \setcaptionmargin{0mm} \captionstyle{normal} \caption{
Dependences of the variability indices $V$ (a), $V_{R}$ (b),
$V_{F}$ (c), and $V_{\chi}$ (d) on angle $dH$ for 14 suspected
variable sources (Tables~3 and 4). } \label{fig6:Majorova_n}
\end{figure*}

\begin{figure}[tb]
\onelinecaptionsfalse
\centerline{
\vbox{
\hbox{
\includegraphics[angle=0,width=0.37\textwidth,clip]{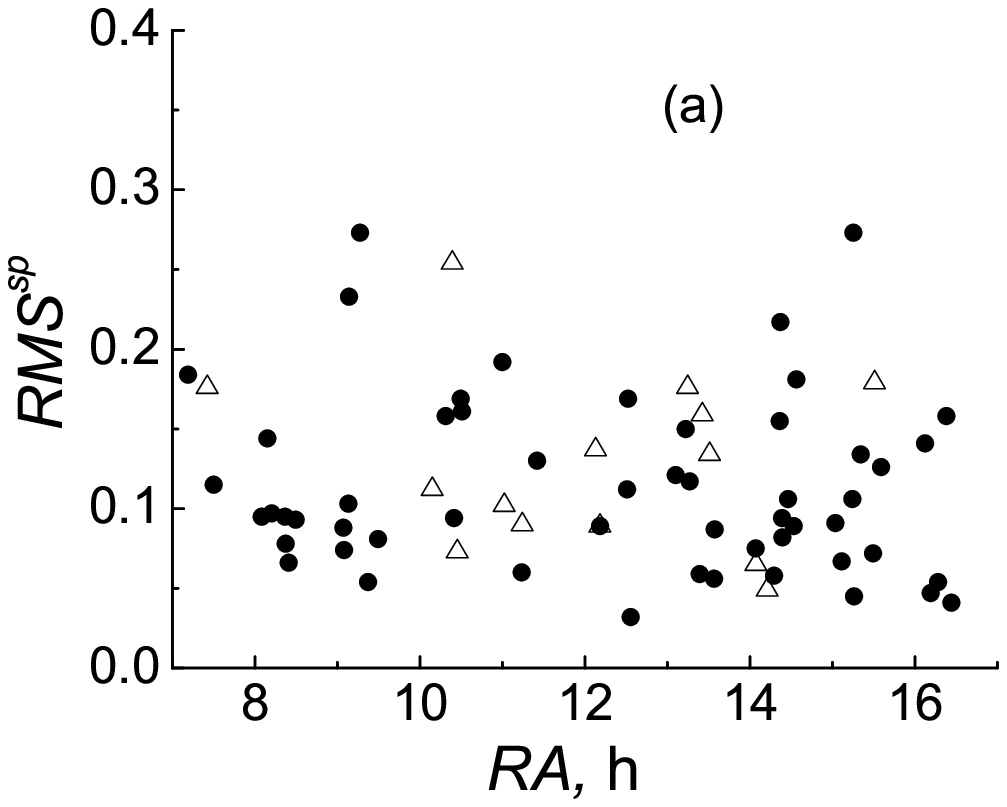}
}
\hbox{
\includegraphics[angle=0,width=0.37\textwidth,clip]{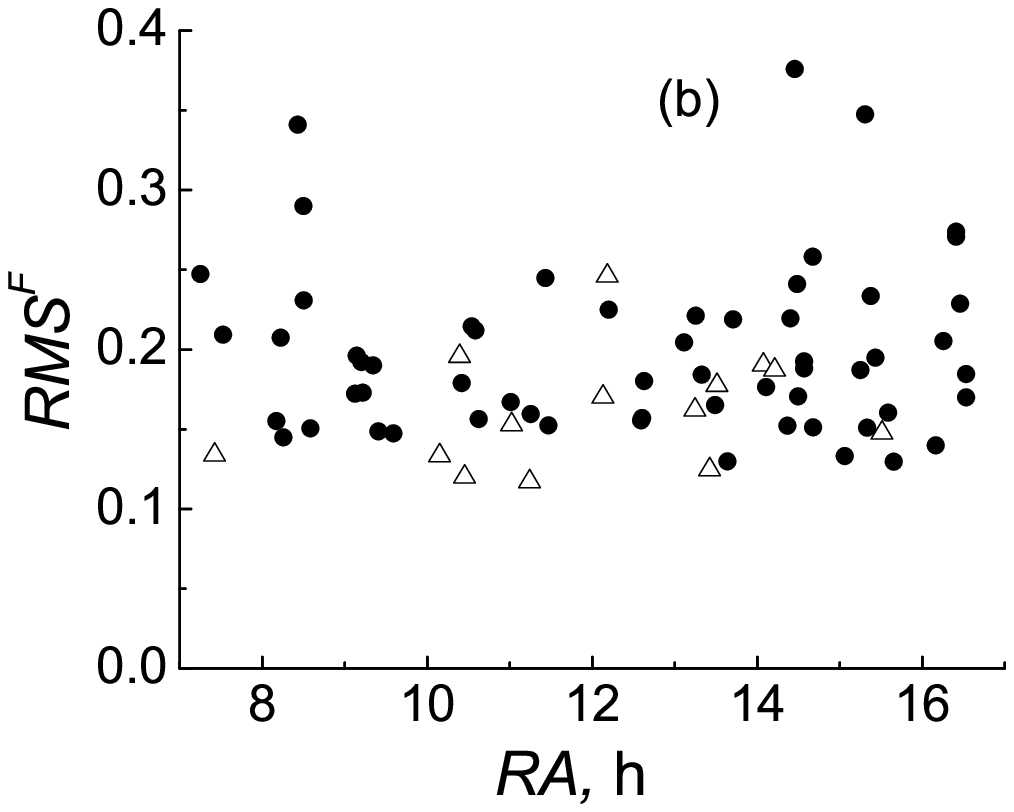}
}
\hbox{
\includegraphics[angle=0,width=0.37\textwidth,clip]{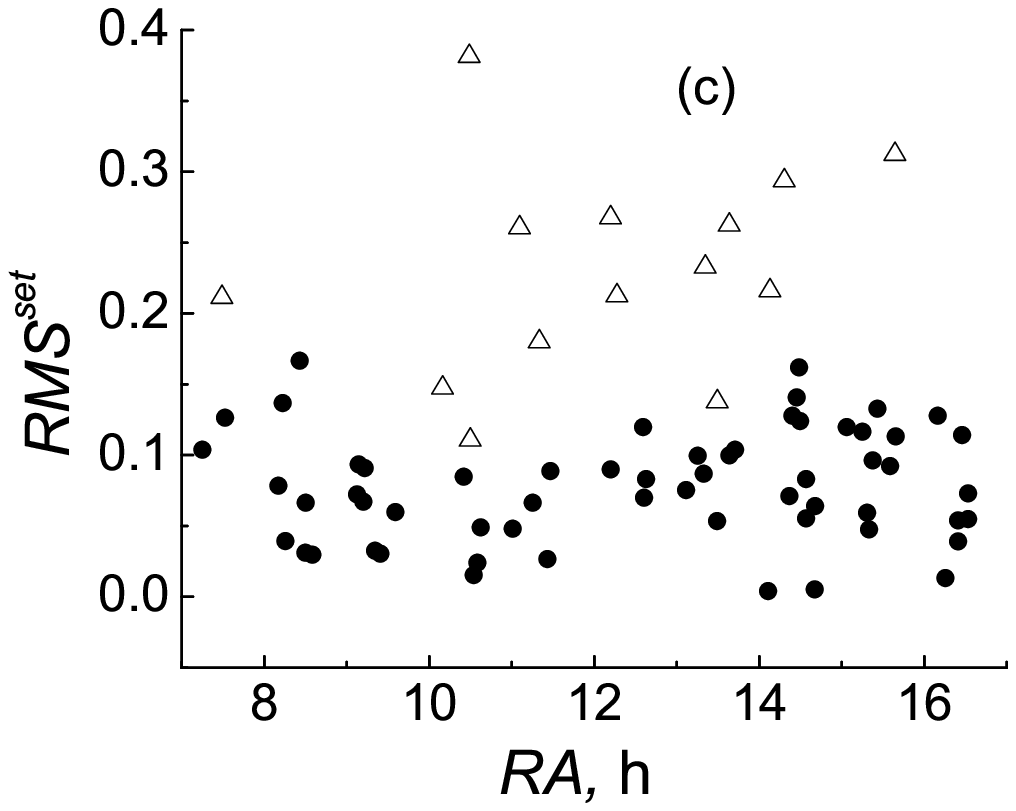}
} } } \setcaptionmargin{0mm} \captionstyle{normal} \caption{
Dependences of the standard errors $RMS^{\,sp}(RA)$ (a),
$RMS^{F}(RA)$ (b), and $RMS^{\,set}(RA)$ (c) for the subsample of
14~suspected variable sources (open triangles) and for the
subsample of ``non-variable'' objects (filled circles). }
\label{fig7:Majorova_n}
\end{figure}

\begin{figure*}[tbp]
\onelinecaptionsfalse
\centerline{
\hbox{
\includegraphics[angle=0,width=0.37\textwidth,clip]{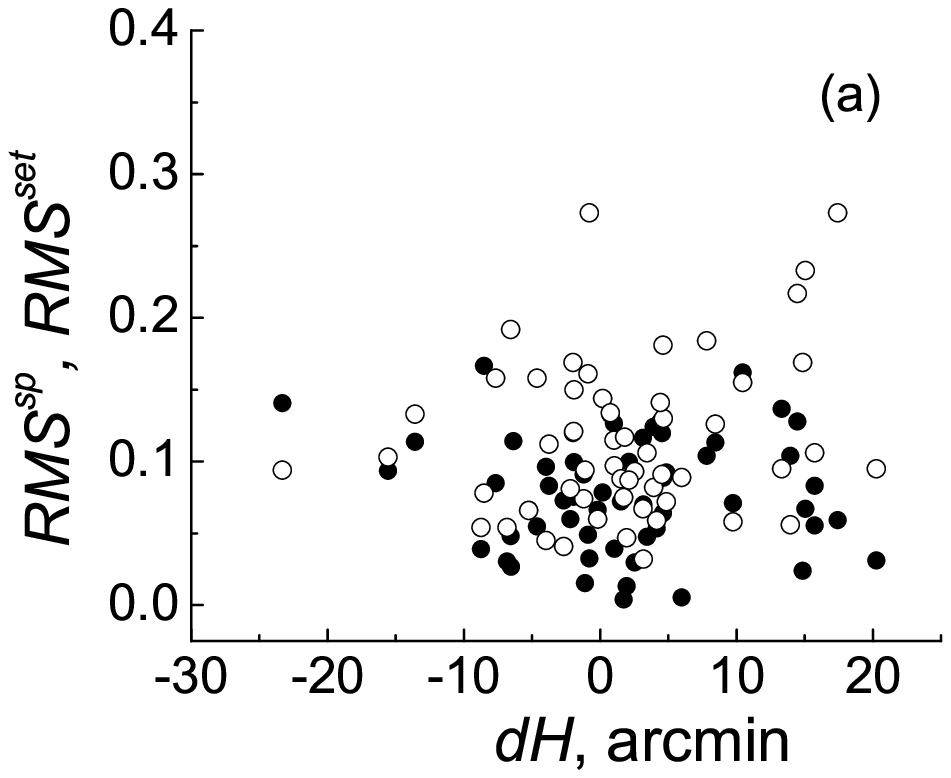}
\includegraphics[angle=0,width=0.37\textwidth,clip]{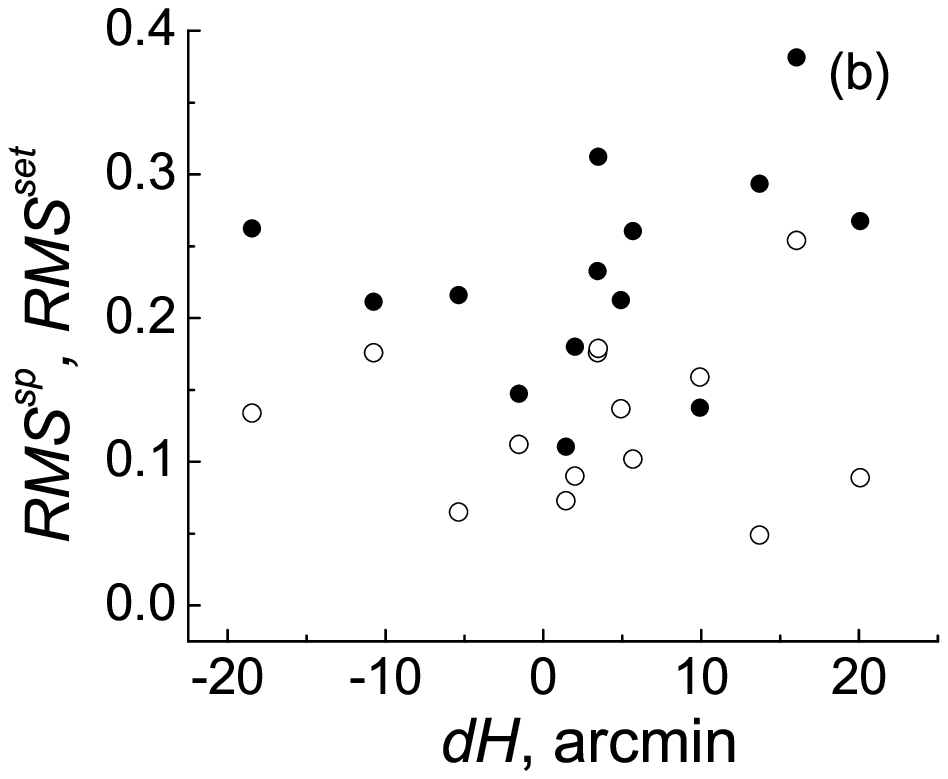}
} } \setcaptionmargin{0mm} \captionstyle{normal} \caption{
Dependences of the standard errors $RMS^{\rm \,set}(dH)$ (filled
circles) and $RMS^{\rm \,sp}(dH)$ (open circles) for
``non-variable'' (a) and suspected variable (b) sources. }
\label{fig8:Majorova_n}
\end{figure*}

\begin{figure*}[tbp]
\onelinecaptionsfalse
\centerline{
\hbox{
\includegraphics[angle=0,width=0.37\textwidth,clip]{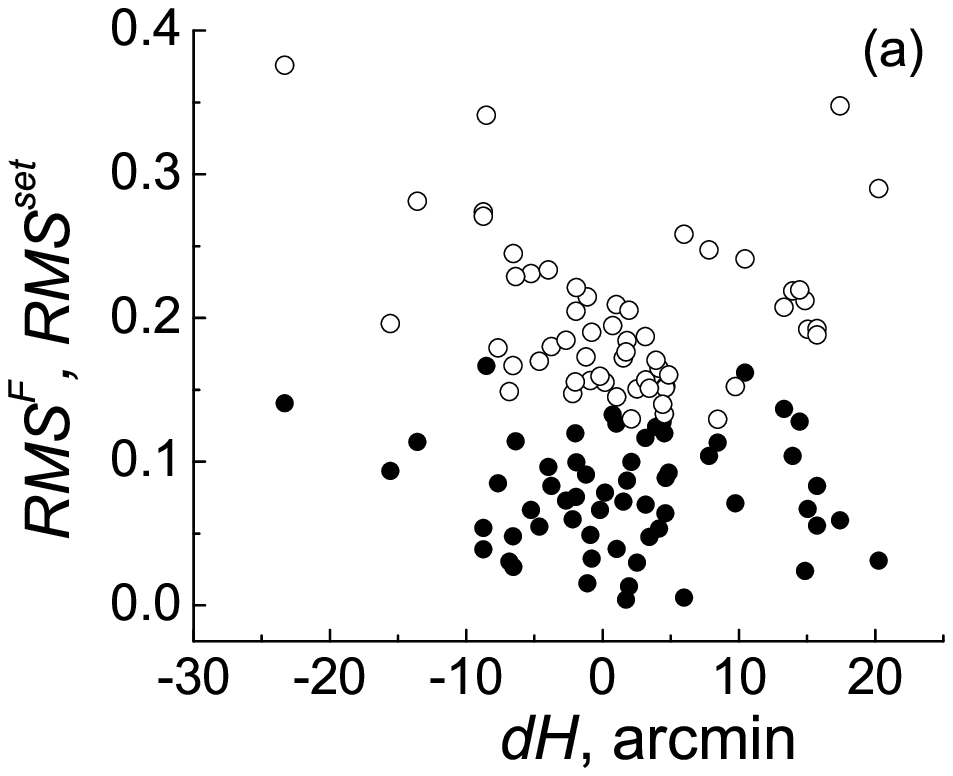}
\includegraphics[angle=0,width=0.37\textwidth,clip]{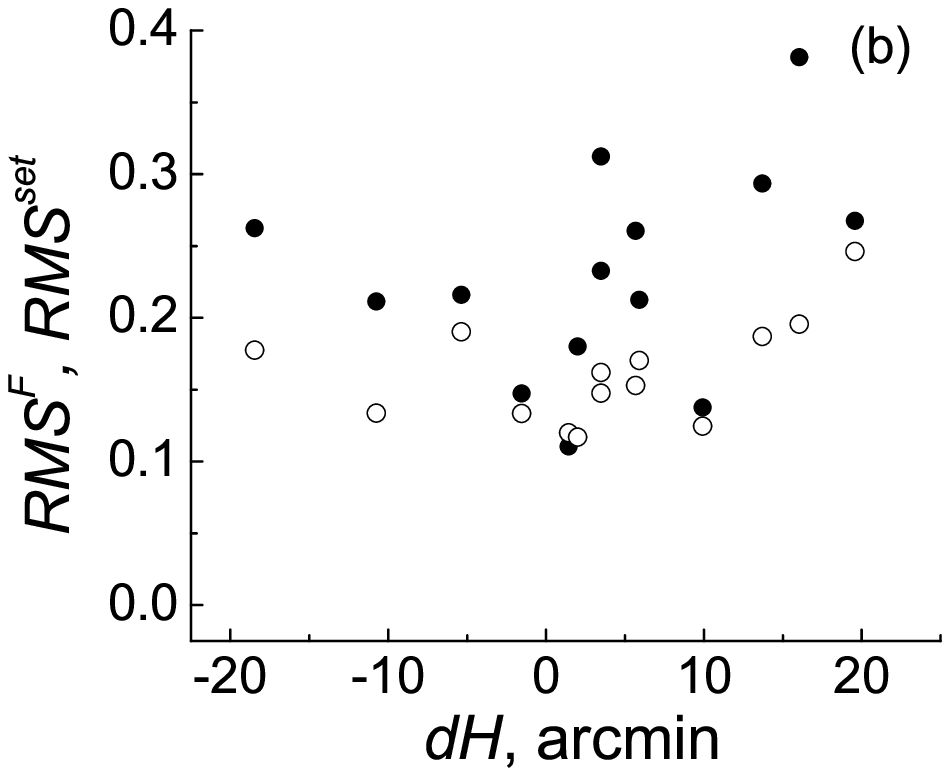}
} } \setcaptionmargin{0mm} \captionstyle{normal} \caption{
Dependences of the standard errors $RMS^{\rm \,set}(dH)$ (filled
circles) and $RMS^{\rm \,F}(dH)$ (open circles) for
``non-variable'' (a) and suspected variable (b) sources. }
\label{fig9:Majorova_n}
\end{figure*}

The same table also lists the $dH$ angles where the source flux
densities reach their maximum ($dH_{1}$) and minimum ($dH_{2}$)
values in the surveys considered (columns~6 and 7). Let us recall
that due to precession, the source declinations and $dH$ varied
from one survey to another. The last column of Table~3 lists the
spectral indices of these objects at 3.94~GHz.

Figure~\ref{fig6:Majorova_n} shows the dependences of the
variability index $V$ on angle $dH$ (a) and the dependences
$V_{R}(dH)$ (b) and $V_{F}(dH)$ (c) for the sources from Table~3.
We discuss the $V_{\chi}(dH)$ dependence shown in
Fig.~\ref{fig6:Majorova_n}d in Section~6. As is evident from these
plots, the radio sources are distributed rather uniformly relative
to the central section of the survey. They are also uniformly
distributed over the observing hours (or in right ascension).

Ten sources with positive long-term variability indices have the
coefficients $V_{R} > 1.5$, and four sources have the $V_{R}$
coefficients in the \mbox{$1.3 <V_{R} < 1.5$} interval. For all
the objects listed in Table~3, $V_{F} \gtrsim 1.5$.

These are mostly sufficiently bright objects \linebreak with flux
densities $\overline{F} > 100$~mJy except for three \linebreak
sources (J\,121328+050009, J\,132448+045758, and \linebreak
J\,140730+044934) whose $\overline{F}$ lie in the
\mbox{$50$--$100$~mJy} interval. The latter pass rather close to
the central section of the survey \mbox{($dH < 7^{'}$)} and show
up well enough on the averaged records obtained by integrating
about 25 transits.

In addition to the above computations, we also estimated the
long-term variability index $V$, eliminating a small systematic
trend in the scatter of the data points of the 1993 and 1994
surveys (Fig.~\ref{fig5:Majorova_n}). As a result, the $V$ indices
became negative for three out of 14 sources. These
objects---J\,121852+051449, J\,134243+050431, and
J\,140730+044934---prove to be the least likely variable source
candidates.

Let us now return to the accuracy of flux density determination
for the calibration sources and compare the relative standard
errors $RMS^{\,set}$, $RMS^{\,sp}$, and $RMS^{F}$ for two
subsamples. One of these subsamples includes 14 suspected variable
sources (Table~3), and the other one includes the ``non-variable''
sources with $V<0$.

Here $RMS^{\,set}$ is the relative standard deviation from the
mean flux density $\overline{F}$, $RMS^{\,sp}$ is the relative
standard error of the scatter of data points in the spectrum of
the source or the error of determination of its flux density from
the approximating curve fitted to its spectrum, and $RMS^{F}$ is
the standard error of the flux density averaged over all surveys.
\begin{equation}
RMS^{\,set} = \sigma^{set}/\overline{F}, \\
\label{9:Majorova_n}
\end{equation}
\begin{equation}
RMS^{F}= \frac{1}{n}\sum_{i}^n RMS_{i}.
\label{10:Majorova_n}
\end{equation}
Here $RMS_{i}$ is the relative standard error of the flux density
of the source in the ${i}$-th survey, computed using
formula~(\ref{5:Majorova_n}).

Figure~\ref{fig7:Majorova_n} shows the  $RMS^{\,sp}$ (a),
$RMS^{F}$ (b), and $RMS^{\,set}$ (c) quantities for the subsample
of suspected variable radio sources (open triangles) and for the
subsample of ``non-variable'' objects (filled circles) as a
function of $RA$.

It is evident from the above dependences that the standard errors
$RMS^{\,sp}$  of the source flux density inferred from the
spectral curves and the standard errors $RMS^{F}$ averaged over
all the surveys practically do not differ for the two subsamples
of the calibration sources.

As for the standard deviations  $RMS^{\,set}$, their values for
the first subsample consisting of sources with  $V > 0$ exceed
significantly the  $RMS^{\,set}$ values for the second subsample
(\mbox{$V < 0$}). Ten out of 14 candidate variable sources have
$RMS^{\,set} > 0.2$. The average $\overline{RMS^{\,set}}$ for the
first and second subsamples of objects are equal to $0.23\pm0.07$
and $0.08\pm0.04$, respectively.

We compare in Figs.~\ref{fig8:Majorova_n}
and~\ref{fig9:Majorova_n} the  $RMS^{\,set}$ values with
$RMS^{\,sp}$ and $RMS^{F}$. The filled circles show the
$RMS^{\,set}(dH)$ dependences, and the open circles---the
$RMS^{\,sp}(dH)$ (Fig.~\ref{fig8:Majorova_n}) and $RMS^{F}(dH)$
(Fig.~\ref{fig9:Majorova_n}) dependences, respectively. Panels (a)
and (b) show the corresponding dependences for the
``non-variable'' sources and the sources listed in Table~3,
respectively.

The relative standard deviations $RMS^{\,set}$ for the
``non-variable'' sources are comparable to the relative standard
errors $RMS^{\,sp}$ of the determination of flux densities from
the spectra and are substantially smaller than the mean standard
errors $RMS^{F}$ of the inferred flux densities averaged over all
the surveys.

The $RMS^{\,set}$ values for objects with positive long-term
variability indices exceed both the $RMS^{F}$ and the
$RMS^{\,sp}$; in the latter case, by almost a factor of two, on
average.

\section{ANALYSIS OF STATISTICAL PROPERTIES OF SUSPECTED VARIABLE SOURCES}

Let us now use statistical techniques to estimate the variability of the sources listed in Table~3,
although, unfortunately, the number of data points is rather small for such an analysis.

To confirm the variability of the objects with positive  $V$, we
performed computations similar to those made by Gorshkov and
Konnikova~\mbox{\cite{g1:Majorova_n}}, Kesteven et
al.~\mbox{\cite{kest:Majorova_n}}, Fanti et
al.~\mbox{\cite{fan:Majorova_n}}, and Seielstad et
al.~\mbox{\cite{sei:Majorova_n}}.

We computed for each of the $n$ surveys the variability amplitude
$\Delta F$ and the parameter $V_{\chi}$, as well as the weighted
average source flux density $\langle F \rangle$, the weighted
average standard error $\langle \sigma \rangle$, and the $\chi^2$
constant for $df=n-1$ degrees of freedom. We computed these
quantities using the following formulas \cite{sei:Majorova_n}:
\begin{equation}
\langle F \rangle  = \sum_{i}^n (F_{i}/\sigma_{i}^2) \Bigl/ \sum_{i}^n\sigma_{i}^{-2}, \\
\label{11:Majorova_n}
\end{equation}
\begin{equation}
\langle \sigma \rangle    = \biggl( \sum_{i}^n (1/\sigma_{i}^2)\biggr)^{-0.5}  , \\
\label{12:Majorova_n}
\end{equation}
\begin{equation}
\chi^2 = \sum_{i}^n \bigl(F_{i} - \langle F \rangle\bigr)^2 \bigl/ \sigma_{i}^2 ,\\
\label{13:Majorova_n}
\end{equation}
\begin{equation}
\Delta F = \biggl( (n-1)[\chi^2 - (n-1)] \Bigl/ \sum_{i}^n (F_{i}/\sigma_{i}^2)\biggr)^{0.5}, \notag \\
\end{equation}
\begin{equation}
V_{\chi} = \Delta F \bigl/ \langle F \rangle .
\label{14:Majorova_n}
\end{equation}

Table~4 summarizes the results of the computations of these
parameters. Column~2 gives the parameter $V_{\chi}$, which
characterizes the relative variation amplitude, and column~3 gives
the probability $p$ of variability according to the  $\chi^2$
criterion. This parameter gives a quantitative estimate of the
probability that a source whose flux densities are distributed as
$\chi^2$ with $n-1$ degrees of freedom may be considered variable
\mbox{($p = 1- \chi^2(n-1)$)}. Column~4 gives the weighted average
source flux densities $\langle F \rangle$, columns~5 and 6---the
absolute ($\langle \sigma \rangle$) and relative ($\langle
\sigma_{otn} \rangle$) weighted average standard errors. Here
$\langle \sigma_{otn} \rangle = \langle \sigma \rangle / \langle F
\rangle$. Column~7 gives the variability amplitudes $\Delta F$ of
the sources, columns~8 and 9---the  $\chi^2$ constant and the
number $df$ of degrees of freedom, respectively, and
column~10---the mean $dH$ angles ($\overline{dH}$) averaged over
all the surveys.

\begin{table*}
\setcaptionmargin{0mm} \onelinecaptionstrue \captionstyle{normal}
\caption{The coefficient $V_{\chi}$ and the probability  $p$}
\medskip
\begin{tabular}{c|@{~}c@{~}|@{~}c@{~}|r|r|c|r|c|c|r}

\hline
~~~~~~~$RA_{2000}$~~~~~$DEC_{2000}$ & $V_\chi$ & $p$ & \multicolumn{1}{c|}{$\langle F \rangle$,} & \multicolumn{1}{c|}{$\langle \sigma \rangle$,} & $\langle \sigma_{otn} \rangle$ & \multicolumn{1}{c|}{$\Delta F$,} & $\chi^2$ & $df$ & \multicolumn{1}{c}{$\overline{dH}$,}  \\
RCR & & & \multicolumn{1}{c|}{mJy} & \multicolumn{1}{c|}{mJy} & & \multicolumn{1}{c|}{mJy} & & & \multicolumn{1}{c}{arcmin} \\
(1) & (2) & (3) & \multicolumn{1}{c|}{(4)} & \multicolumn{1}{c|}{(5)} & (6) & \multicolumn{1}{c|}{(7)} & (8) & (9) & \multicolumn{1}{c}{(10)} \\
\hline
J\,103938.62+051031.3  &0.480  &0.984 &163  &16    &0.100  & 79  &10.7  &3   &$16.04 $ \\
J\,155148.09+045930.5  &0.275  &0.961 & 74  & 7    &0.087  & 20  & 6.9  &2   &$3.47  $\\
J\,135137.56+043542.0  &0.294  &0.960 &353  &30    &0.086  &104  & 6.9  &2   &$-18.46$  \\
J\,142104.21+050845.0  &0.280  &0.928 &169  &19    &0.110  & 48  & 5.3  &2   &$13.69 $ \\
J\,110246.51+045916.7  &0.228  &0.925 & 98  & 9    &0.090  & 22  & 5.2  &2   &$5.67  $\\
J\,112437.45+045618.8  &0.164  &0.895 &446  &24    &0.054  & 73  & 6.1  &3   &$2.01  $\\
J\,101515.53+045305.6  &0.168  &0.864 &112  & 7    &0.060  & 19  & 5.6  &3   &$-1.54 $ \\
J\,074239.34+050704.3  &0.180  &0.855 &336  &22    &0.066  & 60  & 5.4  &3   &$-10.75$  \\
J\,121328.89+050009.9  &0.176  &0.772 & 66  & 6    &0.084  & 12  & 4.5  &3   &$4.92  $\\
J\,132448.14+045758.8  &0.165  &0.762 & 53  & 4    &0.081  &  9  & 4.4  &3   &$3.44  $\\
J\,104551.72+045552.9  &0.083  &0.691 &150  & 8    &0.055  & 12  & 3.7  &3   &$1.43  $\\
J\,121852.16+051449.4  &0.175  &0.681 &217  &27    &0.125  & 38  & 3.6  &3   &$20.07 $ \\
J\,140730.77+044934.9  &0.071  &0.674 & 78  & 8    &0.107  &  6  & 2.2  &2   &$-5.37 $ \\
J\,134243.57+050431.5  &0.026  &0.601 &958  &58    &0.060  & 25  & 3.1  &3   &$9.92  $\\
\hline
\end{tabular}
\end{table*}

A comparison of the data listed in Tables~3 and 4 shows that the
weighted average source flux densities $\langle F \rangle$
computed using formula~(\ref{11:Majorova_n}) practically coincide
with the mean values $\overline{F}$ (formula~(\ref{8:Majorova_n}))
within the errors. The relative standard deviations  $RMS^{\,set}$
exceed significantly the relative weighted standard errors
$\langle \sigma_{otn} \rangle$. The $RMS^{\,set}$ averaged over
all the 14 sources is equal to $0.23\pm0.07$, and the averaged
$\langle \sigma_{otn} \rangle$ to $0.08\pm0.02$.

Figure~\ref{fig6:Majorova_n}d shows  $V_{\chi}$ plotted as a function of angle $dH$.

Let us now see which objects among those listed in Tables~3 and 4
can be considered variable. Kesteven et
al.~\mbox{\cite{kest:Majorova_n}} and Fanti et
al.~\mbox{\cite{fan:Majorova_n}} considered a source to be
possibly variable if its $\chi^2$ probability satisfied the
condition of \mbox{$0.1\% \le 1-p \le 1\%$} and reliably variable
if $ 1-p \le 0.1\%$.

None of the 14 sources listed in Tables~3 and 4 meet these
conditions. In other words, according to the criteria of Kesteven
et al.~\mbox{\cite{kest:Majorova_n}} and Fanti et
al.~\mbox{\cite{fan:Majorova_n}}, our sample of calibration
sources contains neither variable nor likely  variable objects.

Seielstad et al.~\mbox{\cite{sei:Majorova_n}} considered an object
to be variable if $p \ge 0.985$, whereas Gorshkov and
Konnikova~\mbox{\cite{g1:Majorova_n}} considered sources with  $p
\ge 0.98$ and $p \ge 0.95$ to be reliably and possibly variable,
respectively. According to these criteria, the sources
J\,155148+045930 and J\,135137+043542 ($p=0.96$) may be considered
to be possibly variable and the source J\,103938+051031
($p=0.984$) to be reliably variable. Two of these sources
(J\,103938+051031 and J\,135137+043542) are sufficiently bright
objects with flux densities $\overline{F} > 100$~mJy, whose
transits occur at a distance of about 1.5--2 halfwidths of the
vertical PBP from the central section of the survey in
declination; the source J\,155148+045930 is weaker
(\mbox{$\overline{F} = 79$~mJy}), but passes close to the central
section of the survey (\mbox{$\overline{dH} = 3.5^{'}$}).

Wang et al.~\cite{VF:Majorova_n} used the coefficient  $V_{F}$ as
a criterion of variability. They set its threshold value at
$V_{F}=3$ and considered sources reaching this level to be
variable. In other words, they considered the sources whose flux
density difference $\Delta F$ in different surveys exceeds
$3\sigma$ to be variable, where
$\sigma=\sqrt{\sigma_{i}^2+\sigma_{j}^2}$
(formula~(\ref{3:Majorova_n})). Of the 14 suspected variable
sources in our sample only  J\,103938+051031 meets this condition.
Another seven sources meet the condition $\Delta F > 2\sigma$.

Let us now consider the parameter $V_{\chi}$ as a variability
criterion. An analysis of the data reported by Seielstad et
al.~\mbox{\cite{sei:Majorova_n}} shows that the parameters
$V_{\chi}$ of variable sources (with $p \ge 0.985$) mostly exceed
$0.2$. However, there are several objects with
\mbox{$V_{\chi}=0.15$--$0.17$}. Furthermore, some
``non-variable'' objects whose probabilities $p$ are significantly
smaller than $0.985$ have $V_{\chi}$ parameters greater than
$0.2$. In our sample of 14 sources 11 have $V_{\chi} \ge 0.164$
and only three have $V_{\chi} < 0.1$.

To sum up, we can conclude that all the sources listed in Tables~3
and 4 can be considered to be possibly variable, because their
flux density difference determined from the data for different
observing runs exceed the sum of the flux density errors. However,
the confidence level of this variability differs for different
sources. Only one of them can be considered to be reliably
variable according to the criteria of Gorshkov and
Konnikova~\mbox{\cite{g1:Majorova_n}}, Seielstad et
al.~\cite{sei:Majorova_n}, and Wang et

\noindent al.~\cite{VF:Majorova_n}, and two---possibly variable
\footnote{However, even these sources cannot be considered
variable according to the more stringent statistical criteria
adopted by Kesteven et al.~\cite{kest:Majorova_n} and Fanti et
al.~\cite{fan:Majorova_n}.}.

The four objects  (J\,104551+045552, J\,121852+ \linebreak 051449,
J\,134243+050431, and J\,140730+044934) with the lowest positive
long-term variability indices $V$ (\mbox{$V=0.007$--$0.035$}) and
$\chi^{2}$ probabilities (\mbox{$0.6<p<0.7$}) are the least likely
variable source candidates. Furthermore, the latter three sources
change the sign of their variability indices  $V$ to negative if
the small systematic trend in the dependence of \mbox{$G =
(F_{i}/{T_{a}}_{i})/(A/k_{P\!B\!P})$} on $dH$ in the data of the
1993 and 1994 surveys is taken into account.

The remaining seven sources (J\,074239+050704, J\,101515+045305,
J\,110246+045916, J\,112437+ \linebreak 045618, J\,121328+050009,
J\,132448+045758, and \linebreak J\,142104+050845) are intermediate between
the above categories. Although they have positive long-term variability indices,
their  $\chi^{2}$ probabilities are low.

Thus only three out of about 80 selected calibration sources can
be considered to be variable with a probability of $p>0.95$, and
seven more sources can be considered to be possibly variable.

Note that eight of the 14 objects listed in Table~3
(J\,074239+050704, J\,103938+051031, J\,110246+\linebreak 045916,
J\,121328+050009, J\,132448+045758, \linebreak J\,135137+043542,
J\,142104+050845, and J\,155148+\linebreak 045930) have
variability indices $V\gtrsim0.08$, which are comparable with the
long-term variability indices determined by Afanas'ev et
al.~\cite{VV:Majorova_n} for variable sources with flat spectra.

Let us now present the more detailed properties of all of the 14
candidate objects, since they all have optical identifications.

The object J\,074239+050704 (or 4C+05.33) was identified with a
galaxy and its color indices according to the data of the WISE
mid-infrared survey~\cite{W:Majorova_n} are typical of a spiral
galaxy. The NED database confirms this conclusion: the object is
classified as a Seyfert type galaxy (Sy2) with $Z=0.16$. According
to the  GSC~\cite{L:Majorova_n} and USNO-B1~\cite{M:Majorova_n}
catalogs, the scatter of the magnitudes of the object in close
filters amounts to $1\fm2$, and this fact may be indicative of
optical variability. This source was studied by Gorshkov and
Konnikova~\mbox{\cite{g1:Majorova_n}}.

The object J\,101515+045305 (PMN~J\,1015+0452) is a point source.
On the FIRST maps, it was identified with a galaxy, which by its
color index \mbox{$(u-r)>2.22$}~\cite{S:Majorova_n} can be
classified as an early-type galaxy.

The object J\,103939+051031 (or PKS~J\,1037+05) is a binary source
(FIRST) with the so-called \linebreak ``winged'' morphology, which
may be indicative of the interaction of the plasma back flow from
the lobes of the radio source with the nonuniform environment.
Another model explains such a morphology by  ``ageing'' components
that remained from fast reorientation of the black hole and
accretion disk as a result of merging and subsequently resumed
nuclear activity~\cite{C:Majorova_n,Cs:Majorova_n}. The object has
been identified with an optically variable elliptical galaxy
($Z=0.068$) of the Abell\,1066 cluster located close to two other
galaxies. It is the most likely candidate variable object
according to the variability criteria employed.

The J\,110246+045916 object has the same morphology as
J\,103939+051031 and is identified with a possibly optically
variable starlike object, supposedly a quasar.

The J\,112437+045618 (4C+05.50) object is a double source from the
sample of the RC catalog objects with steep spectra. Its optical
identification and spectrum were obtained on the 6-m telescope of
the Special Astrophysical Observatory of the Russian Academy of
Sciences within the framework of the ``Big Trio'' program of the
search for distant galaxies~\cite{trio:Majorova_n}. Parijskij et
al.~\cite{trios:Majorova_n} classified the optical object as a
galaxy ($Z=0.284$) with narrow emission and absorption lines in
its spectrum. In the SDSS survey~\cite{A:Majorova_n} the object is
classified as a Seyfert galaxy (Sy2) with $Z=0.283$, possibly
optically variable.

The J\,121328+050009 (PMN~J\,1213+0500) object is a double source
with a nucleus, which, like J\,112437+045618, is a part of the SS
sample of the RC catalog; it was identified with a galaxy
($Z_{ph}$=0.76).

The J\,132448+045758 object is a double source identified with a
starlike object, probably a quasar.

The J\,135137+043542 (MRC~J\,1349+048) object is a point source, which was studied
by Gorshkov and Konnikova~\mbox{\cite{g1:Majorova_n}}. It is identified in the SDSS
with a faint galaxy. We consider it to be a variable radio source.

The J\,142104+050845 object is a point source identified with a galaxy ($Z=0.455$),
which may be a part of a triplet. It is possibly an optically variable object.

The J\,155148+045930 (PMN~J\,1551+0458) object is a double source; it is,
like J\,112437+045618 and J\,121328+050009, a part of the SS sample of the RC catalog.
The object is identified with a faint $R=$~$23\fm6$ galaxy. We consider it to be
a variable radio source.

We selected the next four sources as candidate variable objects,
but they failed to meet the adopted variability criteria and,
according to the available data, we cannot classify them as
variable.

The J\,104551+045553 (PMN\,J\,1045+0455) object is a likely double source
identified with a galaxy in the SDSS survey.

The J\,121852+051449 object is a double source identified with an optically variable
elliptical galaxy \linebreak (\mbox{$Z=0.078$}), which is the
brightest in the  Abell\,1516 cluster.

The J\,134243+050431 (4C+05.57) object is an FRI-type double
source identified with an optically variable Seyfert galaxy (Sy1),
\mbox{$Z=0.136$}.

The J\,140730+044934 object is a double source identified with a quasar candidate with $Z_{ph}=1.775$,
which is possibly optically variable.

Nine out of 14 sources listed in Table~3 show optical variability:
J\,074239+050704, J\,103938+051031, J\,110246+045916,
J\,112437+045618, J\,121328+ \linebreak 050009, J\,121852+051449,
J\,134243+050431, \linebreak J\,140730+044934, and
J\,142104+050845. The data for the remaining sources is too scarce
to allow any conclusions.

\begin{figure*}[tbp]
\onelinecaptionstrue \centerline{ \vbox{ \hbox{
\includegraphics[angle=0,width=0.37\textwidth,clip]{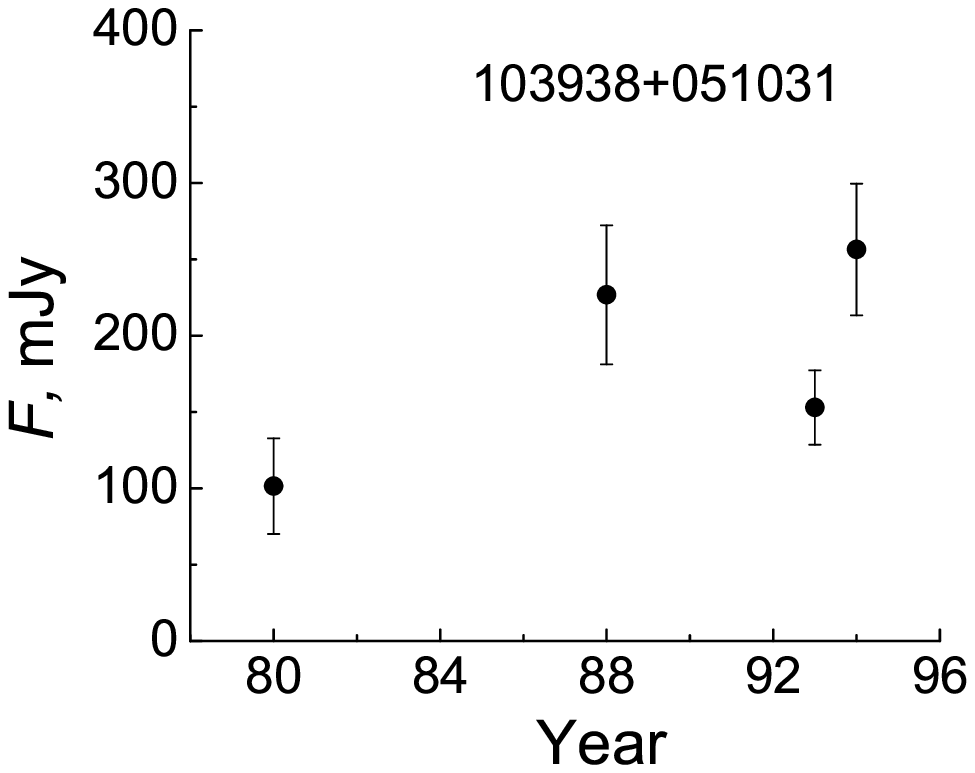}
\includegraphics[angle=0,width=0.37\textwidth,clip]{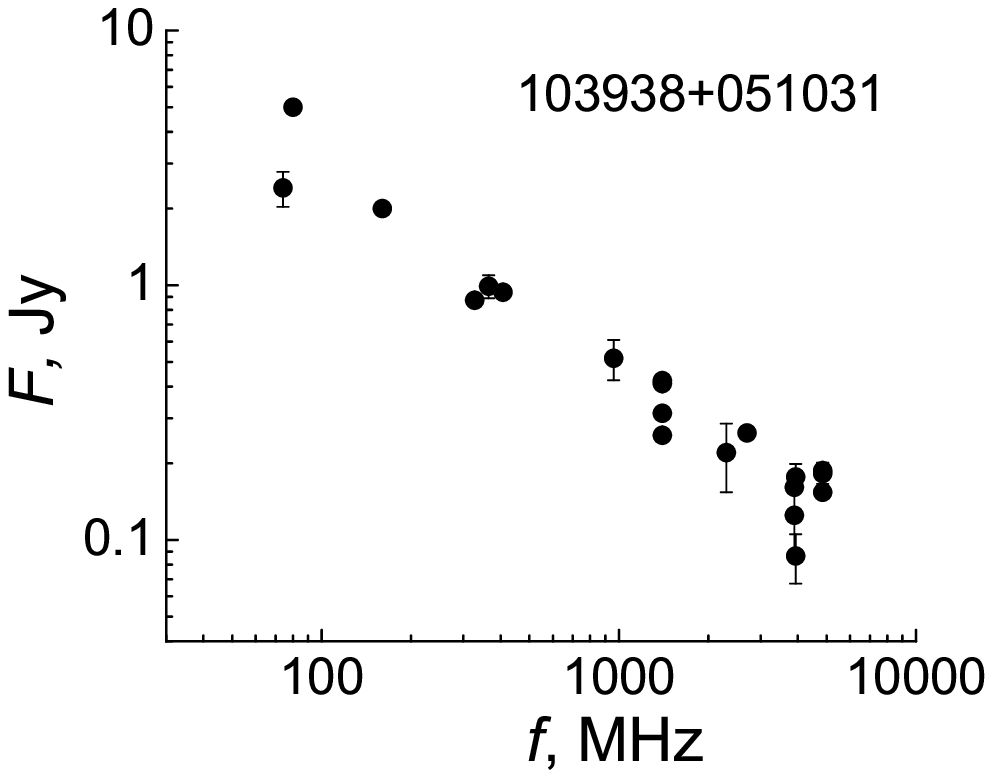}
} \hbox{
\includegraphics[angle=0,width=0.37\textwidth,clip]{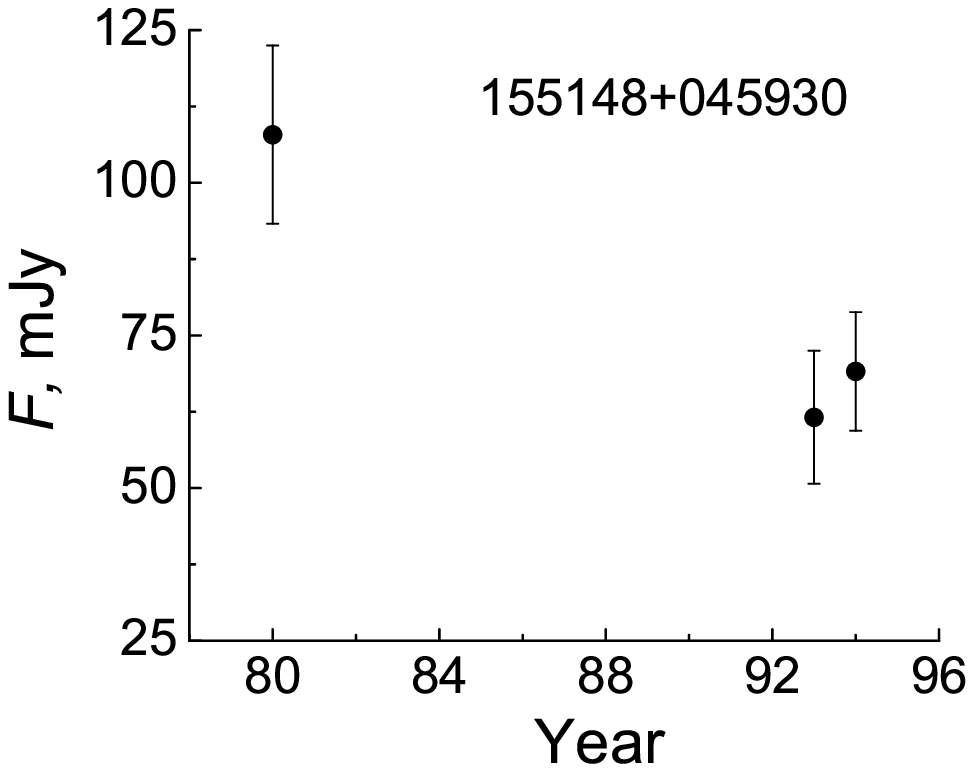}
\includegraphics[angle=0,width=0.37\textwidth,clip]{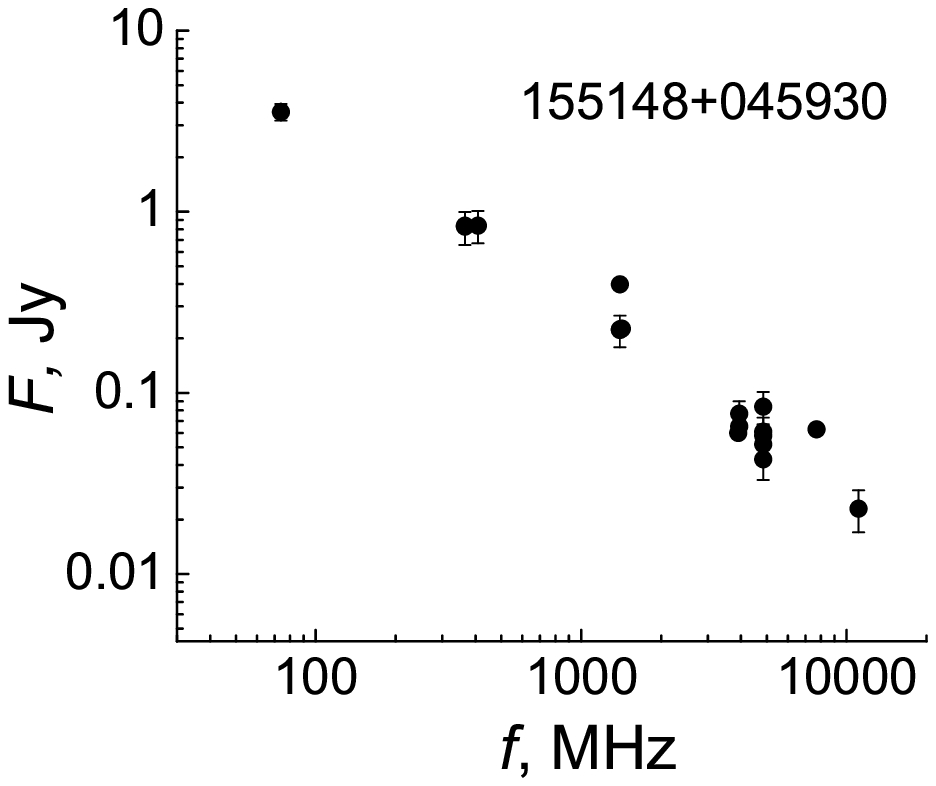}
} \hbox{
\includegraphics[angle=0,width=0.37\textwidth,clip]{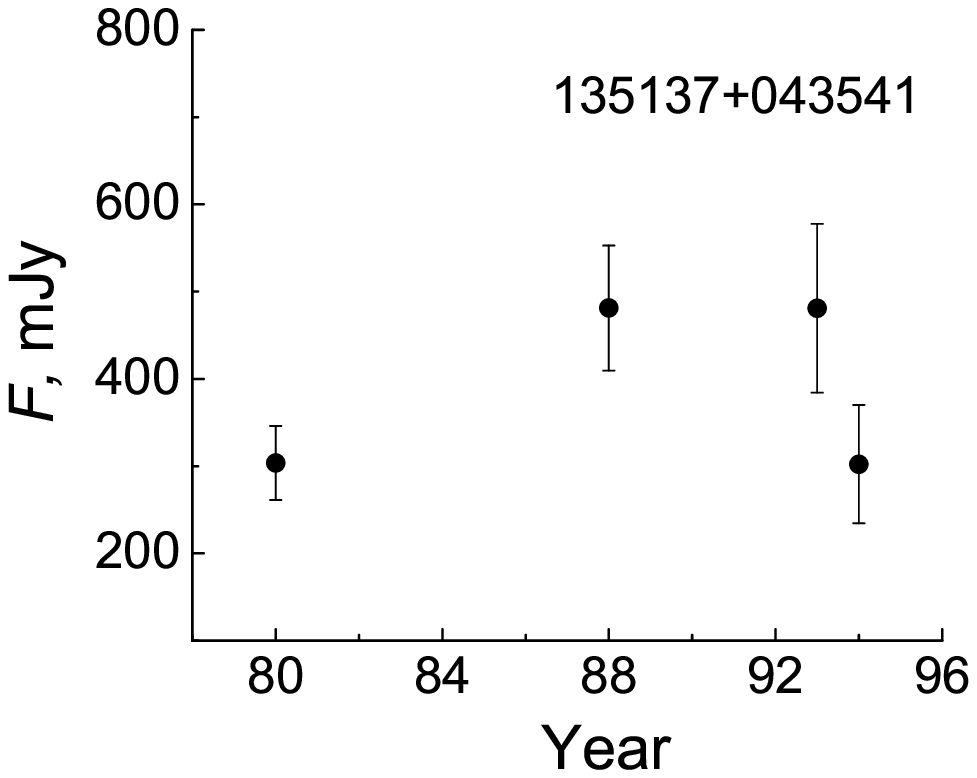}
\includegraphics[angle=0,width=0.37\textwidth,clip]{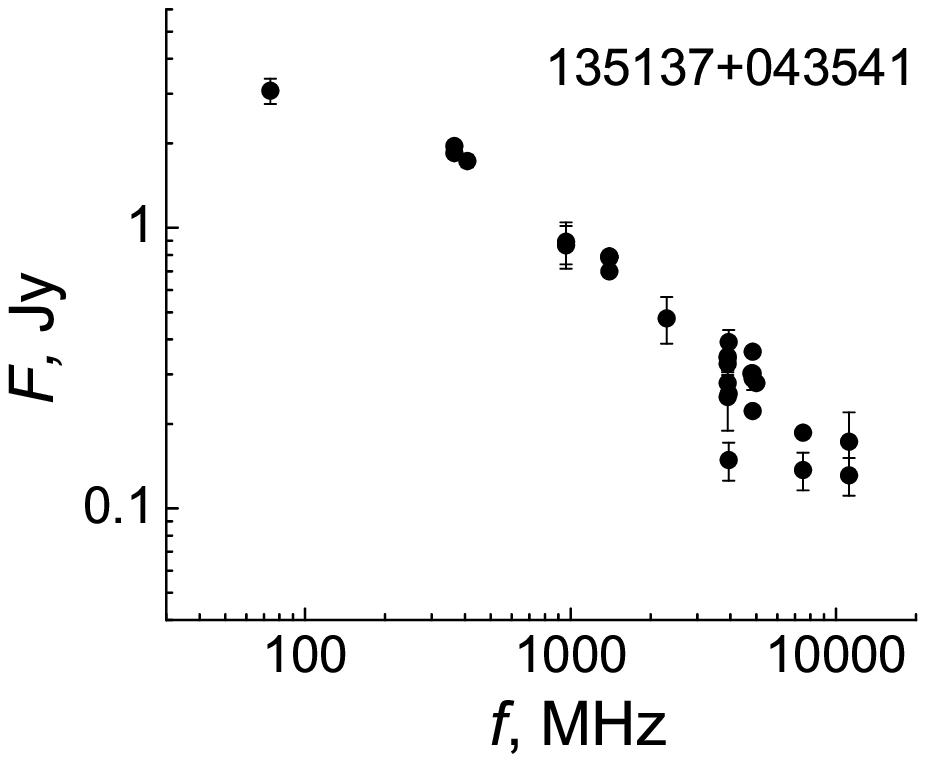}
} } } \setcaptionmargin{0mm} \captionstyle{normal} \caption{Light
curves (left) and spectra (right) of suspected variable
calibration sources ($V > 0$) with the probabilities
$p(\chi^{2})>0.95$. } \label{fig10:Majorova_n}
\end{figure*}

\begin{figure*}[tbp]
\onelinecaptionstrue \centerline{ \vbox{ \hbox{
\includegraphics[angle=0,width=0.37\textwidth,clip]{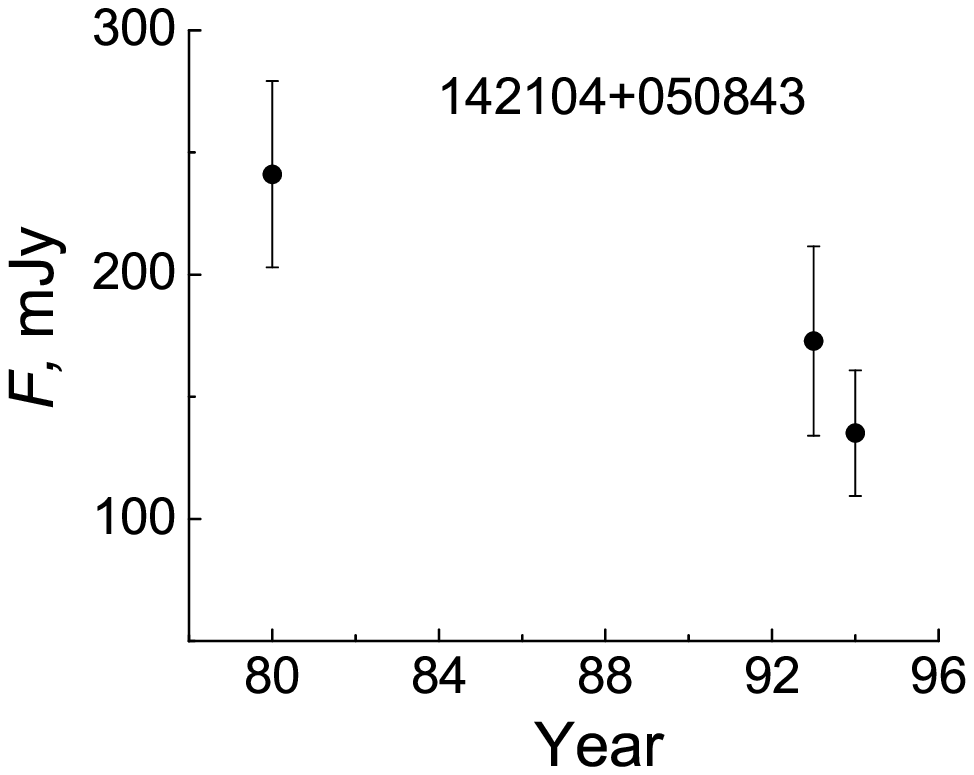}
\includegraphics[angle=0,width=0.37\textwidth,clip]{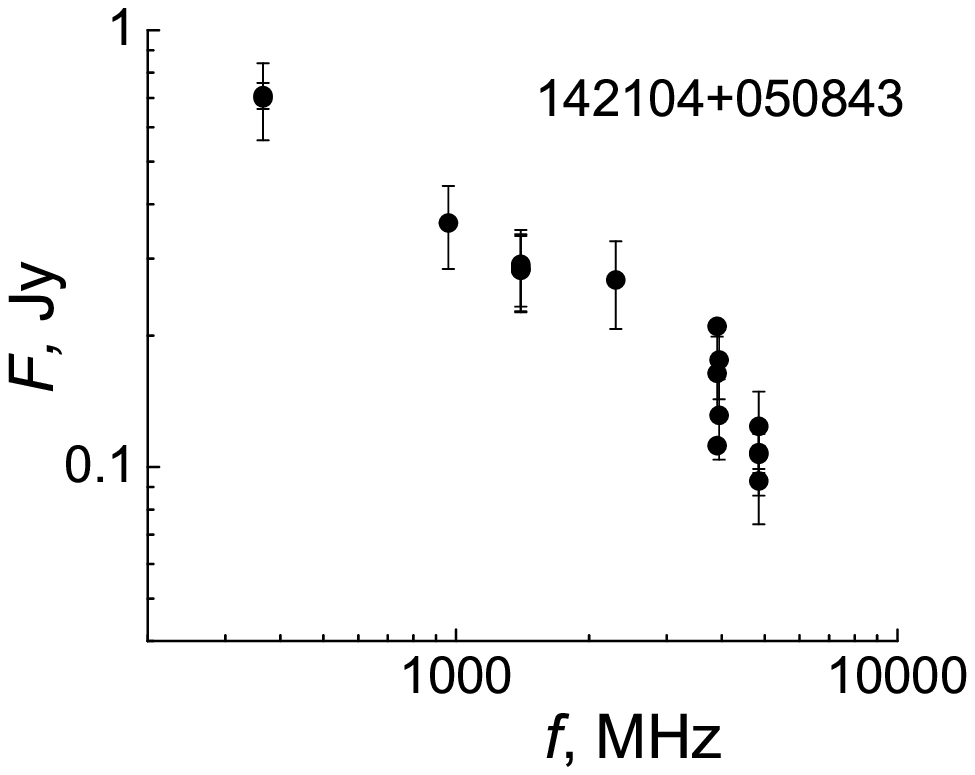}
} \hbox{
\includegraphics[angle=0,width=0.37\textwidth,clip]{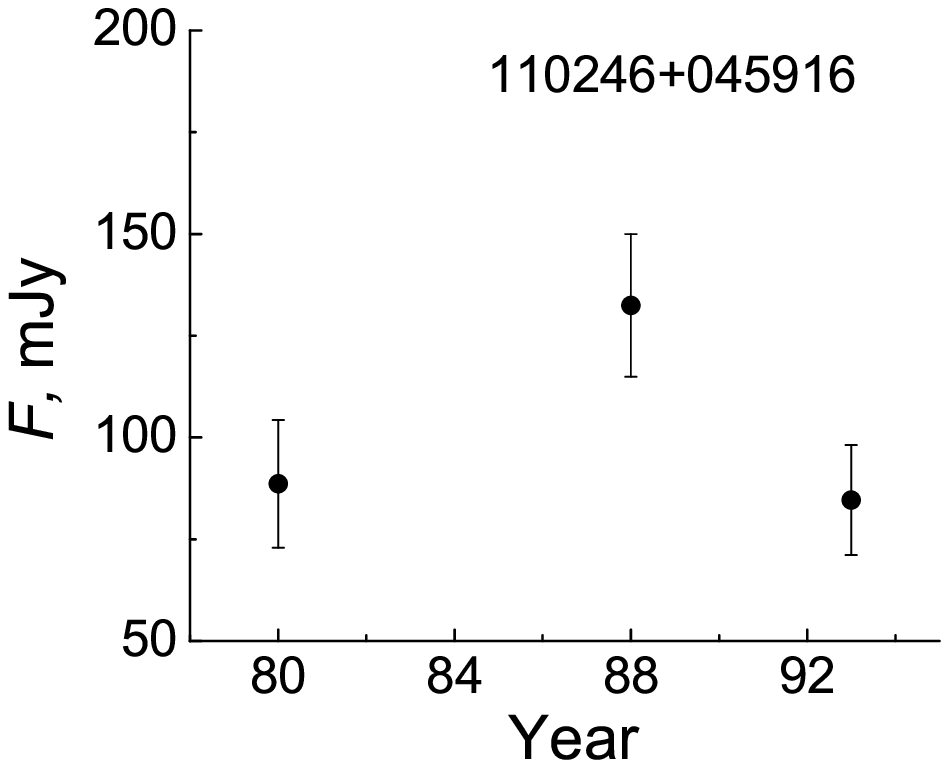}
\includegraphics[angle=0,width=0.37\textwidth,clip]{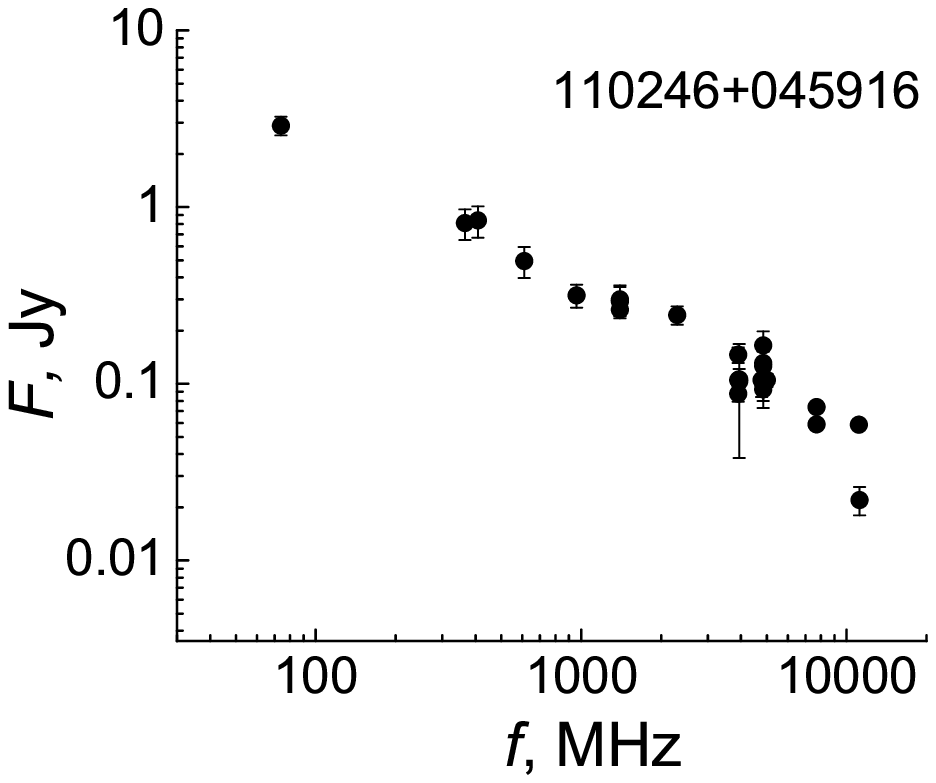}
} \hbox{
\includegraphics[angle=0,width=0.37\textwidth,clip]{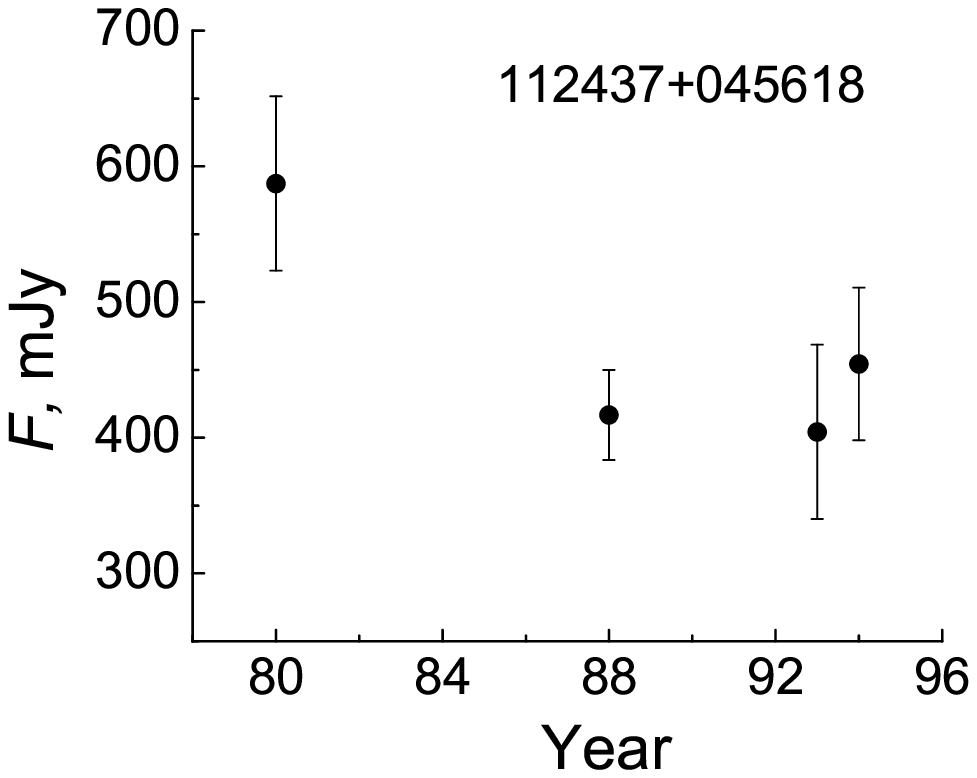}
\includegraphics[angle=0,width=0.37\textwidth,clip]{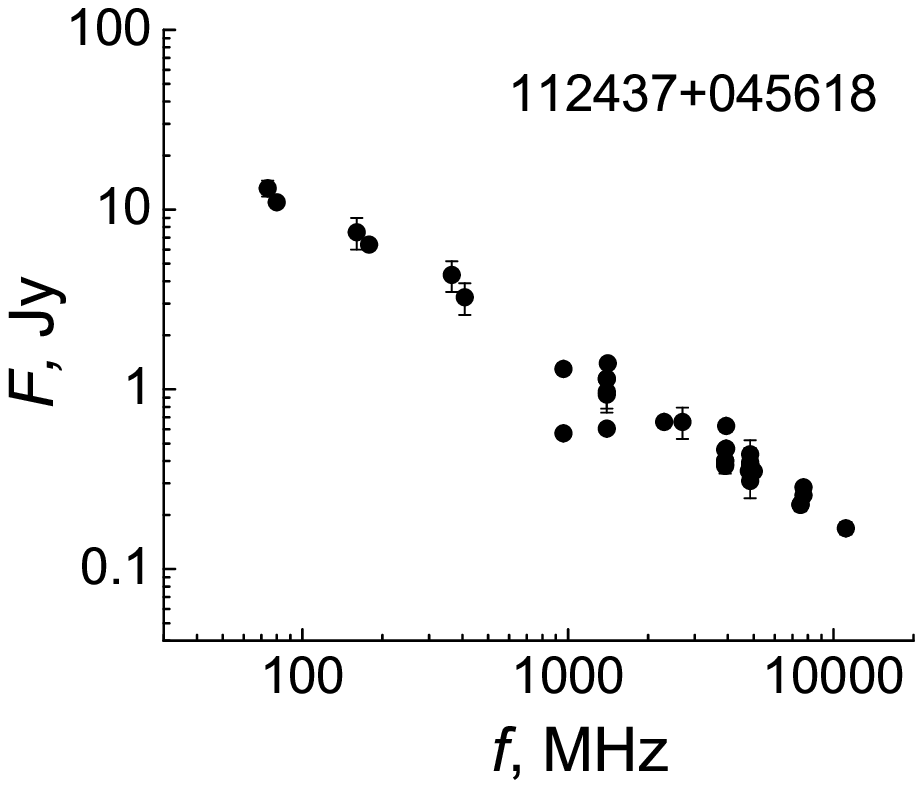}
} } } \setcaptionmargin{0mm} \captionstyle{normal} \caption{ Same
as Fig.~\ref{fig10:Majorova_n} for the sources with $ 0.89 <
p(\chi^{2}) < 0.95$. } \label{fig11:Majorova_n}
\end{figure*}

\begin{figure*}[tbp]
\onelinecaptionstrue \centerline{ \vbox{ \hbox{
\includegraphics[angle=0,width=0.37\textwidth,clip]{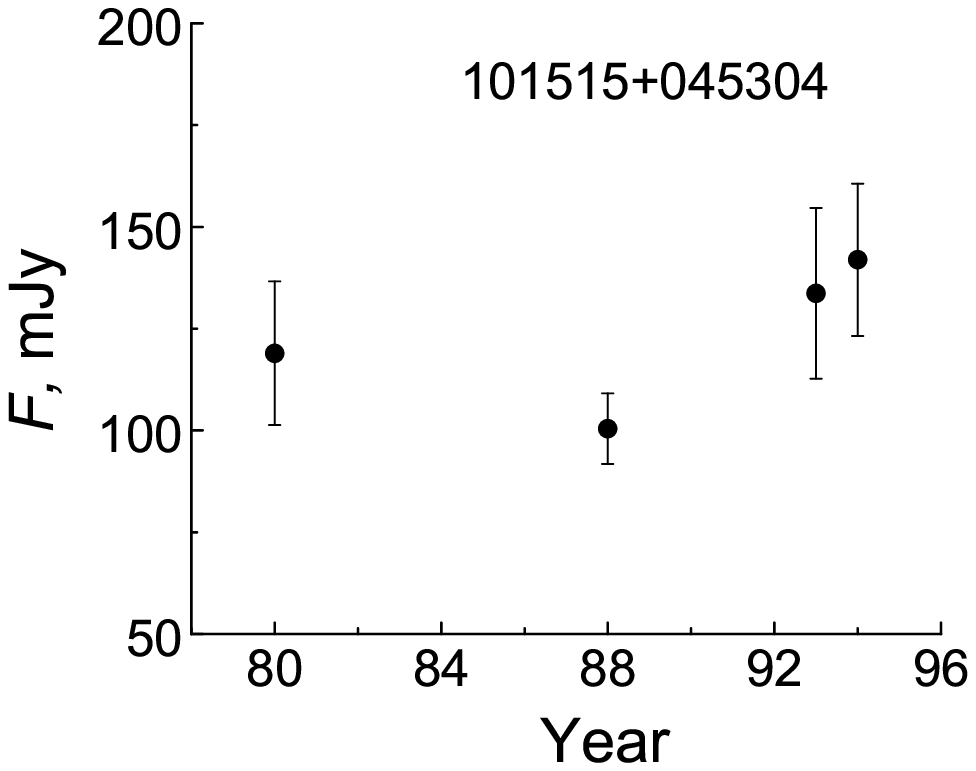}
\includegraphics[angle=0,width=0.37\textwidth,clip]{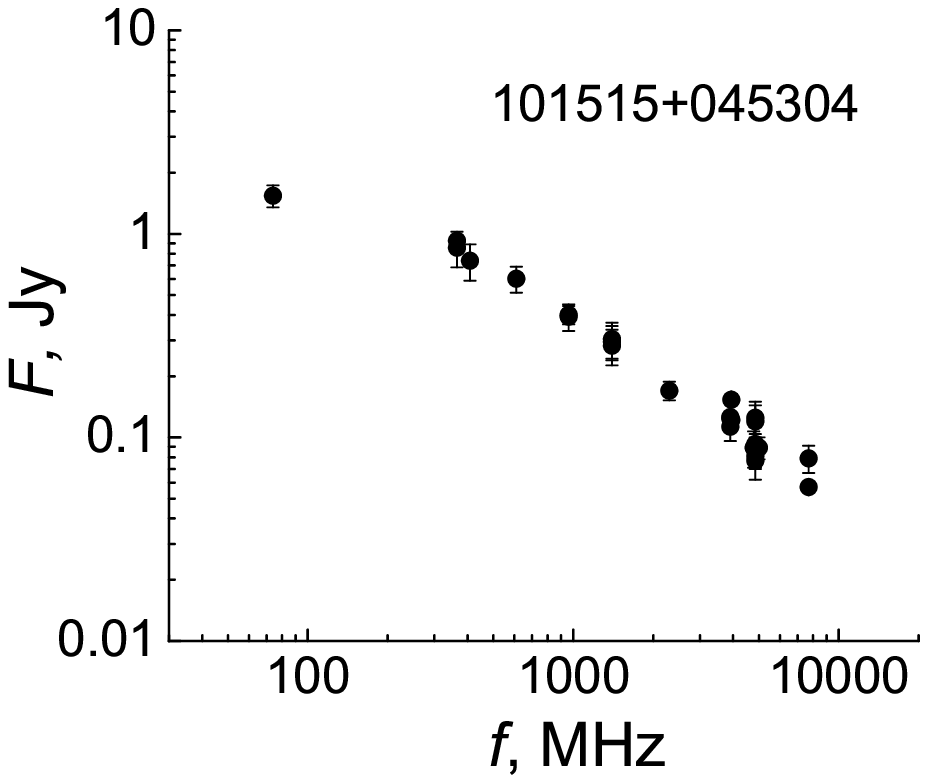}
} \hbox{
\includegraphics[angle=0,width=0.37\textwidth,clip]{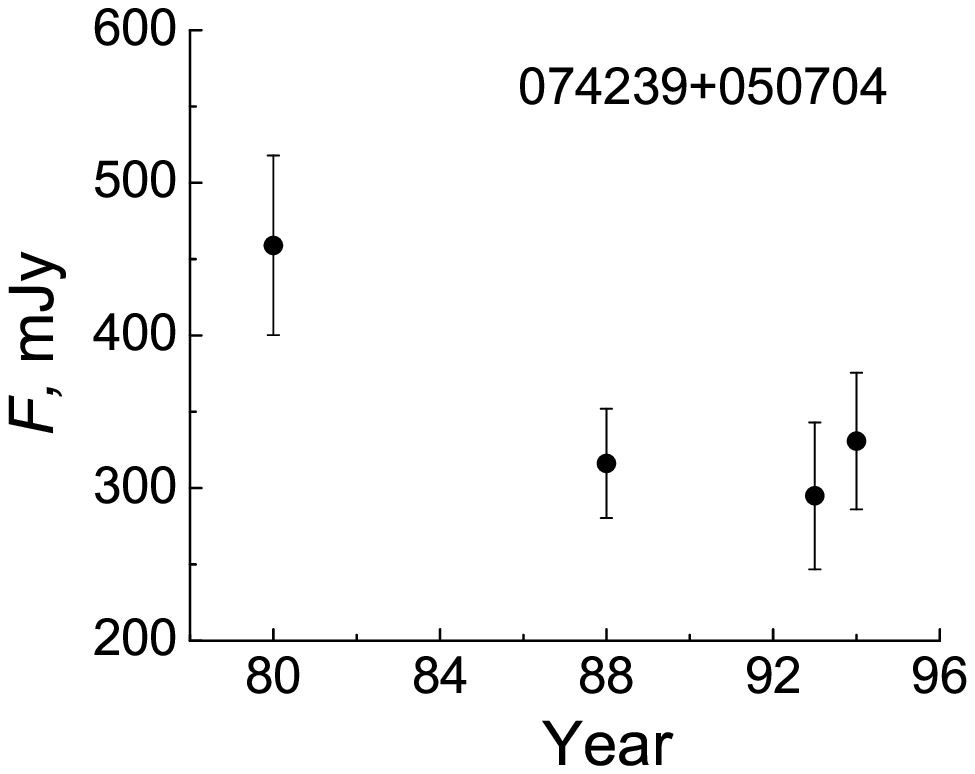}
\includegraphics[angle=0,width=0.37\textwidth,clip]{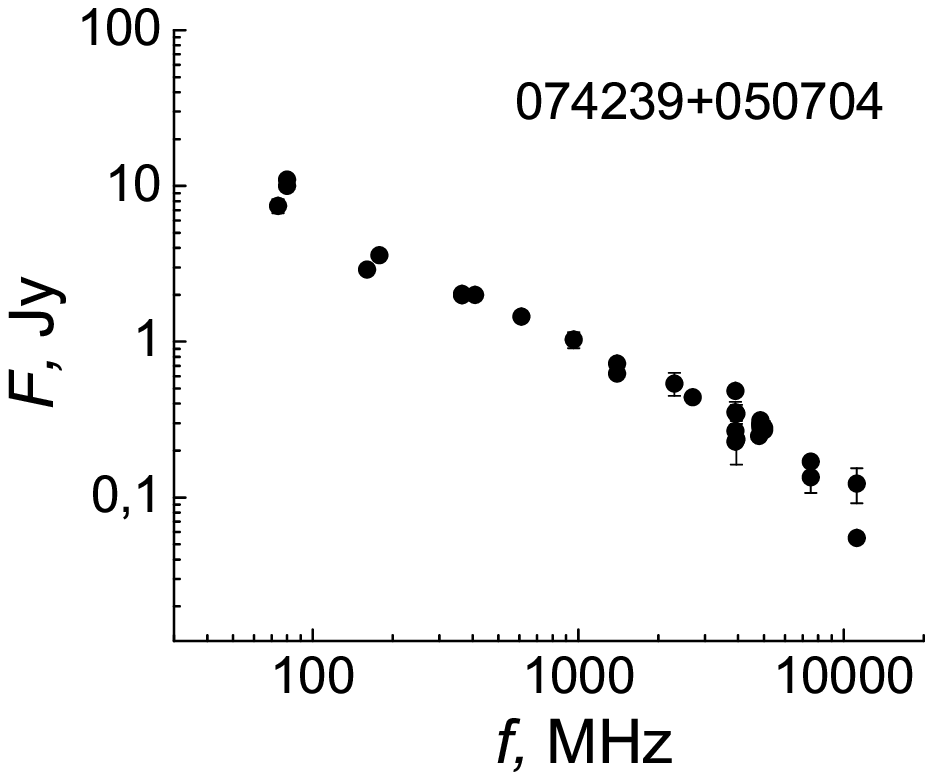}
} \hbox{
\includegraphics[angle=0,width=0.37\textwidth,clip]{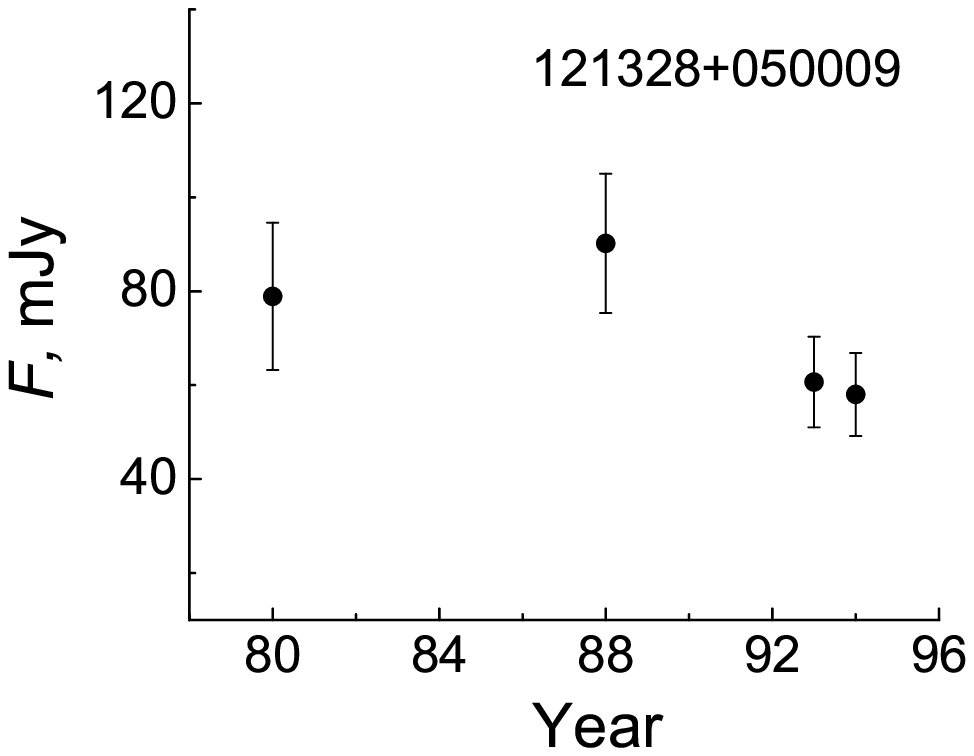}
\includegraphics[angle=0,width=0.37\textwidth,clip]{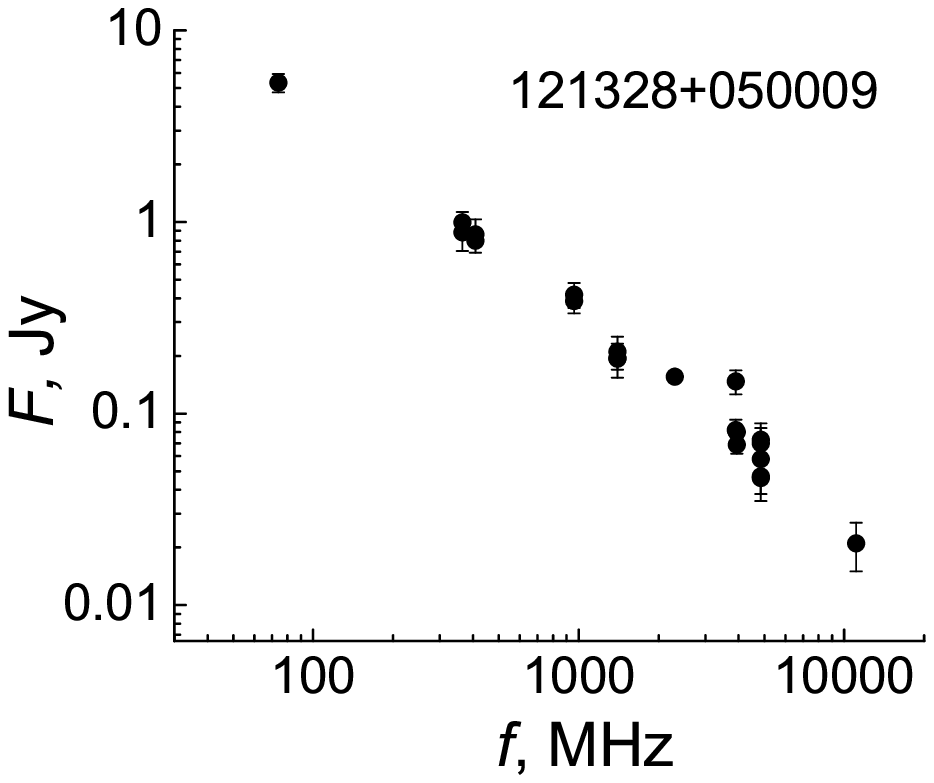}
} \hbox{
\includegraphics[angle=0,width=0.37\textwidth,clip]{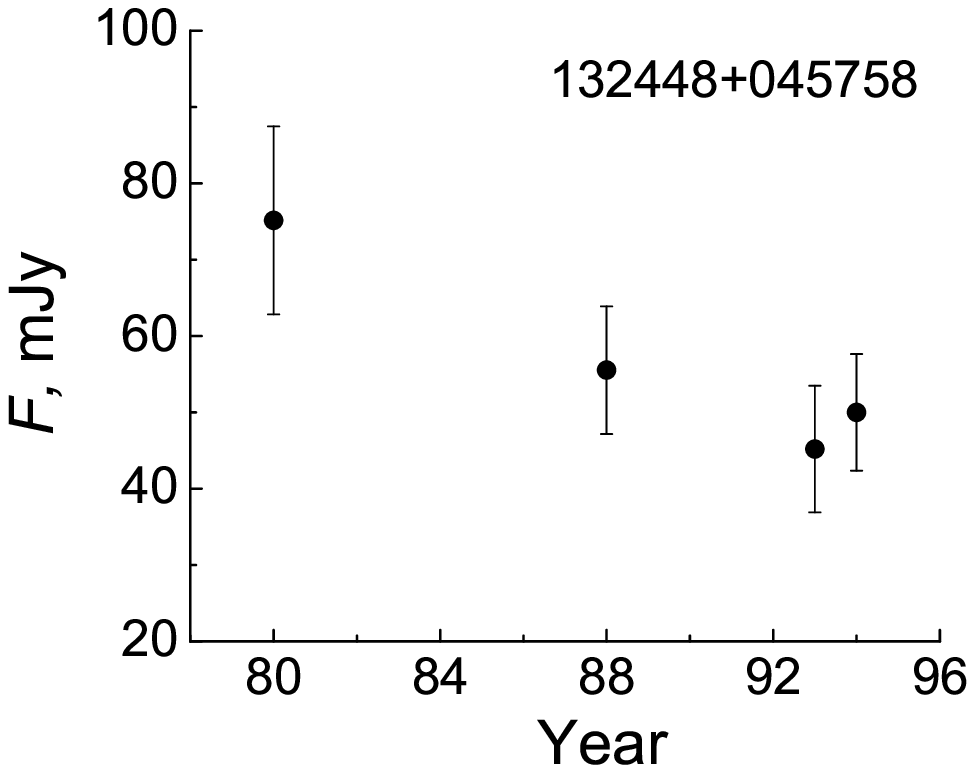}
\includegraphics[angle=0,width=0.37\textwidth,clip]{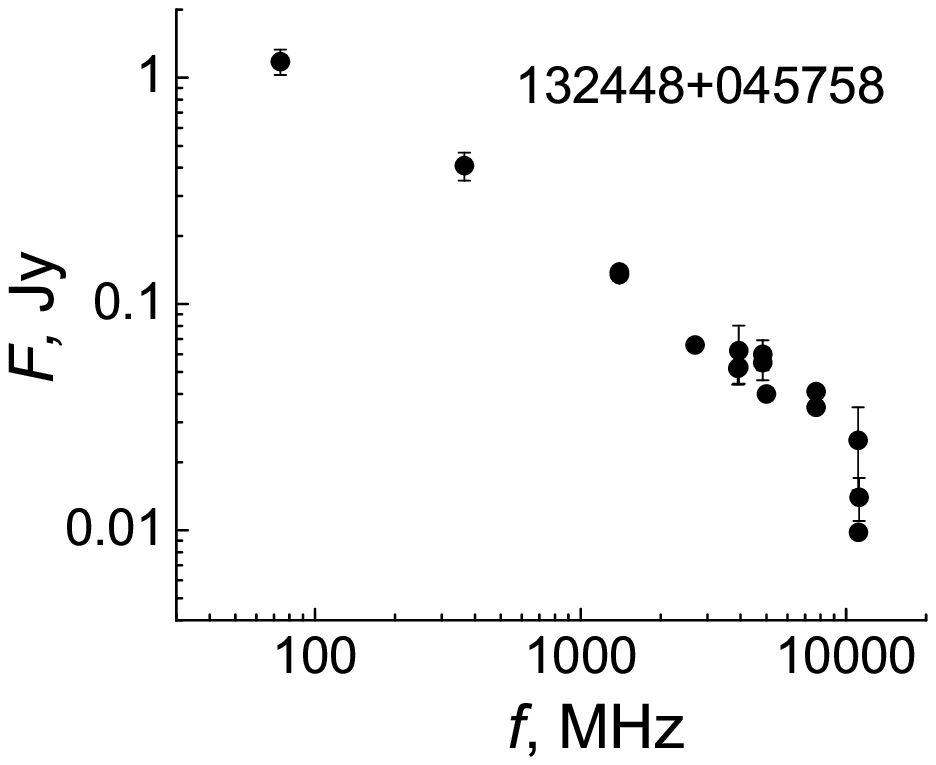}
} } } \setcaptionmargin{0mm} \captionstyle{normal} \caption{Same
as Fig.~\ref{fig10:Majorova_n} for the sources with $0.75 <
p(\chi^{2}) < 0.89$. } \label{fig12:Majorova_n}
\end{figure*}

\begin{figure*}[tbp]
\onelinecaptionstrue \centerline{ \vbox{ \hbox{
\includegraphics[angle=0,width=0.37\textwidth,clip]{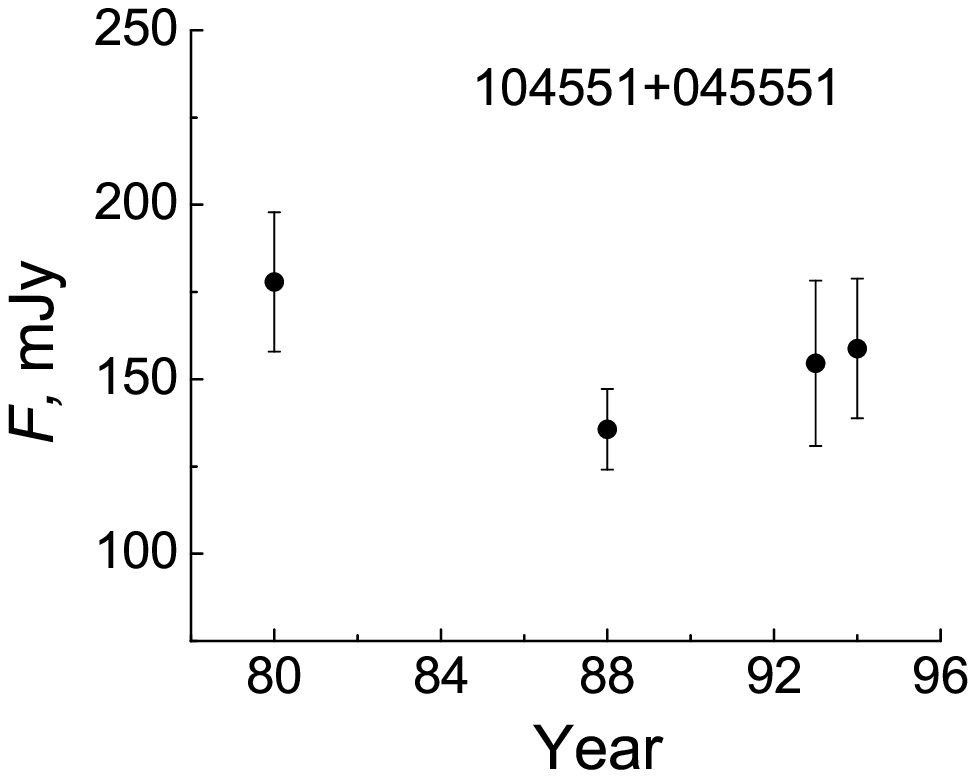}
\includegraphics[angle=0,width=0.37\textwidth,clip]{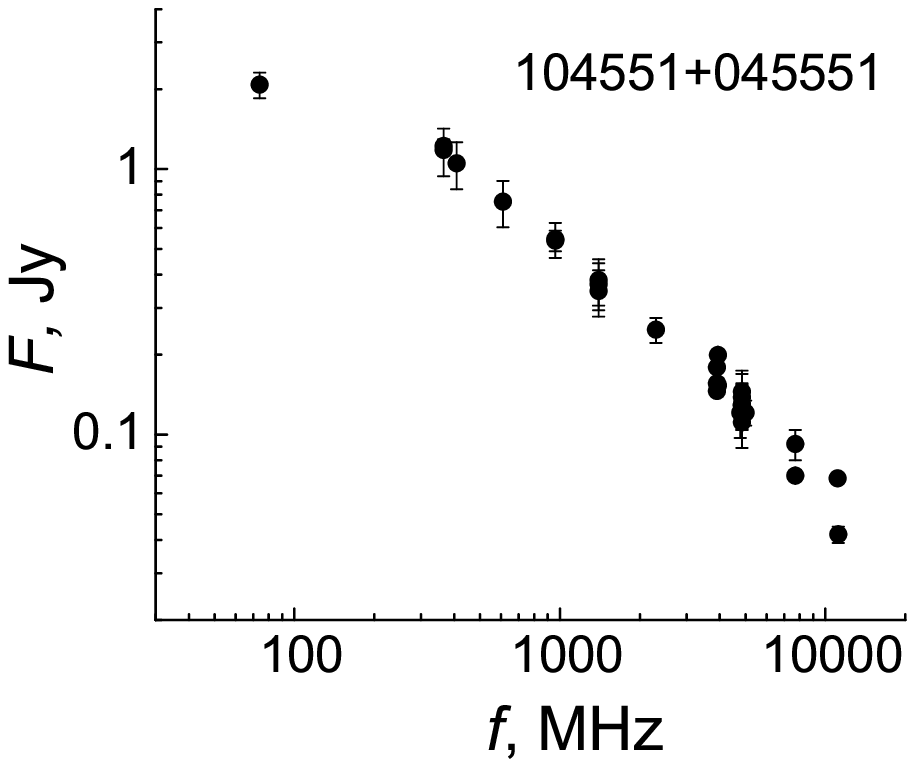}
} \hbox{
\includegraphics[angle=0,width=0.37\textwidth,clip]{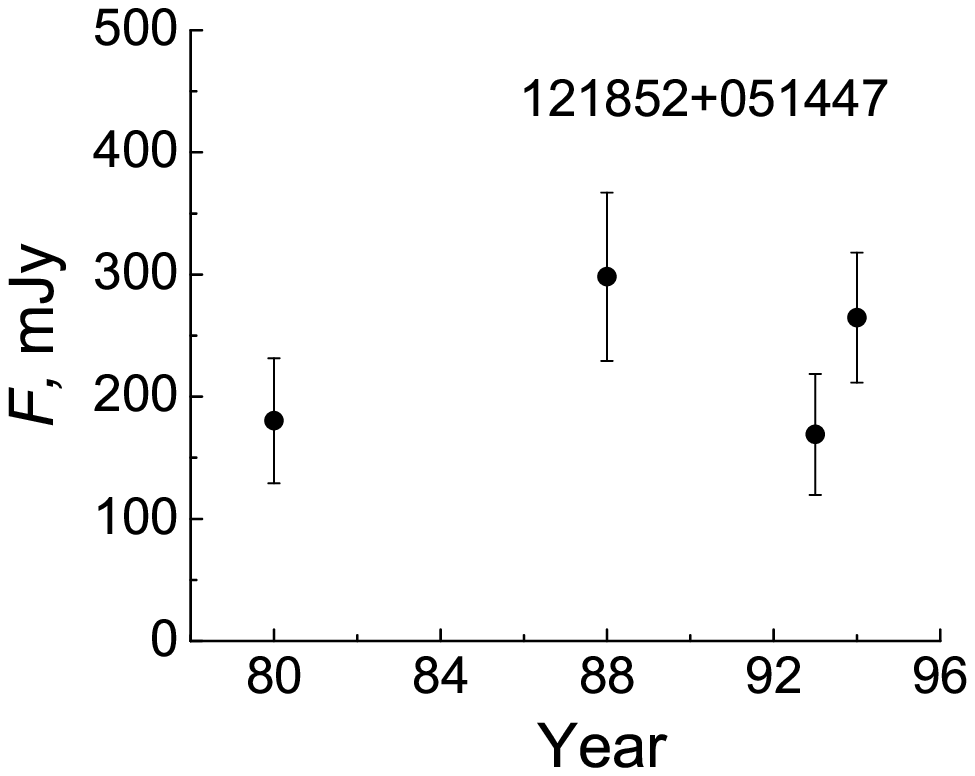}
\includegraphics[angle=0,width=0.37\textwidth,clip]{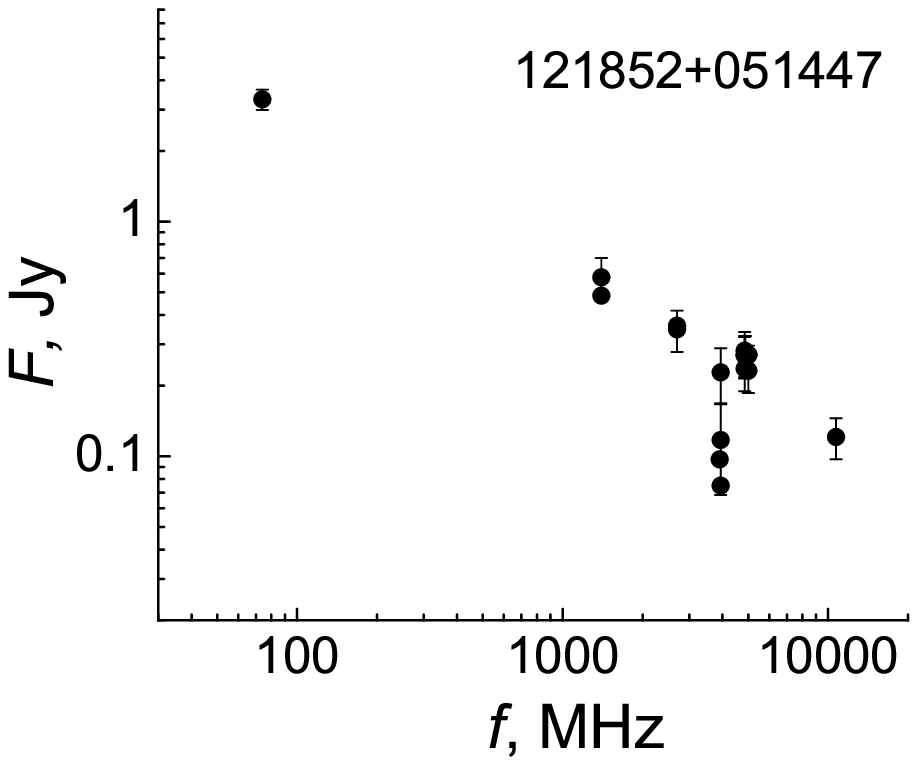}
} \hbox{
\includegraphics[angle=0,width=0.37\textwidth,clip]{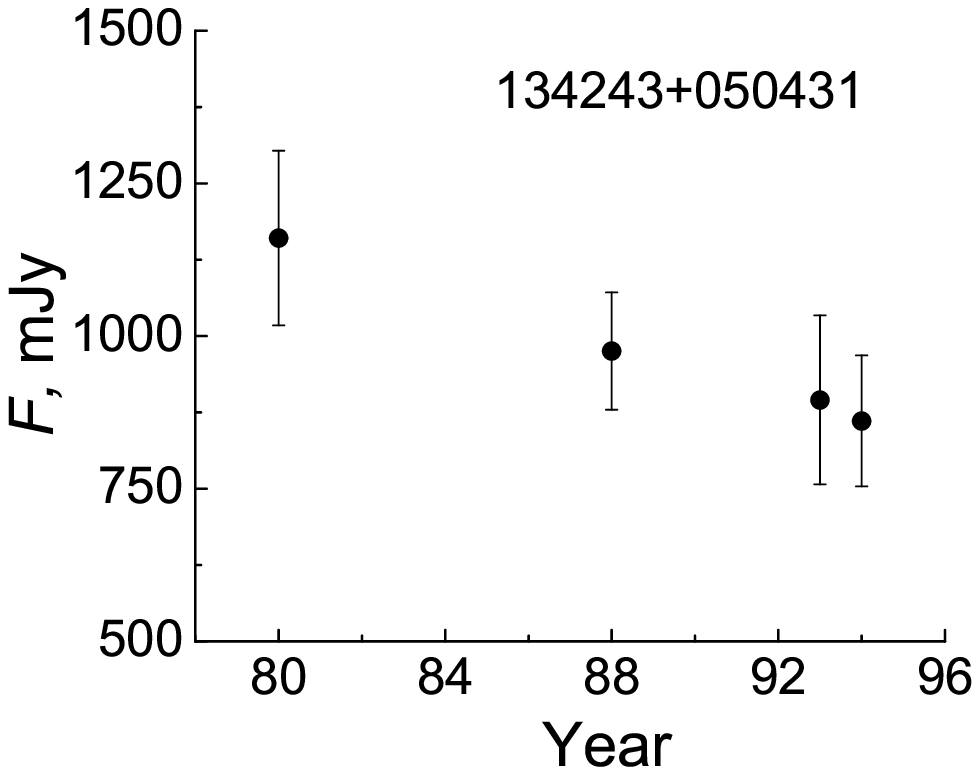}
\includegraphics[angle=0,width=0.37\textwidth,clip]{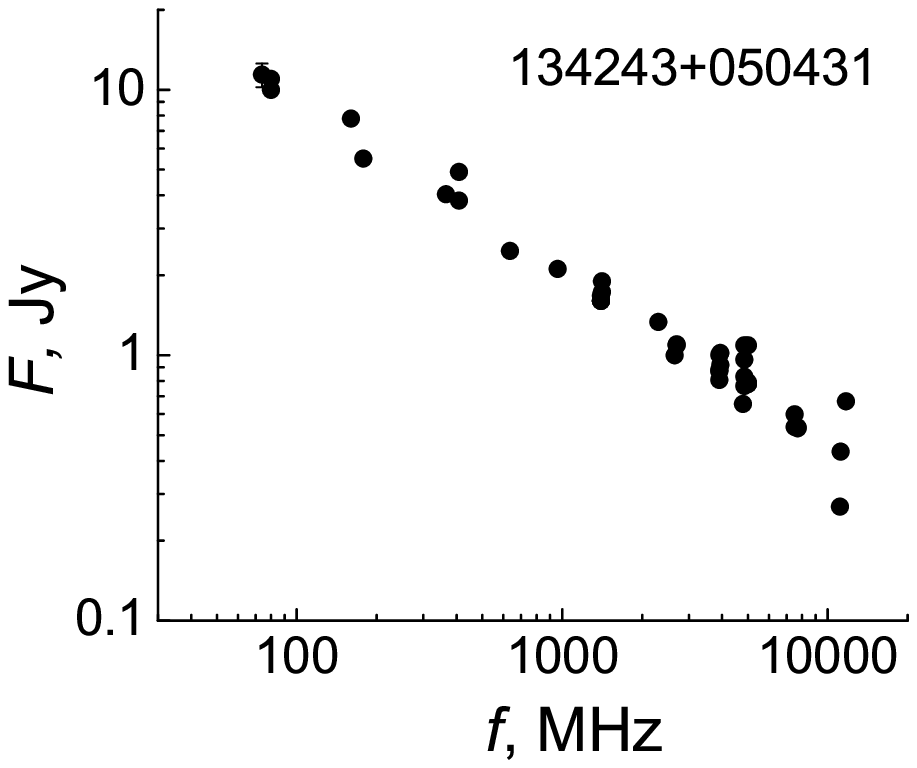}
} \hbox{
\includegraphics[angle=0,width=0.37\textwidth,clip]{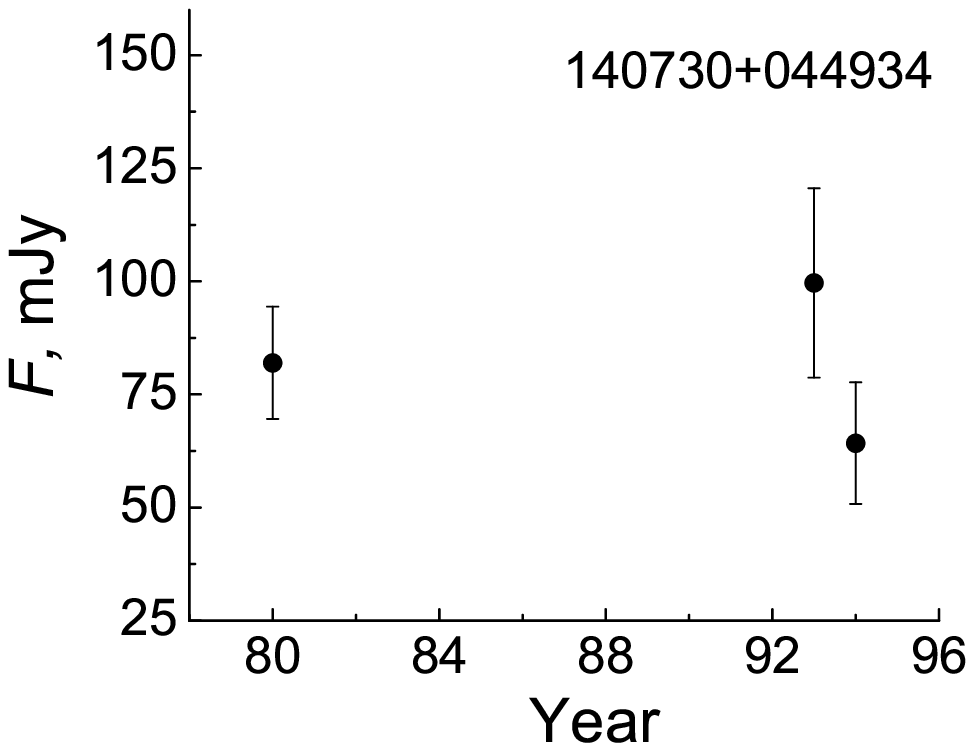}
\includegraphics[angle=0,width=0.37\textwidth,clip]{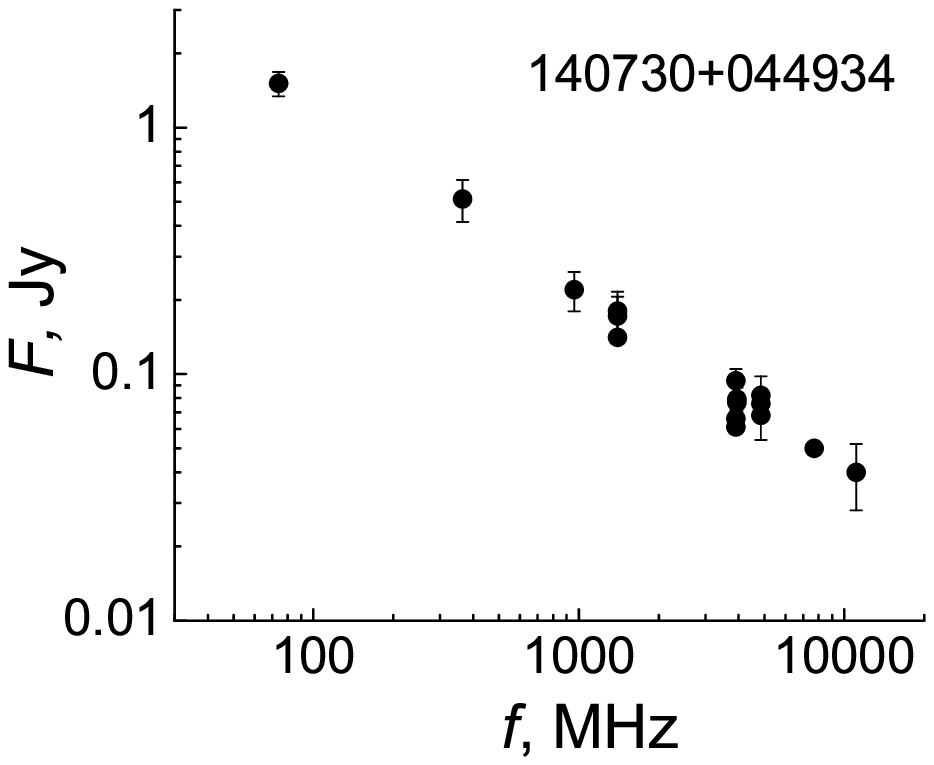}
} } } \setcaptionmargin{0mm} \captionstyle{normal} \caption{ Same
as Fig.~\ref{fig10:Majorova_n} for the sources with $ 0.6 <
p(\chi^{2}) < 0.7$. } \label{fig13:Majorova_n}
\end{figure*}

Figures~\ref{fig10:Majorova_n}--\ref{fig13:Majorova_n} show the
light curves (left panels) and spectra (right panels) of the
calibration sources from Tables~3 and 4:
Fig.~\ref{fig10:Majorova_n} shows the light curves and spectra for
objects with $p > 0.95$, Fig.~\ref{fig11:Majorova_n}---for objects
with $ 0.89 < p < 0.95$, Fig.~\ref{fig12:Majorova_n}---for objects
with $ 0.75 < p < 0.89$, and Fig.~\ref{fig13:Majorova_n}---for
objects with $ 0.6 < p < 0.7$.

\begin{figure*}[tbp]
\onelinecaptionstrue \centerline{ \vbox{ \hbox{
\includegraphics[angle=0,width=0.37\textwidth,clip]{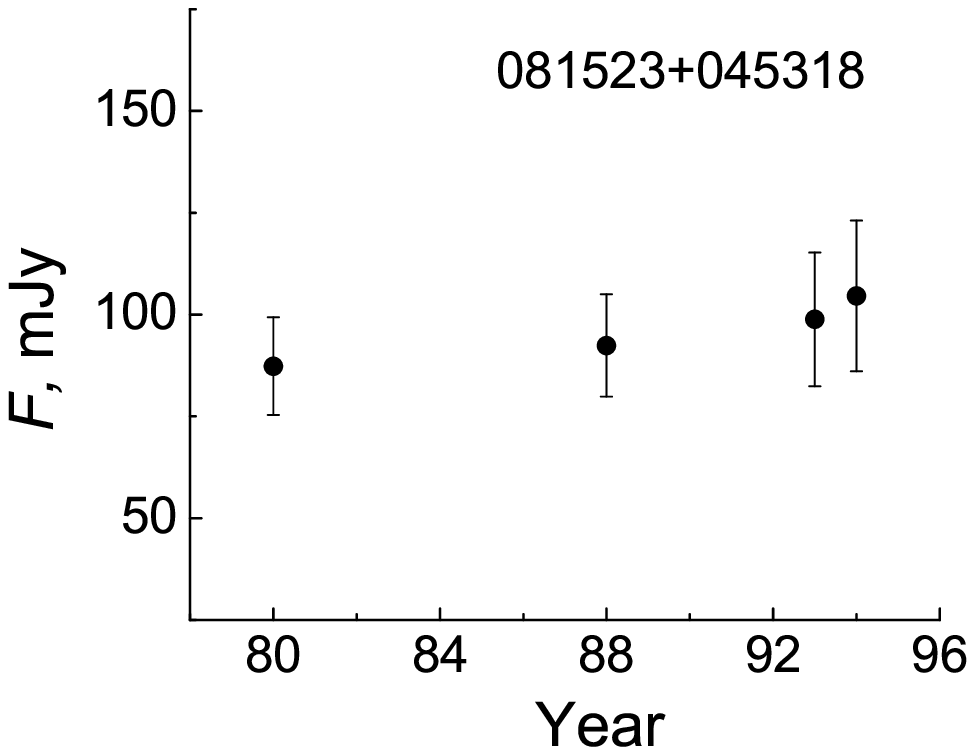}
\includegraphics[angle=0,width=0.37\textwidth,clip]{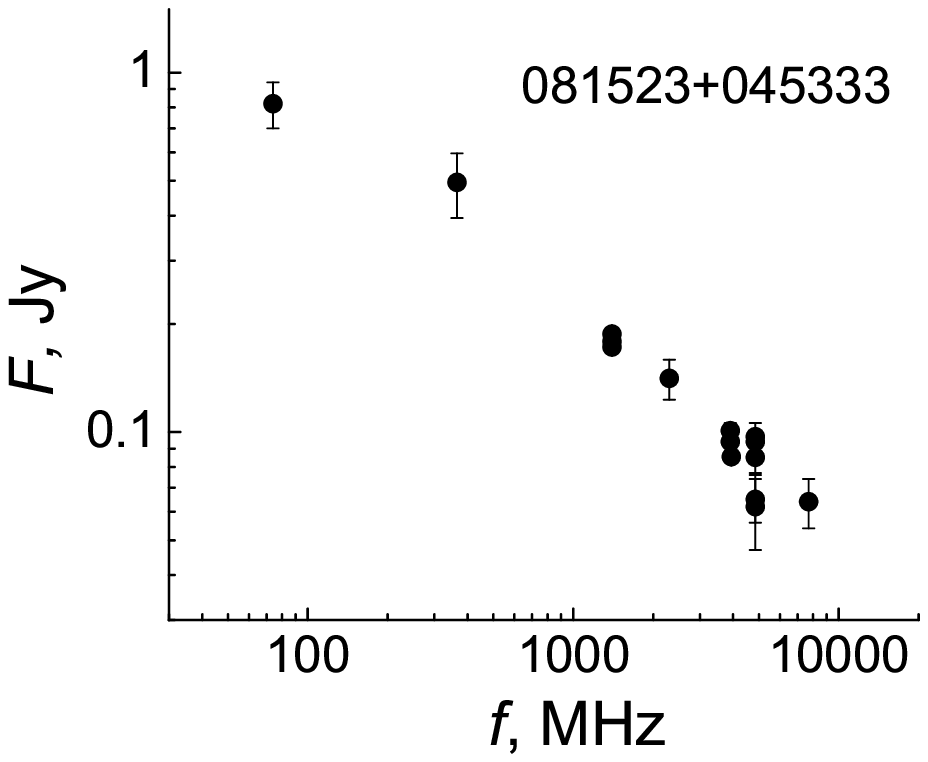}
} \hbox{
\includegraphics[angle=0,width=0.37\textwidth,clip]{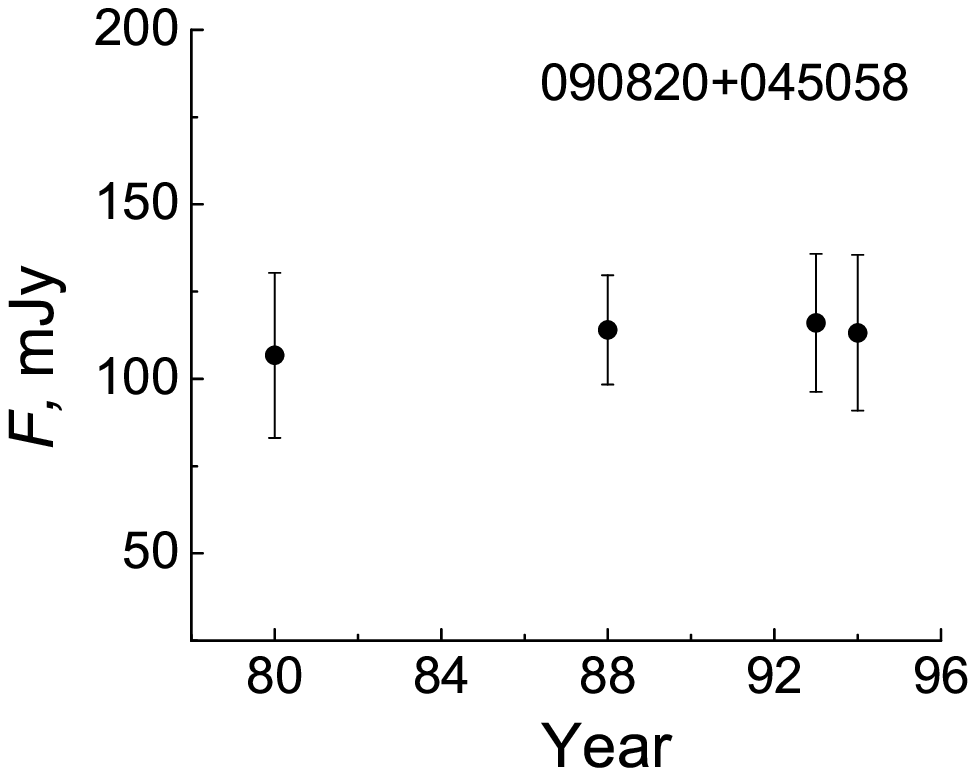}
\includegraphics[angle=0,width=0.37\textwidth,clip]{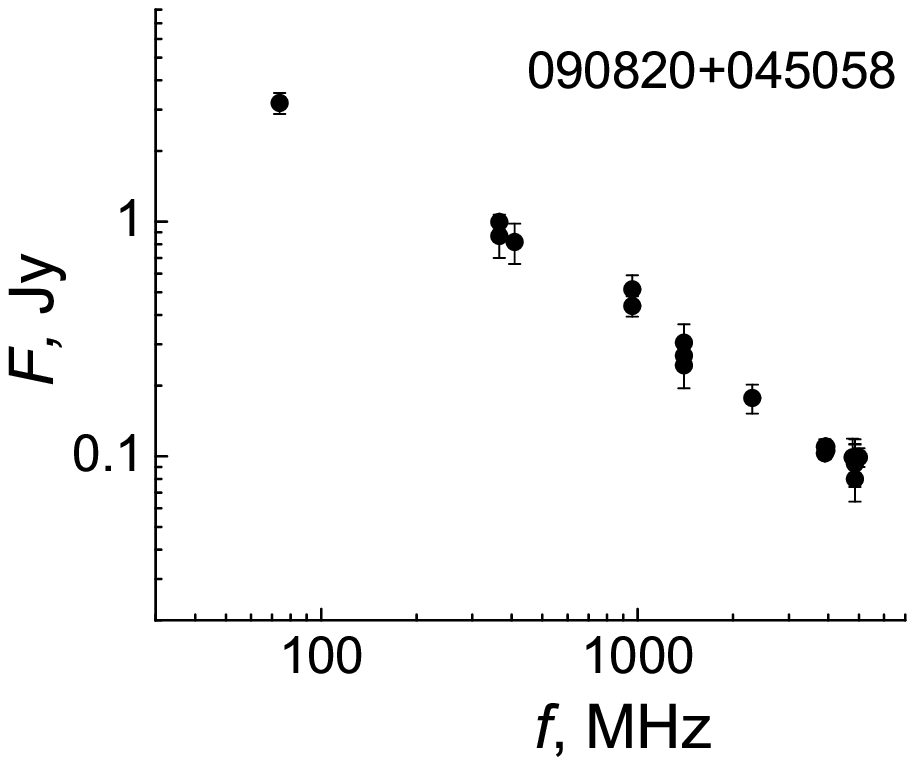}
} \hbox{
\includegraphics[angle=0,width=0.37\textwidth,clip]{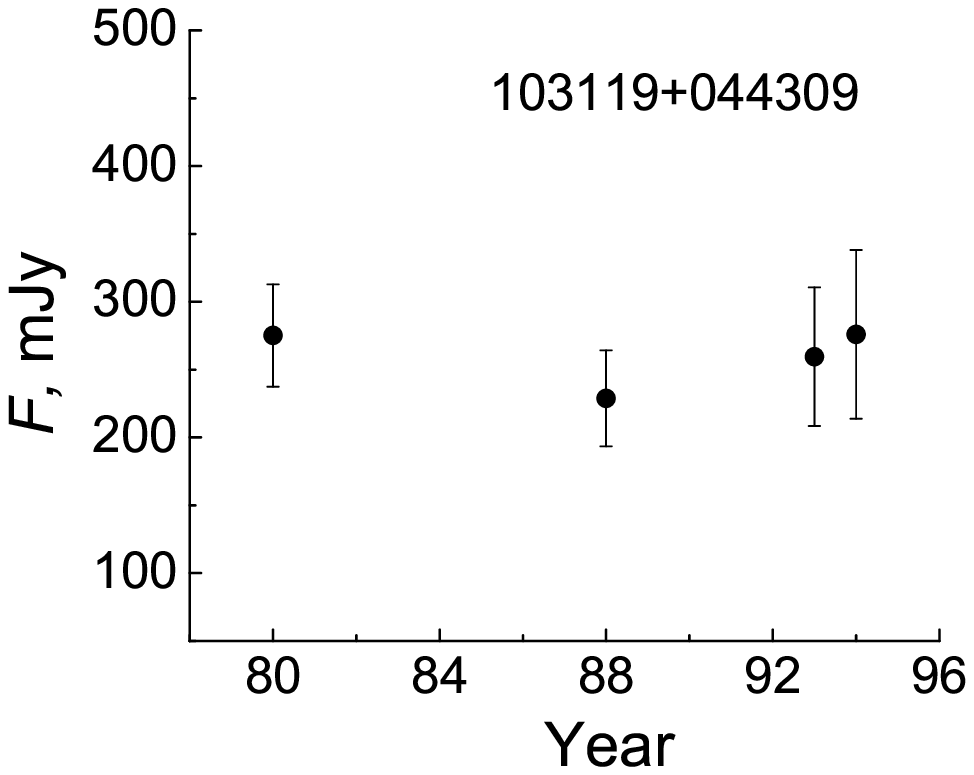}
\includegraphics[angle=0,width=0.37\textwidth,clip]{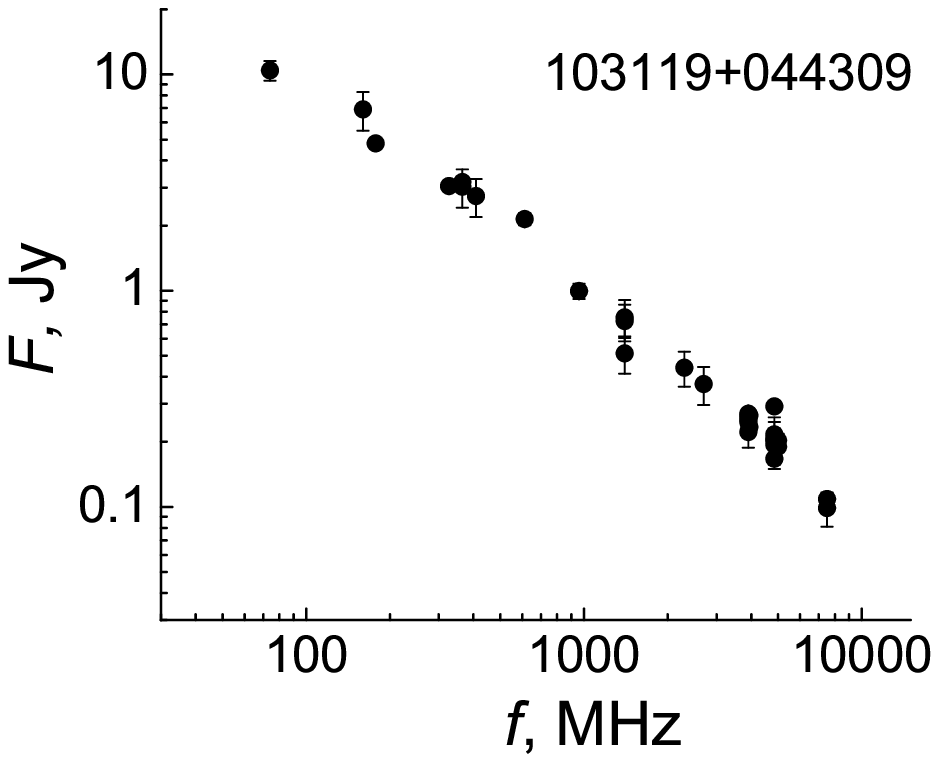}
} } } \setcaptionmargin{0mm} \captionstyle{normal} \caption{ Same
as Fig.~\ref{fig10:Majorova_n}, but for ``non-variable'' objects
with long-term variability indices $V < 0$. }
\label{fig14:Majorova_n}
\end{figure*}

\begin{figure*}[tbp]
\onelinecaptionstrue \centerline{ \vbox{ \hbox{
\includegraphics[angle=0,width=0.37\textwidth,clip]{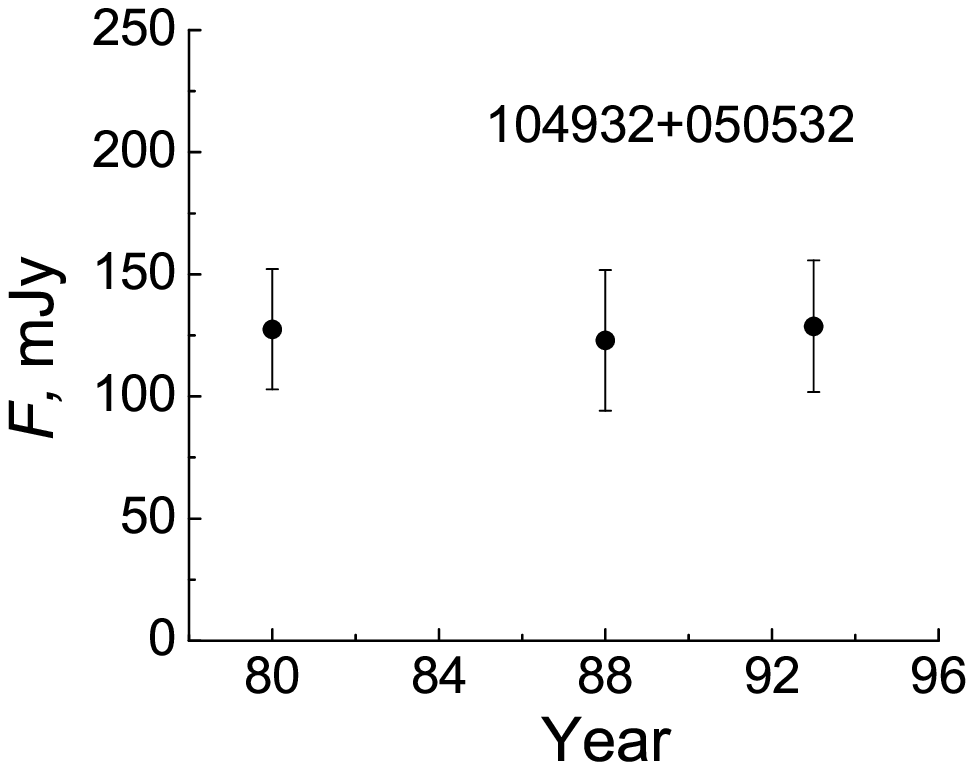}
\includegraphics[angle=0,width=0.37\textwidth,clip]{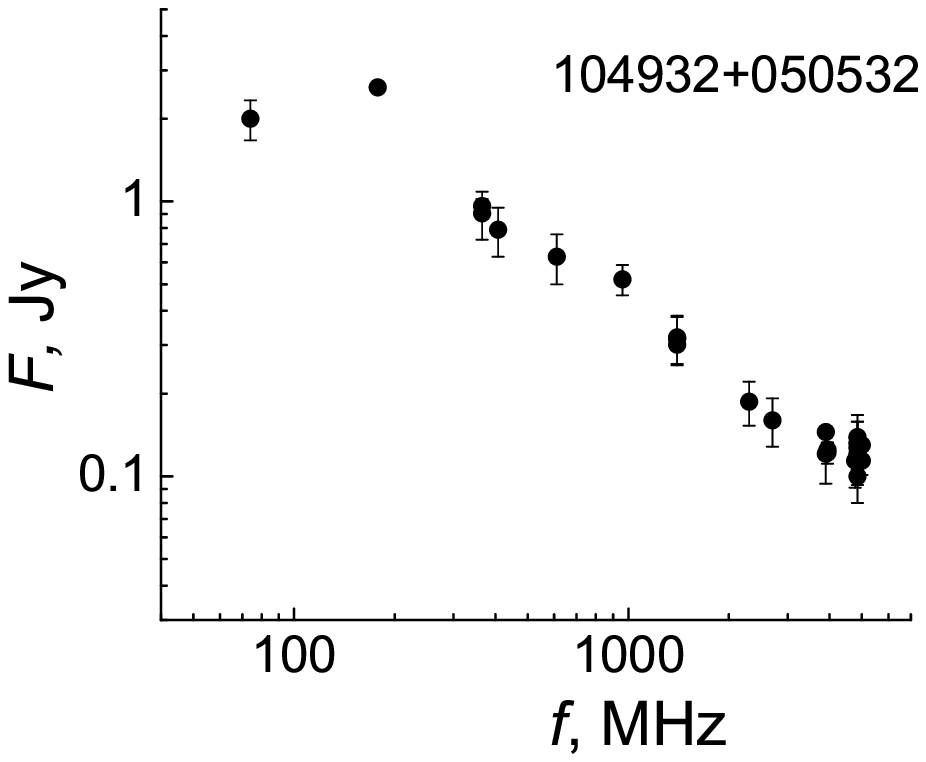}
} \hbox{
\includegraphics[angle=0,width=0.37\textwidth,clip]{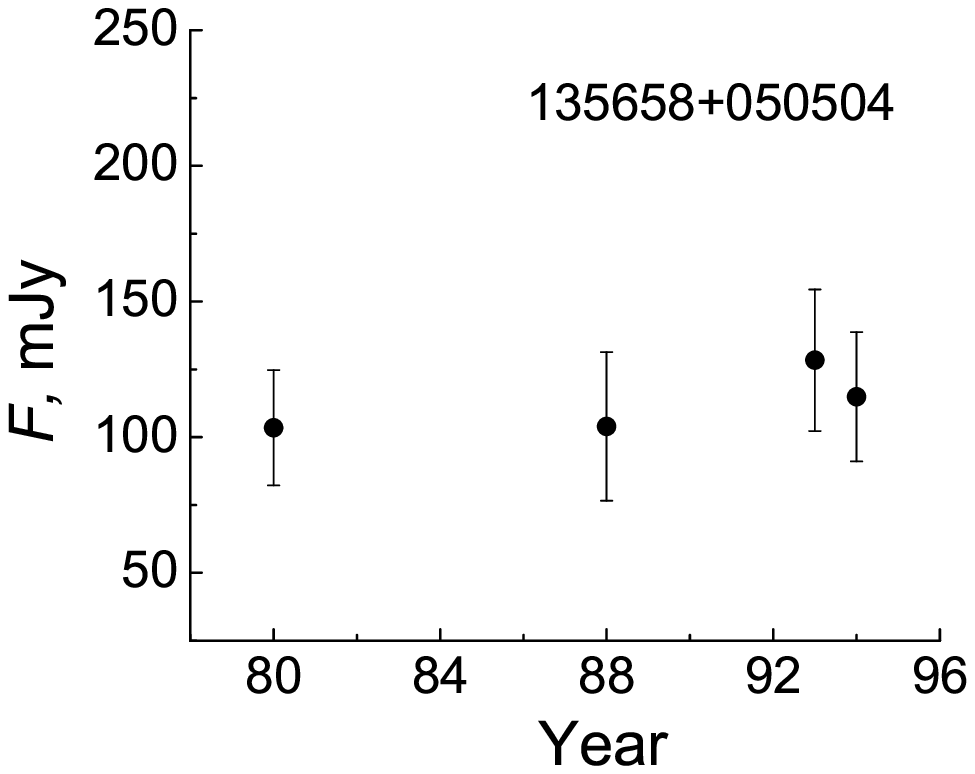}
\includegraphics[angle=0,width=0.37\textwidth,clip]{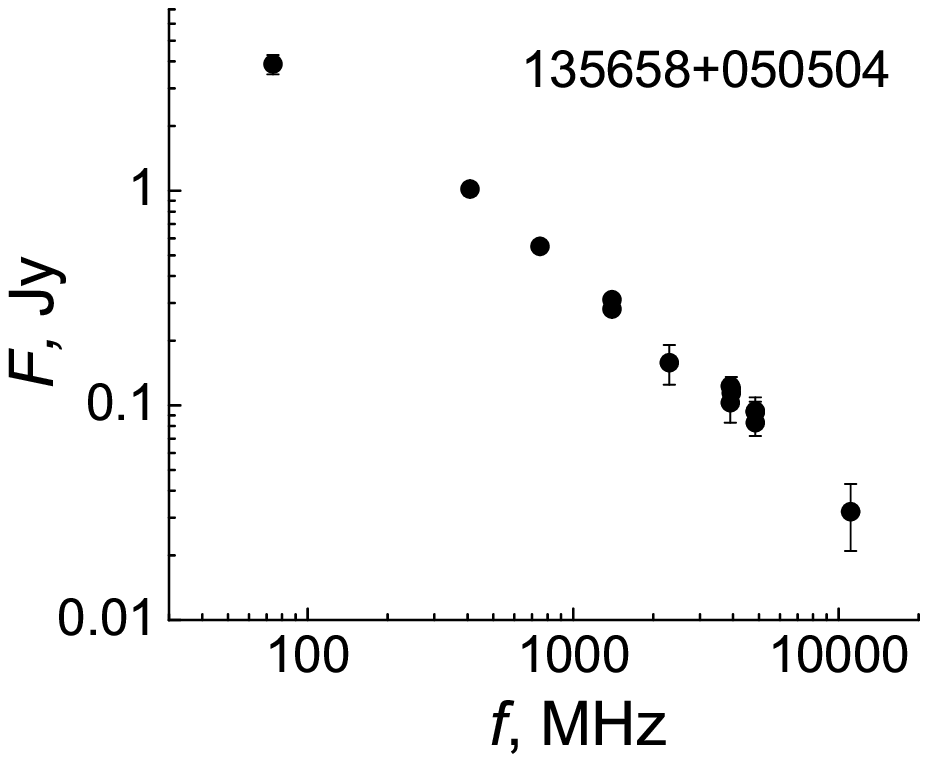} } \hbox{
\includegraphics[angle=0,width=0.37\textwidth,clip]{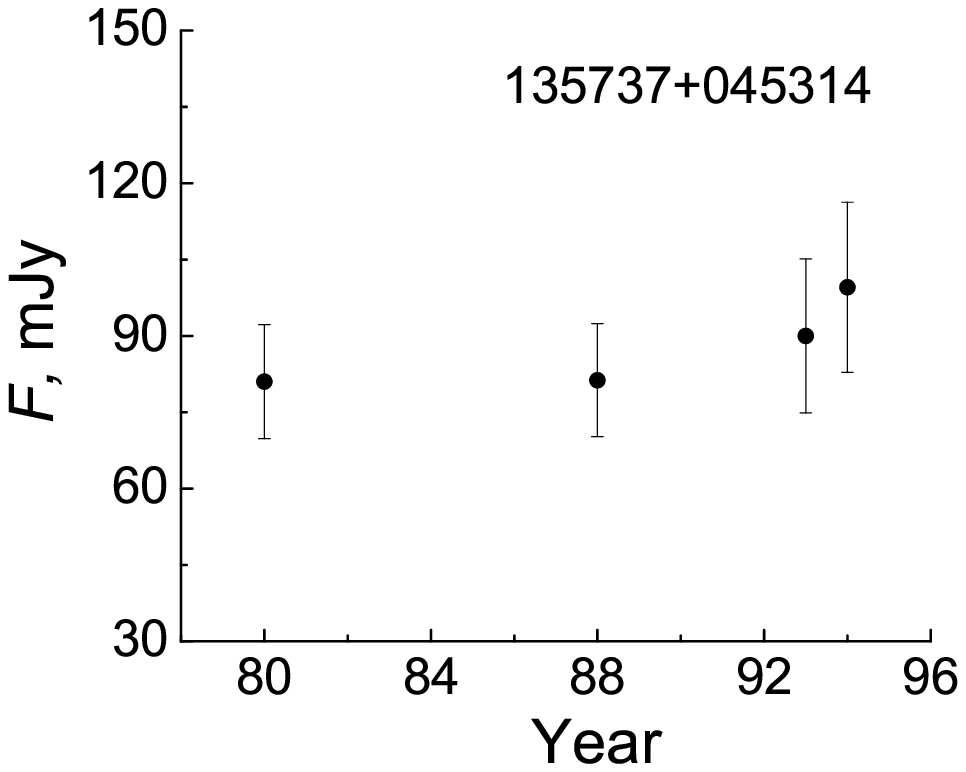}
\includegraphics[angle=0,width=0.37\textwidth,clip]{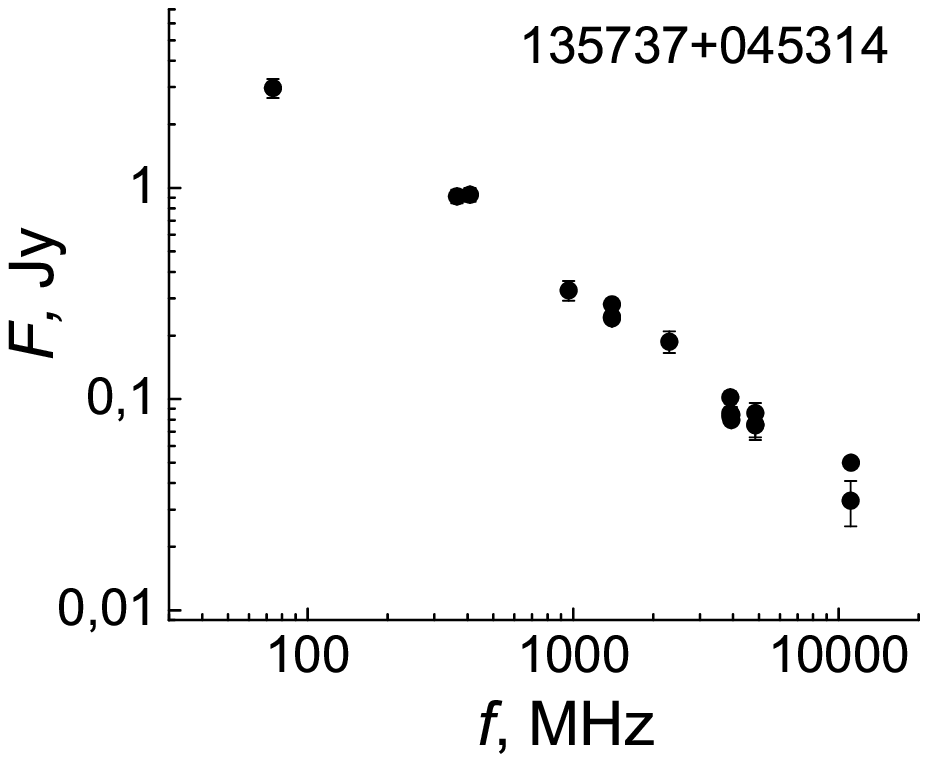}
} } } \setcaptionmargin{0mm} \captionstyle{normal} \caption{ Same
as Fig.~\ref{fig14:Majorova_n}. } \label{fig15:Majorova_n}
\end{figure*}

Figures~\ref{fig14:Majorova_n} and \ref{fig15:Majorova_n} show,
for comparison, the light curves (left panel) and spectra (right
panel) of ``nonvariable'' sources with long-term variability
indices $V < 0$.

\section{CONCLUSIONS}

To find variable sources in the data of the deep surveys carried
out on the RATAN-600 radio telescope in  1980--1994, we performed
a more thorough selection of calibration sources, constructed the
experimental dependences $F/T_{a}$ and the computed calibration
curves, performed a detailed analysis and estimated the relative
standard errors for each  survey.

To test the calibration sources for variability, we performed
quantitative estimates of the parameters that characterize the
variability of objects (our main parameter was the long-term
variability index $V$) and analyzed the statistical properties of
suspected variable objects.

Out of the entire sample of calibration sources  (about 80
objects) 14 had positive long-term variability indices for at
least one pair of surveys. Eight of these sources have the
long-term variability indices $V > 0.08$ and for 10 sources the
maximum flux densities exceed their minimum flux densities by more
than a factor of 1.5.

We estimated the $\chi^2$ variability probabilities $p$ and the
parameter $V_{\chi}$ that characterizes the relative amplitude of
variability for these 14 sources.

We found three objects to be the most likely \linebreak variable
source candidates: J\,155148+045930 \linebreak (\mbox{$p=0.961$}),
J\,135137+043542
 (\mbox{$p=0.960$}), and \linebreak J\,103938+051031 (\mbox{$p=0.984$}).

The  J\,103938+051031 source also meets the variability condition
according to the \mbox{$V_{F} > 3\sigma$} criterion, where
\mbox{$\sigma=\sqrt{\sigma_{i}^2+\sigma_{j}^2}$} ($\sigma_{i}$ and
$\sigma_{j}$ are the root mean square errors of the source flux
density in the $i$-th and $j$-th surveys). Seven sources have
\mbox{$2\sigma < V_{F} < 3\sigma$}, and the probability of
variability \mbox{$p(\chi^2)>0.85$}. The remaining six objects of
this sample have \mbox{$0.6 < p < 0.8$}.

Note, however, that two of the three most likely candidate
variable sources, J\,103938+051031 and J\,135137+043542, in one of
the surveys pass rather far from its central section ($\mid dH
\mid > 18^{'}$). Although they show up conspicuously in the
records, their flux densities are more difficult to determine than
those of the sources that are close to the central section,
primarily because they are extended features.

If we employ the more stringent criteria used by Kesteven et
al.~\mbox{\cite{kest:Majorova_n}} and Fanti et
al.~\mbox{\cite{fan:Majorova_n}}, namely that only the sources
with $p > 0.99$ and $p > 0.999$ can be considered variable and
reliably variable, respectively, then there are no variable
sources in our sample.

Nine out of 14 sources show a scatter of magnitudes in close
filters ranging from $0\fm8$ to $3^{\rm m}$ according to the data
of the GSC and USNO-B1 catalogs and the 2MASS, SDSS, and
LAS~UKIDSS surveys, which is indicative of optical variability.

The estimates of relative standard deviations of flux densities
from their mean values averaged over all the surveys,
$RMS^{\,set}$, for the subsamples with $V > 0$ and $V < 0$ showed
that they differ significantly. The $RMS^{\,set}$ values for
suspected variable sources and for ``non-variable'' sources,
averaged over the entire sample, are equal to
\mbox{$RMS^{\,set}=0.23\pm0.07$} and
\mbox{$RMS^{\,set}=0.08\pm0.04$}, respectively. This leads us to
conclude that the flux densities of the overwhelming majority of
calibration sources varied only slightly from one survey to
another, and that the flux density errors, on average, did not
exceed 10\%.

The calibrating curves and estimates of the relative standard
errors of the inferred flux densities obtained in this study will
make it possible to search for variable sources among a bigger
sample of objects observed in different surveys, as we plan to do
in our forthcoming papers.

\begin{acknowledgments}

This study is our tribute and gratitude to Natalya Sergeevna
Soboleva, her longstanding and fundamental work on deep search
surveys performed on the RATAN-600 radio telescope. This work was
supported in part by the Russian Foundation for Basic Research
(grants nos.~ 11-02-12036, 11-02-00489 and \mbox{10-07-00412}) and
the Ministry of Education and Science of the Russian Federation
(state contracts nos.~16.552.11.7028 and 16.518.11.7062). In this
research we used the VizieR catalogue access tool and the SIMBAD
database, operated at CDS, Strasbourg, France, as well as the
NASA/IPAC Extragalactic Database (NED), operated by the Jet
Propulsion Laboratory, California Institute of Technology, under
the contract with the National Aeronautics and Space
Administration.

\end{acknowledgments}

\end{document}